\documentclass[english,10pt]{article}

\usepackage[english]{babel}
\usepackage[utf8]{inputenc}
\usepackage[T1]{fontenc}

\usepackage{geometry}
\usepackage{graphicx}
\usepackage[colorinlistoftodos]{todonotes}
\usepackage[colorlinks=true, allcolors=blue]{hyperref}
\usepackage{setspace}
\usepackage{enumitem}
\usepackage{titlesec}
\usepackage{comment}
\usepackage{caption}
\usepackage{subcaption}
\usepackage{longtable}
\usepackage[section]{placeins}

\usepackage{hyperref}
\usepackage[
backend=biber,
style=numeric,
citestyle=numeric-comp,
firstinits=true,
eprint=false,
doi=false,
maxbibnames=4,
sorting=none
]{biblatex}
\usepackage{amsmath}
\usepackage{amsthm}
\usepackage{amssymb}
\usepackage{amsfonts}
\usepackage{bbm}
\usepackage{bm}
\usepackage{nicefrac}
\usepackage{mathtools}

\usepackage{tikz}
\usetikzlibrary{shapes}
\usetikzlibrary{fit}
\usetikzlibrary{chains}
\usetikzlibrary{arrows}

\tikzstyle{latent} = [circle,fill=white,draw=black,inner sep=1pt,
minimum size=20pt, font=\fontsize{10}{10}\selectfont, node distance=1]
\tikzstyle{obs} = [latent,fill=gray!25]
\tikzstyle{const} = [rectangle, inner sep=0pt, node distance=1]
\tikzstyle{factor} = [rectangle, fill=black,minimum size=5pt, inner
sep=0pt, node distance=0.4]
\tikzstyle{det} = [latent, diamond]

\tikzstyle{plate} = [draw, rectangle, rounded corners, fit=#1]
\tikzstyle{wrap} = [inner sep=0pt, fit=#1]
\tikzstyle{gate} = [draw, rectangle, dashed, fit=#1]

\tikzstyle{caption} = [font=\footnotesize, node distance=0] %
\tikzstyle{plate caption} = [caption, node distance=0, inner sep=0pt,
below left=5pt and 0pt of #1.south east] %
\tikzstyle{factor caption} = [caption] %
\tikzstyle{every label} += [caption] %

\tikzset{>={triangle 45}}

\newcommand{\edge}[3][]{ %
  \foreach \x in {#2} { %
    \foreach \y in {#3} { %
      \path (\x) edge [->,#1] (\y) ;%
    } ;
  } ;
}

\newcommand{\plate}[4][]{ %
  \node[wrap=#3] (#2-wrap) {}; %
  \node[plate caption=#2-wrap] (#2-caption) {#4}; %
  \node[plate=(#2-wrap)(#2-caption), #1] (#2) {}; %
}

\usepackage{subfiles}
\usepackage{subcaption}

\usepackage{graphicx}
\usepackage{fullpage}
\usepackage{float}
\usepackage{wrapfig}
\usepackage{caption}

\usepackage{algpseudocode}
\usepackage{algorithm}
\usepackage{listings}

\usepackage{amsthm}\usepackage{dsfont}\usepackage{array}\usepackage{mathrsfs}\usepackage{comment}\usepackage{mathrsfs}

\makeatother

\usepackage{babel}
\usepackage{color}
\usepackage{centernot}
\usepackage{hyperref}
\usepackage{babel}
\usepackage{sansmath}

\definecolor{ejc}{RGB}{255,0,0}
\definecolor{ms}{RGB}{0,100,0}

\newcommand{\indep}{\rotatebox[origin=c]{90}{$\models$}}

\newcommand*{\defeq}{\mathrel{\vcenter{\baselineskip0.5ex \lineskiplimit0pt
      \hbox{\scriptsize.}\hbox{\scriptsize.}}}%
  =}

\newcommand{\E}[1]{\mathbb{E}\left[#1\right]}

\renewcommand{\P}[1]{\mathbb{P}\left[#1\right]}

\newcommand{\Pc}[2]{\mathbb{P}\left[\left.#1\right|#2\right]}

\newtheorem{theorem}{Theorem}

\newtheorem{prop}{Proposition}

\newcommand{\reals}{\mathbb{R}}

\newcommand*{\iid}{\mathrm{i.i.d.}}%
\newcommand*{\dIid}{\overset{\iid}{\sim}}%

\allowdisplaybreaks[1]

\RequirePackage{csquotes}

\usepackage{algorithm,algpseudocode}% http://ctan.org/pkg/{algorithms,algorithmx}
\algnewcommand{\Initialize}[1]{%
  \State \textbf{Initialize:}
  \Statex \hspace*{\algorithmicindent}\parbox[t]{.8\linewidth}{\raggedright #1}
}
\algnewcommand{\Inputs}[1]{%
  \State \textbf{Inputs:}
  \Statex \hspace*{\algorithmicindent}\parbox[t]{.8\linewidth}{\raggedright #1}
}

\usepackage{listings}

\definecolor{mygreen}{rgb}{0,0.6,0}
\definecolor{codegray}{rgb}{0.98,0.98,0.98}
\definecolor{mauve}{rgb}{0.58,0,0.82}
\title{%Model-Free Knockoffs for Hidden Markov Models\\ with
Gene Hunting with Knockoffs for Hidden Markov Models}
\author{Matteo Sesia\thanks{Department of Statistics, Stanford
 University, Stanford, CA 94305, U.S.A.}
\and
Chiara Sabatti\footnotemark[1] \thanks{Department of Biomedical Data Science, Stanford
 University, Stanford, CA 94305, U.S.A.}
\and
Emmanuel J. Cand\`es\footnotemark[1] \thanks{Department of Mathematics, Stanford
 University, Stanford, CA 94305, U.S.A.}}

\begin{document}
\maketitle

\begin{abstract}

Modern scientific studies often require the identification of a subset
of relevant explanatory variables, in the attempt to understand an
interesting phenomenon. Several statistical methods have been
developed to automate this task, but only recently has the framework
of model-free knockoffs proposed a general solution that can perform
variable selection under rigorous type-I error control, without
relying on strong modeling assumptions.  In this paper, we extend the
methodology of model-free knockoffs to a rich family of problems where
the distribution of the covariates can be described by a hidden Markov
model (HMM). We develop an exact and efficient algorithm to sample
knockoff copies of an HMM. We then argue that combined with the
knockoffs selective framework, they provide a natural and powerful
tool for performing principled inference in genome-wide association
studies with guaranteed FDR control. Finally, we apply our methodology
to several datasets aimed at studying the Crohn's disease and several
continuous phenotypes, e.g.~levels of cholesterol.


\end{abstract}

\section{Introduction}

\subsection{The need for (more) controlled variable selection} \label{sec:intro_1}

The automatic selection of relevant explanatory variables is a
fundamental challenge in statistics. Its urgency is induced by the
growing reliance of many fields of science on the analysis of large
amounts of data. As researchers are striving to understand
increasingly complex phenomena, the technology of high throughput
experiments now allows them to measure and simultaneously examine
millions of covariates. However, despite the abundance of the
available variables, it is often the case that only a fraction of them
are expected to be relevant to the question of interest.
By discovering which are important, scientists can design a more targeted followup investigation and hope to eventually understand how certain factors influence an outcome. A compelling example is offered by genome-wide association studies (GWAS): here, the goal is to identify which markers of genetic variation influence the risk of a particular disease or a trait, choosing from a pool of hundreds of thousands to millions of single-nucleotide polymorphisms (SNP). \\

In general, a good selection algorithm should be able to detect as many relevant variables as possible using only a small number of samples ($n \ll p$), since these tend to be expensive to acquire. At the same time, it should be sufficiently cautious to ensure that the findings are replicable and not just report spurious correlations or associations. Several statistical techniques have been proposed in an effort to address and balance these two %seemingly
 conflicting needs.
The standard approach adopted in GWAS consists  in controlling the global error when testing a large collection of hypotheses, each probing the effect of one of the typed genetic markers on the phenotype of interest. A p-value for the null hypothesis of no association between a genetic variant and the outcome of interest is obtained using linear models (or generalized linear models for binary traits) with one fixed effect (the genotype of the variant) and possibly random effects capturing the contribution of the rest of the genome.  %separate univariate regressions of the phenotype on each genetic marker. 
 %The problem is thus fashioned into a collection of simple statistical tests, where a null hypothesis asserts that the corresponding variant has no effect on the outcome. 
 %By assuming a generalized linear model, one can then compute p-values and select relevant markers by rejecting the hypotheses that achieve a certain significance threshold. %\ejc{When we have a binary response, I suppose we can still talk about univariate regression?}
 To identify significantly associated variants, the p-values are compared to a threshold that guarantees approximate control of the family-wise error rate (FWER, i.e.~the probability of committing at least one type-I error, across all tests) at the 0.05 level (the standard choice is to perform all individual tests at level $\alpha < 5 \cdot 10^{-8}$).
% The latter is typically determined such that the family-wise error rate (FWER, i.e.~the probability of committing at least one type-I error, across all tests) is below some desired level (i.e.~the standard choice is to perform all individual tests at level $\alpha < 5 \cdot 10^{-8}$).

As it is generally the case, choosing to control the FWER leads to a very conservative selection  of relevant polymorphisms.  Indeed, it has been observed that the variants identified via this strategy---while apparently reproducibly associated with the traits---can typically  only explain a small portion of the genetic variance in the phenotype of interest \cite{Manolio2009}. 
An alternative criterion to evaluate statistical significance is the false discovery rate %ratio 
(FDR) \cite{benjamini1995}. The FDR is a particularly attractive concept when one expects a multiplicity of true discoveries. This has led to its adoption in studies involving  gene expression   and  many other genomic measurements \cite{Storey2003}, including the  study of expression quantitative trait loci (eQTL).  A broader adoption of the FDR has been advocated as a natural strategy to improve the power of GWAS \cite{Sabatti2003,Storey2003,Brzyski2017} for complex traits.
%since it is known that the FWER can be excessively stringent unless only very few features are truly relevant. In practice, it has been observed that often the variables identified through GWAS can only explain a small portion of the genetic variance in the phenotype of interest \cite{Manolio2009}. Therefore, a broader adoption of the FDR has been advocated as a natural strategy to bridge this gap by improving the power of GWAS \cite{Sabatti2003,Storey2003,Brzyski2017}. 
%On the other hand, controlling the FDR is already the standard approach in other types of genomic analysis (e.g. the study of expression quantitative trait loci). \\

Controlled variable selection is an inherently difficult task in high dimensions, but  GWAS present at least two specific %peculiar MATTEO: I think peculiar is a word that we Italian like: the family makes often fun of me for using it a bit too often :)
 challenges. First, many polygenic phenotypes  depend on the genetic variants through mechanisms that are mostly unknown \cite{Zuk2012} and may involve the interaction of different genetic polymorphisms \cite{Carlborg2004}.  Unfortunately, the current analysis methods neglect the possibility that the response depends on the explanatory variables in a non linear fashion and through complicated interactions. Clearly, methods based on marginal testing are ill-equipped to detect interactions and the few approaches that simultaneously analyze the role of multiple variants rely on strong linearity assumptions.
% 
% Unfortunately, methods based on marginal testing are unable to capture these effects, while also being unsatisfactory from other points of view \cite{candes2016}.
%\color{red} CHIARA:  I do not know what these point of views are. I think we cannot leave this suspended in this fashion. Also, I am not clear why methods based on marginal testing have a harder time with non linear relation that multivariate methods. Or perhaps this is not what you are trying to say \color{black}
% Similarly, most of their available multivariate alternatives also rely on strong modeling assumptions (e.g. generalized linear models) whose validity in a GWAS is disputable (and in general statistically untestable whenever $p>n$, to the best of our knowledge). \color{red} CHIARA: What methods/application to GWAS you have in mind? I can think of geneSLOPE that have restrictive assumption, or lasso-like models for logistic regression, but I do not think these calculate p-values. What are the "statistically untestable" things? \color{black}
The second prominent obstacle arises from the presence of correlations among the covariates. The expression 
%In the field of GWAS, 
\textit{linkage disequilibrium} is used in genetics to 
 %the commonly used term that 
 indicate the tight dependence between the alleles at polymorphisms that occupy nearby positions in the genome. This association is due to the process with which the DNA is transmitted in humans and it is a fundamental characteristics of the explanatory variables in GWAS. Methods aiming for valid inference in this settings should certainly take it into account.

%  Even though it is mostly infeasible to deal with this type of correlations in the traditional statistical analysis, a method aiming for valid inference in GWAS should certainly take them into account. \ejc{I do not understand what this sentence means.}

These issues motivate the need for the development of new statistical methodologies that can identify important variables for complex phenomena, while providing rigorous guarantees of type-I error control under milder and well-justified assumptions.

\subsection{The assumptions of model-free knockoffs}
\textit{Model-free knockoffs}, recently introduced in
\cite{candes2016}, partially address the aforementioned issues by
taking a radically different path from the traditional literature on
high-dimensional variable selection. They provide a powerful and
versatile method that enjoys rigorous FDR control, under no modeling
assumptions on the conditional distribution $F_{Y|X}$ of the response
$Y$ given the covariates $X$. In fact, $F_{Y|X}$ may remain completely
arbitrary and unspecified. The suprising result is achieved by
considering a setting in which the distribution $F_X$ of the
covariates is presumed to be known. When this is the case, the latter
can be used to generate appropriate ``negative control'' variables
(the \textit{knockoff copies}). These knockoffs are created
independently of the measured outcome and they allow to distinguish
the relevant from the unnecessary variables. As a consequence, it
becomes possible to estimate and, further, control the FDR.

In many circumstances, the premises of model-free knockoffs can be argued to be more principled than those of its traditional counterparts. Intuitively, it is reasonable to shift the central burden of assumptions from $F_{Y|X}$ to $F_X$, since the former is the essentially the object of inference. In a GWAS, an agnostic approach to the conditional distribution of the response is especially valuable, due to the possibly complex nature of the relations between genetic variants and phenotypes. Moreover, the presumption of knowing $F_X$ is well grounded. On the one hand, geneticists have at their disposal a rich set of models for  how DNA variants arise and spread across human populations over time. 
%does not lack fundament. 
On the other hand, genome-wide variation has been assessed in large collections of  individuals: the \textit{UK Biobank} (\url{http://www.ukbiobank.ac.uk}) %CITATION http://www.ukbiobank.ac.uk
contains the genotypes of 500,000 subjects, the RPGEH (\url{https://www.dor.kaiser.org/external/DORExternal/rpgeh/index.aspx})
has similar information for over 100,000 individuals,  and hundreds of thousands of additional samples are available via \textit{dbGaP} (\url{https://www.ncbi.nlm.nih.gov/gap}),
to cite a few examples. The combination of theoretical understanding and data gives us a good handle on $F_X$.
% the \textit{1000 Genomes Project} or the \textit{dbGaP}) are available and offer valuable insight into the structure of the covariate distribution. \\

In general, the fundamental difficulty with the method of model-free
knockoffs is related to the construction of those knockoff
copies. This task requires knowledge of the underlying distribution of
the original variables, which can rarely be expected to be accessible
exactly. In some cases a good approximation is available, but a
separate computational issue emerges. Even if the true $F_X$ were
known, it may still be unfeasible to create the knockoff copies
required by this procedure. Until now, the only special case for which
an algorithm has been developed is that of multivariate normal
covariates \cite{candes2016}. In this sense, model-free knockoffs have
not yet fully resolved the second crucial difficulty of GWAS that we
mentioned earlier. A multivariate normal approximation cannot fully
take advantage of the precious prior information that we have on the
sequential structure of allele frequencies across SNPs \cite{Wall2003}. % \ms{I tried to make this statetement less strong. However, here I don't feel like dwelling on the subtleties of covariance matrix estimation with shrinkage. I think it's obvious that genotypes aren't multivariate normal.} Moreover, the number of parameters in an unconstrained covariance matrix is extremely large in the usual scenarios faced in this field (i.e.~$p > 10^5$).
It thus seems important to develop new techniques that can exploit
some of the advances in the study of linkage disequilibrium and
population genetics, and exploit accurate parametric models for
$F_X$. 

% in order to incorporate a more suitable description of $F_X$ into the framework of model-free knockoffs.

\subsection{Our contributions}
In this paper, we introduce a new algorithm to sample knockoff copies of variables distributed as a hidden Markov model (HMM). To the best of our knowledge, this result is the first extension of model-free knockoffs beyond the special case of a Gaussian design and it involves a class of covariate distributions that is of great practical interest. In fact, HMMs are widely employed in a variety of fields to describe sequential data with complex correlations. \\

 While many applications of HMMs are found in the context of speech processing \cite{juang1991} and video segmentation \cite{Boreczky1998}, their presence has also become nearly ubiquitous in the statistical analysis of biological sequences. Important instances include protein modeling \cite{krogh1994}, sequence alignment \cite{Hughey1996}, gene prediction \cite{Krogh1997}, copy number reconstruction \cite{WetB07}, segmentation of the genome into diverse functional elements \cite{Ernst2012} identification of ancestral DNA segments and population history \cite{Falush2003,TetR06, Li2011b}. 
  %and population genetics \cite{Hobolth2007}. 
 Of special interest to us, following the empirical observation that variation along the human genome could be described by blocks of limited diversity \cite{PetC01}, HMMs have been broadly adopted to describe haplotypes---the sequence of alleles at a series of markers along one chromosome. The literature  is too extensive to recapitulate: we simply recall that taking the move from some initial formulations \cite{SetD01,ZetS02,QetL02,Li2003}, there are now a vast set of models and algorithms that are used routinely
and effectively to reconstruct haplotypes (phase) and to impute missing genotype values. Some of the most common software implementations include  \texttt{fastPHASE} \cite{Scheet2006}, \texttt{Impute} \cite{Marchini2007,Marchini2010}, \texttt{Beagle} \cite{Browning2007,Browning2011},  \texttt{Bimbam} \cite{guam2008} and  \texttt{MaCH} \cite{Li2010}.
% 
%   a with modern applications to genotype imputation \cite{Marchini2010} and haplotype phasing \cite{Browning2011}. We focus our attention on methods developed for the latter problem, in which the goal is to infer missing SNPs from a partially observed genotype sequences. Today, most of the best imputation techniques are based on HMMs and they can achieve very high accuracy (i.e. the software \texttt{Impute} \cite{Marchini2007,Marchini2010}, \texttt{MaCH} \cite{Li2010}, \texttt{fastPHASE} \cite{Scheet2006}, \texttt{Bimbam} \cite{guam2008} and \texttt{Beagle} \cite{Browning2007,Browning2011}). 
%  
  The success of these algorithms in reconstructing partially observed genotypes can be tested empirically and their realized accuracy is a testament to the fact that 
 %  Even though a complete scientific understanding of genetic variation may still be elusive, at this point it is clear that 
   HMMs offer a good phenomenological description of the dependence between the explanatory variables in GWAS. 

By developing a knockoff contruction for HMMs, we can incorporate the prior knowledge on patterns of genetic variation. As a result, we obtain a new variable selection method that addresses all the critical issues of GWAS discussed in Section \ref{sec:intro_1} and enjoys:
\begin{enumerate}[leftmargin=*]
\item \textbf{Agnostic} conditional characterization of the response given the covariates. As in the general model-free knockoff framework, no assumptions are made here. We are completely free from the rather questionable restrictions of linear models and other parametric alternatives.
\item \textbf{Principled} description of the distribution of the covariates. A sensible model inspired by prior scientific knowledge naturally deals with the correlations across SNPs.
\item \textbf{Powerful} performance inherited from the framework of
  model-free knockoffs. Sophisticated machine learning tools can be
  used to assess variable importance, without losing any control over
  the FDR. In addition, any side information about the likelihood of
  $Y$ given $X$ can be leveraged to improve power. 
\item \textbf{Computationally efficient} construction of knockoff copies, derived from the mathematically amenable properties of hidden Markov models. The complexity of the entire procedure can be shown to be $O(np)$.
\end{enumerate}

% Hidden Markov models are widely used to describe patterns of genetic variation, such as SNP data. Li and Stephen's introduced a HMM (\cite{Li2003}) to represent the chromosome of an individual as a sequence of chunks from the other ``donor'' individuals. This model has been adopted and extended in a variety of subsequent works, with applications ranging from the inference of genetic ancestry \cite{Falush2003, Lawson2012} to the imputation of missing data \cite{Marchini2010}. \\

% \subsection{Knockoff for Non-Gaussian Variables}
% We have seen in Section [?] that it is easy to sample MF knockoffs if the distribution of $X$ is multivariate normal. However, there are many scientific problems in which this assumption clearly does not hold. In particular, our work is motivated by genome-wide association studies, in which the variables typically consist of single-nucleotide polymorphisms that can only take a few discrete values. We have argued in Section [?] that these data are more accurately described by a Hidden Markov Model (HMM). How should one then proceed in this setting? We address this problem by developing a practical and efficient algorithm for knockoffs when $X$ is distributed as a HMM.

\subsection{Related works}

This paper is most closely related to \cite{candes2016}, which has introduced the framework of model-free knockoffs. Their work focuses on the special case of multivariate Gaussian variables, while ours extends their results to HMMs. On the other hand, earlier instances of the knockoff method \cite{barber2015, barber2016} are focused on the linear regression problem with a fixed design matrix. 

Traditional multivariate variable selection techniques have been applied in GWAS on numerous occasions. Some works have employed penalized regression, but they either lack type-I error control \cite{Hoggart2008, Wu2009} or require very restrictive modeling assumptions \cite{Brzyski2017}. Similarly, their Bayesian alternatives \cite{Li2011, guan2011} do not provide finite-sample guarantees. Some have tried to control the type-I errors of standard penalized regression methods through stability selection \cite{Alexander2011}, but they have observed that the resulting procedure does not correctly account for variable correlations and is less powerful than marginal testing.
  Others have employed non-parametric machine learning tools
  \cite{Bureau2005} that can produce variable importance measures, but
  no valid inference. In theory, some inferential guarantees have been
  obtained for the Lasso \cite{Zhao2006, candes2009}, GLMs
  \cite{vandegeer2014} and even random forests \cite{wager2015}, but
  they only hold under rather stringent sparsity assumptions. %large-sample asymptotic conditions that are clearly not satisfied in the case of GWAS (where $n \ll p$). 

Hidden Markov models have appeared before as part of a variable selection procedure for GWAS, in order to combine marginal tests of association from correlated SNPs \cite{sun2009,Wei2012}. However, this approach is fundamentally different from ours, since it is neither multivariate nor model-free.

\section{Model-free controlled variable selection via knockoffs}

%Since this work builds upon and extends the model-free knockoff methodology of Candes et. al \cite{Candes2016}, we begin by reviewing the essentials of this method. \\

\subsection{Problem statement}

The controlled variable selection problem can be naturally stated in formal terms by adopting the general setting of \cite{candes2016}.
Suppose that we can observe a response $Y \in \reals$ and a vector of covariates $X = (X_{1}, \ldots, X_{p}) \in \reals^p$. Given $n$ such samples $(X^{(i)},Y^{(i)})_{i=1}^{n}$ drawn from a population, we would like to know which variables are associated with the response. This can be made more precise by assuming that
\[
  (X^{(i)},Y^{(i)}) \dIid F_{XY}, \qquad i \in \{1,\ldots,n\},
\]
for some joint distribution $F_{XY}$. Here, the concept of a \textit{relevant} variable can be understood by first defining its opposite. We say that $X_j$ is \textit{null} if and only if $Y$ is independent of $X_j$, conditionally on all other variables $X_{-j} = \{X_1, \ldots, X_p\} \setminus \{X_j\}$. This uniquely defines the set of null variables $\mathcal{H}_0 = \{j: X_j \text{ is null}\}$ and its complement $\mathcal{S} = \{j: X_j \text{ is relevant}\} = \{1,\ldots,p\} \setminus \mathcal{H}_0$. Our goal is to obtain an estimate $\hat{\mathcal{S}}$ of $\mathcal{S}$, while controlling the false discovery ratio, that is now defined as:
\begin{align*}
  \text{FDR} \defeq \E{\frac{|\hat{S} \cap \mathcal{H}_0|}{|\hat{S}|}}.
\end{align*}
We emphasize the natural logic of this definition: a variable is null
if it has no predictive power whatsoever once we take into account all the
other variables; i.e.~it does not influence the response in any
way. To relate this with model-based inference, \cite{candes2016}
shows that in a logistic model, being null is 
equivalent---under an extremely mild condition---to saying that the
corresponding regression coefficient vanishes.

\subsection{The method of knockoffs} \label{sec:MFknokckoffs_summary}

The main idea of the model-free knockoffs methodology \cite{candes2016} is to generate a new set of artificial covariates, the \textit{knockoff copies} of $X$, so that they have the same structure as the original ones but are known to be null. These can then be used as ``negative controls'' to estimate the FDR with almost any variable selection algorithm of choice. Model-free knockoffs can thus be seen as a versatile wrapper that allows one to extend rigorous statistical guarantees, under very mild assumptions, to powerful practical methods that would otherwise be too complex for a traditional theoretical analysis. A detailed description of this procedure would fall outside the scope of this paper, but we nonetheless begin with a brief summary because our work builds upon this and extends its applicability.\\

\textbf{Knockoff variables.} For each variable $X_j$, suppose that we can construct a knockoff copy $\tilde{X}_j$ in such a way that the original variables $X=(X_1,\ldots,X_p)$ and their knockoffs $\tilde{X}=(\tilde{X}_1,\ldots,\tilde{X}_p)$ satisfy the following two conditions: 
\begin{align} \label{eq:knock_cond_1}
  & \tilde{X} \indep \, Y | X,
\end{align}
and
\begin{align} \label{eq:knock_cond_2}
  \left( X, \tilde{X} \right)_{\text{swap}(S)} \overset{d}{=} \left( X, \tilde{X} \right)
  \qquad \text{for any } S \subset \{1,\ldots,p \}.
\end{align}
Above, $(X, \tilde{X})_{\text{swap}(S)}$ denotes the vector produced
by swapping the entries $X_j$ and $\tilde{X}_j$, for each $j \in
S$. The \textit{pairwise exchangeability} condition in
\eqref{eq:knock_cond_2} requires the distribution of $(X,\tilde{X})$
to be invariant under this transformation. This property is essential
and we will discuss later how it is not always easy to obtain a
non-trivial\footnote{$\tilde{X} = X$ would obviously satisfy this, but
  it would be of no use.} vector $\tilde{X}$ that satisfies it. We
refer to the other \eqref{eq:knock_cond_1} as the \textit{nullity} condition, since it immediately implies that all knockoffs are null. This  clearly holds whenever $\tilde{X}$ is constructed without looking at $Y$ and is necessary for the knockoff copies to be used as negative controls. \\

\textbf{Feature importance measures.} Once the knockoff copies of $X$ are created, one proceeds by computing two vectors of ``feature importance statistics'': $T=(T_1, \ldots, T_p)$ and $\tilde{T}=(\tilde{T}_1, \ldots, \tilde{T}_p)$. For each $j$, $T_j$ and $\tilde{T}_j$ measure the importance of $X_j$ and $\tilde{X}_j$, respectively, in predicting $Y$. These can be estimated in almost any arbitrary way from the available data. As an example, we can think of letting $T_j$ and $\tilde{T}_j$ be the magnitude of the Lasso coefficients for $X_j$ and $\tilde{X}_j$, obtained by regressing $Y$ on $(X, \tilde{X})$. However, this is just the simplest example from a multitude of potentially more powerful alternatives. Nothing prevents us from computing our estimates by exploiting some form of cross-validation, applying boosting, training a random forest or even a neural network. The only constraint is that $X$ and $\tilde{X}$ should always be treated ``fairly'', i.e.~disregarding which one is a knockoff and which one is not. In mathematical terms, we say that swapping any subset $S$ of the original variables with their knockoff copies should have the only effect of swapping the corresponding elements of $T$ with $\tilde{T}$.  \\

\textbf{The knockoff filter.} The estimated importance measures of the original variables are then compared to those of their corresponding knockoff copies. If $X_j$ is truly relevant, one would expect $T_j$ to be larger than $\tilde{T}_j$. Conversely, they will tend to behave similarly when $X_j$ is null. Formally, one calculates statistics $W_j = w_j(T_j, \tilde{T}_j)$, for some anti-symmetric\footnote{It is required that $w_j(T_j, \tilde{T}_j) = - w_j(\tilde{T}_j, T_j)$. A typical choice is $W_j = |T_j| - |\tilde{T}_j|$ or $W_j = \max(T_j, \tilde{T}_j) \operatorname{sgn}(T_j - \tilde{T}_j)$.}  function $w_j$. Properties \eqref{eq:knock_cond_1} and \eqref{eq:knock_cond_2} imply that all the null $W_j$ satisfy the \textit{flip-sign condition}\footnote{For all $j \in \mathcal{H}_0$, $\text{sign}(W_j)$ are i.i.d. coin flips, conditionally on $(|W_1|, \ldots, |W_p|)$.} required to apply the knockoff filter of \cite{barber2015}. Finally, the latter selects a set $\hat{\mathcal{S}}$ of relevant variables while controlling the FDR at the desired target level $\alpha$.\\

\subsection{Constructing knockoffs}
Fundamental ingredients of the knockoff method are, of course, the
artificial variables $\tilde{X}$. In Section
\ref{sec:MFknokckoffs_summary} we saw that they need to obey the
pairwise exchangeability \eqref{eq:knock_cond_1} and strong nullity
\eqref{eq:knock_cond_2} properties, but we have not discussed how to
construct them.\footnote{A special case considered in
  \cite{candes2016} assumes that $X$ has a multivariate normal
  distribution, where it is possible to derive a simple regression
  formula for sampling $\tilde{X}$.} 
% However, the normal distribution does not always approximate well real data and this assumption can be hard to justify in practice. This motivates us to find new ways of generating knockoff copies for different families of variables. \\
A possible direction is suggested by the Sequential Conditional Independent Pairs (SCIP) ``algorithm'' in \cite{candes2016}. For any known covariate distribution, a knockoff copy $\tilde{X}$ can be obtained by sequentially sampling each of its components according to:
\begin{algorithm}
  \caption{SCIP characterization of knockoffs \label{alg:SCIP}}
  \begin{algorithmic}[1] 
    \For{$j=1$ \textbf{to} $p$}
    \State{\textbf{sample} $\tilde{X}_j$ from $p(X_j| X_{-j}, \tilde{X}_{1:(j-1)})$}, independently of $X_j$
    \EndFor
  \end{algorithmic}
\end{algorithm}\\
Above, $p(X_j| X_{-j}, \tilde{X}_{1:(j-1)})$ denotes the conditional
distribution of $X_j$ given $(X_{-j}, \tilde{X}_{1:(j-1)})$. At first
sight it may appear that the SCIP algorithm is a ``universal''
knockoff generator and our problem is already solved. Unfortunately,
the conditional distribution $p(X_j| X_{-j}, \tilde{X}_{1:(j-1)})$
depends on the knockoff variables $\tilde{X}_{1:(j-1)}$ generated by
the SCIP itself during the previous iterations. This distribution can
be very difficult or impossible to compute in general, even though the
distribution of $X$ is known. Therefore, the SCIP algorithm appears
only to be an abstract recipe and remains totally impractical as it stands. \\

This is where our research begins. In this paper, we draw inspiration from Algorithm \ref{alg:SCIP} and develop new exact and computationally efficient procedures for creating knockoff copies when the model that well describes $X$ is a Markov chain or a hidden Markov model. In particular, the latter has the most interesting scientific applications, but for the sake of simplicity we begin by considering the simpler case of a Markov chain.

% Construction of MF Knockoffs
%\section{Methodology}

\section{Knockoffs for Markov chains} \label{sec:dmc}

In this section, we show how to generate knockoffs if $X$ is distributed as a Markov chain, using a practical procedure derived from the SCIP algorithm; this will be useful later when we deal with a broader class of covariate models. In the interest of simplicity, we focus our attention to discrete Markov chains. Formally, we say that a vector of random variables $X = (X_1, \ldots, X_p)$, each taking values in a finite state space $\mathcal{X}$, is distributed as a discrete Markov chain if its joint probability mass function (pmf) can be written as 
\begin{align} \label{eq:def_dmc}
  \P{X_1=x_1, \ldots, X_p=x_p} = q_1(x_1) \prod_{j=2}^{p} Q_j(x_{j} | x_{j-1}).
\end{align}
Above, $q_1(x_1) = \P{X_1 = x_1}$ denotes the marginal distribution of the first element of the chain, while the transition matrices between consecutive variables are $Q_j(x_{j} | x_{j-1}) = \Pc{X_{j} = x_{j}}{X_{j-1}=x_{j-1}}$. \\
%In the following, we assume that the parameters $q_1$ and $Q=(Q_2, \ldots, Q_{p})$ are known.\\

Before presenting the general result, we propose a simple example to clarify why it is feasible to generate knockoffs for distributions in the form of \eqref{eq:def_dmc}. Suppose that we have $p=3$ variables. In order to create a vector of knockoffs $\tilde{X} = (\tilde{X}_1, \tilde{X}_2, \tilde{X}_3)$, according to the SCIP algorithm, one should proceed in three steps.
\begin{enumerate}
  \item First, we must sample $\tilde{X}_1$ from $p(X_1| X_{2}, X_3)$, independently of the observed value of $X_1$. By the Markov property, we can also forget about $X_3$ since $p(X_1| X_{2}, X_3)=p(X_1| X_{2})$. The pmf of this conditional distribution is $p(X_1| X_{2}) \propto q_1(X_1) \, Q_2(X_2|X_1)$. Therefore, we can easily sample $\tilde{X}_1$ from:
\begin{align*}
  \Pc{\tilde{X}_1 = \tilde{x}_1}{X_{-1}=x_{-1}} \propto q_1(\tilde{x}_1) \, Q_2(x_2|\tilde{x}_1),
 \end{align*}
 since we only need to compute the normalization constant. For reasons that will become clear in a moment, we make the dependence of the normalization constant $\mathcal{N}_1(X_2)$ on $X_2$ explicit, by defining a ``normalization function'' $\mathcal{N}_1(k) = \sum_{l \in \mathcal{X}} q_1(l) \, Q_2(k|l)$.
\item %\ejc{I would like to make a suggestion. Above, here and below, wouldn't it make more sense to say: the SCIP algorithm asks us to sample $\tilde{X}_2$ from the distribution of $X_2$ conditional on $X_{1}, X_3, \tilde{X}_1)$. From the previous point, it follows that $p(\tilde{X}_2| X_{1}, X_3, \tilde{X}_1) \propto Q_2(\tilde{X}_2|X_1) \, Q_3(X_3|\tilde{X}_2) \, p(\tilde{X}_1 | \tilde{X}_2)$.}
    Now, the SCIP algorithm asks us to sample $\tilde{X}_2$ from $p(X_2| X_{1}, X_3, \tilde{X}_1)$. From the previous point, it follows that $p(X_2| X_{1}, X_3, \tilde{X}_1) \propto Q_2(X_2|X_1) \, Q_3(X_3|X_2) \, p(\tilde{X}_1 | X_2)$. Since we are only interested in the terms that contain $X_2$, we can use the normalization function $\mathcal{N}_1(X_2)$ to rewrite this as $p(X_2| X_{1}, X_3, \tilde{X}_1) \propto Q_2(X_2|X_1) \, Q_3(X_3|X_2) \, \frac{Q_2(X_2|\tilde{X}_1)}{\mathcal{N}_1(X_2)}$. Therefore, we sample $\tilde{X}_2$ according to:
\begin{align*}
  \Pc{\tilde{X}_2 = \tilde{x}_2}{X_{-2}=x_{-2}, \tilde{X}_1 = \tilde{x}_1} \propto Q_2(\tilde{x}_2|x_1) \, Q_3(x_3|\tilde{x}_2) \, \frac{Q_2(\tilde{x}_2|\tilde{x}_1)}{\mathcal{N}_1(\tilde{x}_2)}.
 \end{align*}
Note that, from this expression, it is clear that we should have evaluated the normalization function $\mathcal{N}_1(k)$ of the previous step for all $k \in \mathcal{X}$. Similarly, we now need to compute the new normalization function $\mathcal{N}_2(k) = \sum_{l \in \mathcal{X}} Q_2(l|X_1) \, Q_3(k|l) \, \frac{Q_2(l|\tilde{X}_1)}{\mathcal{N}_1(l)}$ in order to sample $\tilde{X}_2$ and proceed to the final step. 
  \item By the same argument, it is easy to verify that $p(X_3| X_{2}, X_1, \tilde{X}_1, \tilde{X}_2) \propto Q_3(X_3|X_2) \, \frac{Q_3(X_3|\tilde{X}_2)}{\mathcal{N}_2(X_3)}$. Again, the normalization constant is straightforward to compute and does not depend on $\mathcal{N}_1(\cdot)$. Thus, we can also sample the last knockoff variable $\tilde{X}_3$ from
\begin{align*}
  \Pc{\tilde{X}_3 = \tilde{x}_3}{X_{-3}=x_{-3}, \tilde{X}_{1:2} = \tilde{x}_{1:2}} \propto Q_3(\tilde{x}_3|x_2) \, \frac{Q_3(\tilde{x}_3|\tilde{x}_2)}{\mathcal{N}_2(\tilde{x}_3)}.
 \end{align*}
  \end{enumerate}

\begin{figure}[h!]
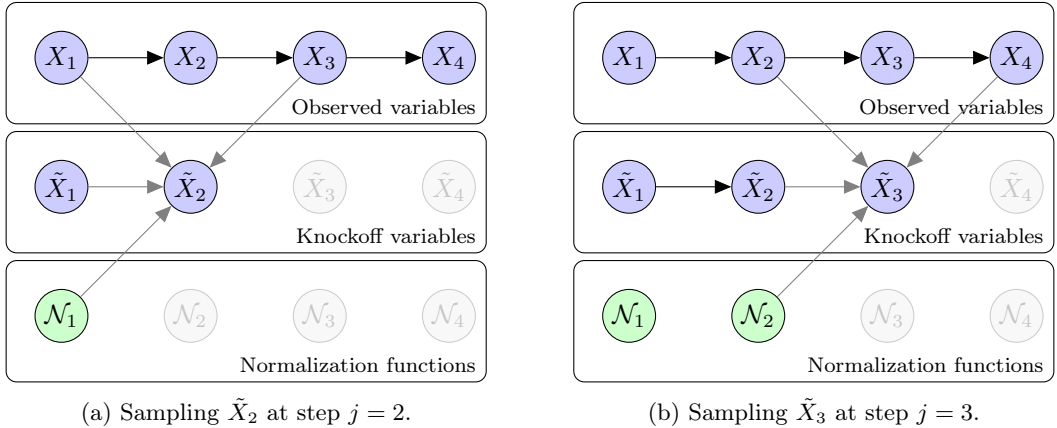

  \centering
  \begin{subfigure}[b]{0.45\textwidth}      \captionsetup{justification=centering}
      \centering
      \tikz{ %
        \node[obs, fill=blue, fill opacity=0.2, text opacity=1] (X1) {$X_{1}$} ; %
        \node[obs, right=of X1, fill=blue, fill opacity=0.2, text opacity=1] (X2) {$X_{2}$} ; %
        \node[obs, right=of X2, fill=blue, fill opacity=0.2, text opacity=1] (X3) {$X_{3}$} ; %
        \node[obs, right=of X3, fill=blue, fill opacity=0.2, text opacity=1] (X4) {$X_{4}$} ; %
        \node[obs, below=of X1, fill=blue, fill opacity=0.2, text opacity=1] (Xt1) {$\tilde{X}_{1}$} ; %
        \node[obs, right=of Xt1, fill=blue, fill opacity=0.2, text opacity=1] (Xt2) {$\tilde{X}_{2}$} ; %
        \node[obs, right=of Xt2, opacity=0.2] (Xt3) {$\tilde{X}_{3}$} ; %
        \node[obs, right=of Xt3, opacity=0.2] (Xt4) {$\tilde{X}_{4}$} ; %
        \node[obs, below=of Xt1, fill=green, fill opacity=0.2, text opacity=1] (N1) {$\mathcal{N}_{1}$} ; %
        \node[obs, right=of N1, opacity=0.2] (N2) {$\mathcal{N}_{2}$} ; %
        \node[obs, right=of N2, opacity=0.2] (N3) {$\mathcal{N}_{3}$} ; %
        \node[obs, right=of N3, opacity=0.2] (N4) {$\mathcal{N}_{4}$} ; %
        \plate[inner sep=0.25cm, xshift=-0.12cm, yshift=0.12cm] {plateX} {(X1) (X2) (X3) (X4)} {\textcolor{black}{Observed variables}}; 
        \plate[inner sep=0.25cm, xshift=-0.12cm, yshift=0.12cm] {plateX} {(Xt1) (Xt2) (Xt3) (Xt4)} {\textcolor{black}{Knockoff variables}}; 
        \plate[inner sep=0.25cm, xshift=-0.12cm, yshift=0.12cm] {plateX} {(N1) (N2) (N3) (N4)} {\textcolor{black}{Normalization functions}}; 
        \edge {X1} {X2}; %
        \edge {X2} {X3}; %
        \edge {X3} {X4}; %
        \edge[draw=gray] {Xt1} {Xt2}; %
        \edge[draw=gray] {X1} {Xt2}; % 
        \edge[draw=gray] {X3} {Xt2}; %
        \edge[draw=gray] {N1} {Xt2}; %
     }
     \caption{Sampling $\tilde{X}_2$ at step $j=2$.}
  \end{subfigure}
  \begin{subfigure}[b]{0.45\textwidth}      \captionsetup{justification=centering}
      \centering
      \tikz{ %
        \node[obs, fill=blue, fill opacity=0.2, text opacity=1] (X1) {$X_{1}$} ; %
        \node[obs, right=of X1, fill=blue, fill opacity=0.2, text opacity=1] (X2) {$X_{2}$} ; %
        \node[obs, right=of X2, fill=blue, fill opacity=0.2, text opacity=1] (X3) {$X_{3}$} ; %
        \node[obs, right=of X3, fill=blue, fill opacity=0.2, text opacity=1] (X4) {$X_{4}$} ; %
        \node[obs, below=of X1, fill=blue, fill opacity=0.2, text opacity=1] (Xt1) {$\tilde{X}_{1}$} ; %
        \node[obs, right=of Xt1, fill=blue, fill opacity=0.2, text opacity=1] (Xt2) {$\tilde{X}_{2}$} ; %
        \node[obs, right=of Xt2, fill=blue, fill opacity=0.2, text opacity=1] (Xt3) {$\tilde{X}_{3}$} ; %
        \node[obs, right=of Xt3, opacity=0.2] (Xt4) {$\tilde{X}_{4}$} ; %
        \node[obs, below=of Xt1, fill=green, fill opacity=0.2, text opacity=1] (N1) {$\mathcal{N}_{1}$} ; %
        \node[obs, right=of N1, fill=green, fill opacity=0.2, text opacity=1] (N2) {$\mathcal{N}_{2}$} ; %
        \node[obs, right=of N2, opacity=0.2] (N3) {$\mathcal{N}_{3}$} ; %
        \node[obs, right=of N3, opacity=0.2] (N4) {$\mathcal{N}_{4}$} ; %
        \plate[inner sep=0.25cm, xshift=-0.12cm, yshift=0.12cm] {plateX} {(X1) (X2) (X3) (X4)} {\textcolor{black}{Observed variables}}; 
        \plate[inner sep=0.25cm, xshift=-0.12cm, yshift=0.12cm] {plateX} {(Xt1) (Xt2) (Xt3) (Xt4)} {\textcolor{black}{Knockoff variables}}; 
        \plate[inner sep=0.25cm, xshift=-0.12cm, yshift=0.12cm] {plateX} {(N1) (N2) (N3) (N4)} {\textcolor{black}{Normalization functions}}; 
        \edge {X1} {X2}; %
        \edge {X2} {X3}; %
        \edge {X3} {X4}; %
        \edge[draw=black] {Xt1} {Xt2}; %
        \edge[draw=gray] {Xt2} {Xt3}; %
        \edge[draw=gray] {X2} {Xt3}; % 
        \edge[draw=gray] {X4} {Xt3}; %
        \edge[draw=gray] {N2} {Xt3}; %
     }
     \caption{Sampling $\tilde{X}_3$ at step $j=3$.}
   \end{subfigure}
   \caption{Graphical representation of the SCIP algorithm applied to a Markov chain, in the case $p=4$. At the $j$th step, $\tilde{X}_j$ is sampled using the values of the variables $(X_{j-1}, X_{j+1}, \tilde{X}_{j-1})$ and the normalization function $\mathcal{N}_{j-1}$ computed at the previous step. The algorithm begins from $\tilde{X}_1$ and proceeds sequentially until it reaches $\tilde{X}_p$. At each stage, a new knockoff variable is sampled and a normalization function is evaluated. The final outcome is a new Markov chain that is a knockoff copy of the original $X$.} \label{fig:sampling_mc}
 \end{figure}

In this example, we see that each conditional law $p(X_j| X_{-j}, \tilde{X}_{1:(j-1)})$ takes a tractable closed form. This simplification of the SCIP algorithm is a rather natural consequence of the Markov property and it holds for any number of variables $p$. A graphical sketch of the general procedure is provided in Figure \ref{fig:sampling_mc}.

  With this intuition clear in mind, we are now ready to formally state the main result of this section, whose proof is in the Appendix. 
  
\begin{prop}[Knockoff copies of a Markov chain] \label{prop:MC_knock}
  The SCIP algorithm applied to a discrete Markov chain generates the $j$th knockoff variable $\tilde{X}_j$ by sampling from
\begin{align} \label{eq:MC_knock_P}
  \Pc{\tilde{X}_j = \tilde{x}_j}{X_{-j}=x_{-j}, \tilde{X}_{1:(j-1)}=\tilde{x}_{1:(j-1)}} = 
  \begin{dcases}
    \frac{q_1(\tilde{x}_1) \, Q_2(x_2|\tilde{x}_1)}{\mathcal{N}_1(x_2)},
  &  j=1, \\
      \frac{Q_{j}(\tilde{x}_j|x_{j-1}) \, Q_{j}(\tilde{x}_j|\tilde{x}_{j-1}) \, Q_{j+1}(x_{j+1}|\tilde{x}_{j})}{\mathcal{N}_{j-1}(\tilde{x}_{j}) \, \mathcal{N}_{j}(x_{j+1})},
      &  1 < j < p, \\
      \frac{Q_{p}(\tilde{x}_p|x_{p-1}) \, Q_{p}(\tilde{x}_p|\tilde{x}_{p-1})}{\mathcal{N}_{p-1}(\tilde{x}_{p}) \, \mathcal{N}_{p}(1)},
      &  j = p,
    \end{dcases}
\end{align}
with the normalization functions $\mathcal{N}_j: \mathcal{X} \mapsto \reals_+$ defined recursively as
% \begin{define}[Normalization Functions for Markov Chain Knockoffs]
%   For any $j \in \{1,\ldots, p\}$, let $x = (x_1, \ldots, x_p)$ be the observed values of the original variables and $\tilde{x}_{1:(j-1)} = (x_1, \ldots, x_{j-1})$ be the sampled values of the first $j-1$ knockoff variables. The function $\mathcal{N}_j $ is defined as
\begin{align} \label{eq:MC_knock_Z}
\mathcal{N}_j(k) =
\begin{dcases}
  \sum_{l \in \mathcal{X}} q_1(l) \, Q_2(k|l),
  & j = 1, \\
  \sum_{l \in \mathcal{X}} \frac{Q_{j}(l|x_{j-1}) \, Q_{j}(l|\tilde{x}_{j-1}) \, Q_{j+1}(k|l)}{\mathcal{N}_{j-1}(l)},
  & 1 < j < p, \\
  \sum_{l \in \mathcal{X}} \frac{Q_{p}(l|x_{p-1}) \, Q_{p}(l|\tilde{x}_{p-1})}{\mathcal{N}_{p-1}(l)},
  & j = p.
\end{dcases}
\end{align}
\end{prop}
This result allows us to summarize the SCIP algorithm for a Markov chain as follows:
\begin{algorithm}
  \caption{Knockoff copies of a discrete Markov chain \label{alg:MC_knock}}
  \begin{algorithmic}[1] 
%    \Inputs{$x=(x_1, \ldots, x_p)$, $q_1$, $Q = (Q_1, \ldots, Q_{p-1})$}
    \For{$j=1$ \textbf{to} $p$}
    \For{$k$ \textbf{in} $\mathcal{X}$}
    \State{\textbf{compute} $\mathcal{N}_j(k)$ according to \eqref{eq:MC_knock_Z}}
    \EndFor
    \State{\textbf{sample} $\tilde{X}_j$ according to \eqref{eq:MC_knock_P}}
    \EndFor
  \end{algorithmic}
\end{algorithm}\\
At each step $j$, the evaluation of the normalization function $\mathcal{N}_j(k)$ involves a sum over all elements of the finite state space $\mathcal{X}$ and it only depends on the previous $\mathcal{N}_{j-1}(\cdot)$. Since this operation must be repeated for all values of $k$, sampling the $j$th knockoff variable requires $O(|\mathcal{X}|^2)$ time, where $|\mathcal{X}|$ is the number of possible states of the Markov chain. This procedure is sequential, generating one knockoff variable at a time. Therefore, the total computation time is $O(p|\mathcal{X}|^2)$, while the required memory is $O(|\mathcal{X}|)$. It is also trivially parallelizable if one wishes to construct a knockoff copy for each of $n$ independent Markov chains. These features make Algorithm \ref{alg:MC_knock} efficient and suitable for high-dimensional applications.

\section{Knockoffs for hidden Markov models} \label{sec:hmm_knock}

We have seen that a consequence of the memoryless property of Markov
chains is that the SCIP algorithm simplifies sufficiently to become
practically implementable. In this section, we build on this insight
and develop am efficient method to sample knockoff copies for the more 
general class of hidden Markov models.

\subsection{Hidden Markov models}

\begin{figure}[!htb]
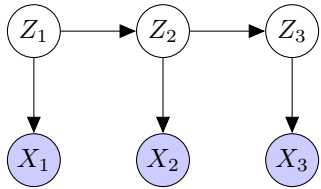

      \centering
      \tikz{ %
        \node[latent] (Z1) {$Z_{1}$} ; %
        \node[latent, right=of Z1] (Z2) {$Z_{2}$} ; %
        \node[latent, right=of Z2] (Z3) {$Z_{3}$} ; %
        \node[obs, below=of Z1, fill=blue, fill opacity=0.2, text opacity=1] (X1) {$X_{1}$} ; %
        \node[obs, below=of Z2, fill=blue, fill opacity=0.2, text opacity=1] (X2) {$X_{2}$} ; %
        \node[obs, below=of Z3, fill=blue, fill opacity=0.2, text opacity=1] (X3) {$X_{3}$} ; %
        \edge[opacity=1] {Z1} {Z2}; %
        \edge[opacity=1] {Z2} {Z3}; %
        \edge[opacity=1] {Z1} {X1}; %
        \edge[opacity=1] {Z2} {X2}; %
        \edge[opacity=1] {Z3} {X3}; %
      }
      \caption{Graphical representation of an HMM with $p=3$ observed variables. The latent chain $Z=(Z_1,\ldots,Z_p)$ cannot be directly observed and it is marginally distributed as a Markov chain. Conditional on $Z$, each observed variable $X_j$ is independently distributed according to some emission law $f_j(x_j|z_j) = \Pc{X_j=x_j}{Z_j=z_j}$.} \label{fig:hmm_3}
    \end{figure}

A hidden Markov model (HMM) assumes the presence of a latent Markov chain whose states are not directly visible. Instead, to each hidden state corresponds an emission distribution from which, conditional on the Markov chain, the observations are independently sampled. Of course, in the extreme case in which all emission distributions are deterministic, this model reduces to a Markov chain. 
%Despite the apparent similarity with the latter, the additional layer of complexity makes HMMs very versatile. \\
%Some examples from their wide range of successful applications have been mentioned in the introduction.
%Some of the earliest instances of their use are found in the context of speech processing \cite{juang1991,gales2007} and face recognition \cite{nefian1998}. In more recent times, the presence of HMMs has become almost ubiquitous in the analysis of biological sequence data. Important examples include protein modeling \cite{krogh1994}, sequence alignment \cite{yoon2009}, gene prediction \cite{stanke2003} and population genetics \cite{kern2010}. 
%For the present work, we are mostly interested the description of the correlations between alleles (\textit{linkage disequilibrium}) on different sites along a genome. Hidden Markov models have been broadly adopted for this task, arguably starting from the seminal paper of \cite{Li2003} and continuing with modern applications to genotype imputation \cite{Marchini2010} and haplotype phasing \cite{Browning2011}.  \\
%A more detailed list of references supporting the use of these models to describe the distribution of covariates in genome-wide association studies is postponed to Section \ref{sec:HMM_SNP}. \\
Formally, 
% We now formally define the class of HMMs for which we will construct
% knockoff copies.  For the sake of simplicity, we consider a latent
% chain with a finite state space and discrete emission
% distributions. At the price slightly more involved notation, our
% results can be easily extended to the case of continuous emission
% distributions.
% \begin{define}[Discrete hidden Markov model]
we say that a vector of random variables $X = (X_1, \ldots, X_p)$
taking values in a finite state space $\mathcal{X}$ is distributed as
a discrete hidden Markov model (HMM) with $K$ hidden states if there
exists a vector of latent random variables $Z = (Z_1, \ldots, Z_p)$
such that
\begin{align}
\begin{split} \label{eq:HMM_def}
\begin{cases}
  Z  \sim \text{MC}\left(q_1, Q\right) & \text{(latent discrete Markov chain)}, \\
  X_j | Z  \sim X_j | Z_j \overset{\text{ind.}}{\sim} f_j(X_j|Z_j) & \text{(emission distribution)}.
\end{cases}
\end{split}
\end{align}
% \end{define}
Above, $\text{MC}\left(q_1, Q\right)$ indicates the law of a discrete Markov chain as in \eqref{eq:def_dmc}. The structure of an HMM can be intuively understood with a graphical model, as shown in Figure \ref{fig:hmm_3} in the case $p=3$.

    We emphasize that we are restricting our attention to these
    discrete distributions solely for the sake of simplicity.  At the
    price of a slightly more involved notation, the knockoffs
    construction can be easily extended to the case of continuous
    emission distributions.

\subsection{Generating knockoffs for an HMM}
In an HMM, the observed variables no longer satisfy the Markov property. In fact, computing the conditional distributions $p(X_j| X_{-j}, \tilde{X}_{1:(j-1)})$ from Algorithm \ref{alg:SCIP} would involve a sum over the possible states of all latent variables. The complexity of this operation is exponential in the number of variables $p$, thus making the na\"ive approach unfeasible even for moderately large datasets. \\

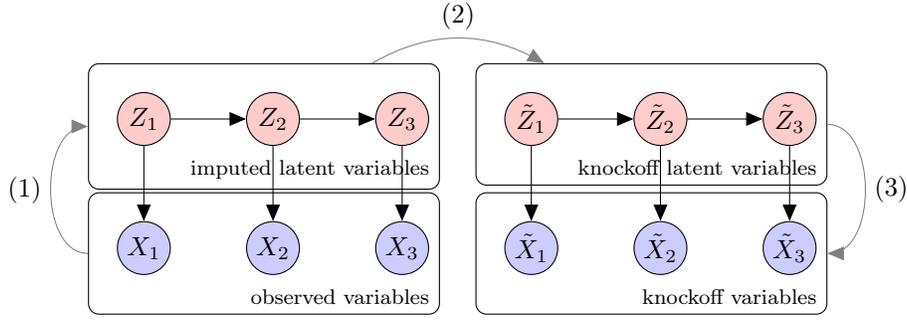
\begin{figure}[h!]
      \centering      
      \begin{tikzpicture}
        \node[latent, fill=red, fill opacity=0.2, text opacity=1] (Z1) {$Z_{1}$} ; %
        \node[latent, right=of Z1, fill=red, fill opacity=0.2, text opacity=1] (Z2) {$Z_{2}$} ; %
        \node[latent, right=of Z2, fill=red, fill opacity=0.2, text opacity=1] (Z3) {$Z_{3}$} ; %
        \node[obs, below=of Z1, fill=blue, fill opacity=0.2, text opacity=1] (X1) {$X_{1}$} ; %
        \node[obs, below=of Z2, fill=blue, fill opacity=0.2, text opacity=1] (X2) {$X_{2}$} ; %
        \node[obs, below=of Z3, fill=blue, fill opacity=0.2, text opacity=1] (X3) {$X_{3}$} ; %
        \node[latent, right=of Z3, fill=red, fill opacity=0.2, text opacity=1] (Zt1) {$\tilde{Z}_{1}$} ; %
        \node[latent, right=of Zt1, fill=red, fill opacity=0.2, text opacity=1] (Zt2) {$\tilde{Z}_{2}$} ; %
        \node[latent, right=of Zt2, fill=red, fill opacity=0.2, text opacity=1] (Zt3) {$\tilde{Z}_{3}$} ; %
        \node[obs, below=of Zt1, fill=blue, fill opacity=0.2, text opacity=1] (Xt1) {$\tilde{X}_{1}$} ; %
        \node[obs, below=of Zt2, fill=blue, fill opacity=0.2, text opacity=1] (Xt2) {$\tilde{X}_{2}$} ; %
        \node[obs, below=of Zt3, fill=blue, fill opacity=0.2, text opacity=1] (Xt3) {$\tilde{X}_{3}$} ; %
        \plate[inner sep=0.25cm, xshift=-0.12cm, yshift=0.12cm] {plateX} {(X1) (X2) (X3)} {observed variables}; 
        \plate[inner sep=0.25cm, xshift=-0.12cm, yshift=0.12cm] {plateZ} {(Z1) (Z2) (Z3)} {imputed latent variables}; 
        \plate[inner sep=0.25cm, xshift=-0.12cm, yshift=0.12cm] {plateZt} {(Zt1) (Zt2) (Zt3)} {knockoff latent variables}; 
        \plate[inner sep=0.25cm, xshift=-0.12cm, yshift=0.12cm] {plateXt} {(Xt1) (Xt2) (Xt3)} {knockoff variables}; 
        \edge[opacity=1] {Z1} {Z2}; %
        \edge[opacity=1] {Z2} {Z3}; %
        \edge[opacity=1] {Z1} {X1}; %
        \edge[opacity=1] {Z2} {X2}; %
        \edge[opacity=1] {Z3} {X3}; %
        \edge[opacity=1] {Zt1} {Zt2}; %
        \edge[opacity=1] {Zt2} {Zt3}; %
        \edge[opacity=1] {Zt1} {Xt1}; %
        \edge[opacity=1] {Zt2} {Xt2}; %
        \edge[opacity=1] {Zt3} {Xt3}; %
        \edge[bend left=90, draw=gray, edge label=(1)] {plateX} {plateZ}; %
        \edge[bend left, draw=gray, edge label=(2)] {plateZ} {plateZt}; %
        \edge[bend left=90, draw=gray, edge label=(3)] {plateZt} {plateXt}; %
      \end{tikzpicture}
      \caption{Sketch of the three stages of Algorithm \ref{alg:HMM_knock} for generating knockoff copies of an HMM, in the case $p=3$.} \label{fig:alg_hmm_knock}
    \end{figure}

Our solution is inspired by the traditional forward-backward methods for hidden Markov models. Having observed a vector $x$ of observations from an HMM, we propose to construct a knockoff copy $\tilde{x}$ as follows:
\begin{algorithm}[!htb]
  \caption{Knockoff copies of a hidden Markov model \label{alg:HMM_knock}}
  \begin{algorithmic}[1] 
%    \Inputs{$x=(x_1, \ldots, x_p)$, $q_1$, $Q = (Q_1, \ldots, Q_{p-1}), f=(f_1, \ldots, f_p)$}
    \State \textbf{sample} $z=(z_1,\ldots,z_p)$ from $\Pc{Z}{X=x}$, using a forward-backward procedure
    \State \textbf{sample} a knockoff copy $\tilde{z}=(\tilde{z}_1,\ldots,\tilde{z}_p)$ of $z=(z_1,\ldots,z_p)$, using Algorithm \ref{alg:MC_knock}
    \State \textbf{sample} $\tilde{x}$ from the conditional distribution of $X$ given $Z=\tilde{z}$.
  \end{algorithmic}
\end{algorithm}\\
A graphical representation of this algorithm is shown in Figure \ref{fig:alg_hmm_knock}. In the first stage of Algorithm \ref{alg:HMM_knock}, the unobserved values of the latent Markov chain are imputed by sampling from the conditional distribution of $Z$ given $X$. This can be done efficiently with a forward-backward iteration similar to the Viterbi algorithm, as discussed in the next subsection. It turns out that the computation time required by this operation is $O(pK^2)$. Once the vector $Z$ is sampled, a knockoff copy $\tilde{Z}$ can be obtained by applying Algorithm \ref{alg:MC_knock}. We already know that the complexity of this stage is also $O(pK^2)$. % This approach is justifed by the HMM assumption that $Z$ is marginally distributed as a Markov chain.
     Finally, we only have to sample $\tilde{X}$ from the conditional distribution of $X$ given $Z=\tilde{z}$. This final task is trivial because the emission distributions are conditionally independent given the latent Markov chain. Since the third step is trivially $O(p|\mathcal{X}|)$, it follows that the whole procedure runs in $O(p(K^2 \lor |\mathcal{X}|))$ time.

    Our next result proves the validity of this approach.

\begin{theorem}[Knockoff copies of an HMM] \label{thm:HMM_knock}
  If the vector of random variables $X=(X_1,\ldots,X_p)$ is distributed as the HMM in \eqref{eq:HMM_def}, then $(\tilde{X}, \tilde{Z})$ generated by Algorithm \ref{alg:HMM_knock} is a knockoff copy of $(X,Z)$. That is, for any subset $S \subseteq \{1,\ldots,p\}$, 
\begin{align} \label{eq:swapXZ}
  \left( (X, \tilde{X})_{\text{swap}(S)}, (Z, \tilde{Z})_{\text{swap}(S)}\right)
  & \overset{d}{=} \left( (X, \tilde{X}), (Z, \tilde{Z}) \right).
\end{align}
In particular, this implies that $\tilde{X}$ is a knockoff copy of $X$.
\end{theorem}

\begin{proof} %[Proof of Theorem \ref{thm:HMM_knock}]
  It suffices to prove \eqref{eq:swapXZ}, since marginalizing over
  $(Z, \tilde Z)$ implies
  $(X, \tilde{X})_{\text{swap}(S)} \overset{d}{=} (X, \tilde{X})$. By
  conditioning on the values of the latent variables, one can write
\begin{align*}
  & \P{(X, \tilde{X}) = (x, \tilde{x})_{\text{swap}(S)}, (Z, \tilde{Z}) = (z, \tilde{z})_{\text{swap}(S)}} \\
  & = \Pc{(X, \tilde{X}) = (x, \tilde{x})_{\text{swap}(S)}}{(Z, \tilde{Z}) = (z, \tilde{z})_{\text{swap}(S)}} 
    \P{(Z, \tilde{Z}) = (z, \tilde{z})_{\text{swap}(S)}} \\
  & = \Pc{(X, \tilde{X}) = (x, \tilde{x})}{(Z, \tilde{Z}) = (z, \tilde{z})} 
    \P{(Z, \tilde{Z}) = (z, \tilde{z})_{\text{swap}(S)}} \\
  & = \Pc{(X, \tilde{X}) = (x, \tilde{x})}{(Z, \tilde{Z}) = (z, \tilde{z})} 
    \P{(Z, \tilde{Z}) = (z, \tilde{z})} \\
  & = \P{(X, \tilde{X}) = (x, \tilde{x}), (Z, \tilde{Z}) = (z, \tilde{z})}.
\end{align*}
The second equality above follows from the conditional independence of the emission distributions in the hidden Markov model given the latent variables (Algorithm \ref{alg:HMM_knock}, line 3). The third equality follows from the fact that $\tilde{Z}$ is a knockoff copy of $Z$ (Algorithm \ref{alg:HMM_knock}, line 2).
\end{proof}

% As a consequence of Proposition \ref{prop:MC_knock}, the complexity of the second step of this algorithm is $O(K^2p)$. The same complexity is also required for the first step, as discussed in Section \ref{sec:sampling_hmm}. Since the third step is trivially $O(Kp)$, it follows that the whole procedure runs in $O(K^2p)$ time.

\subsection{Sampling hidden paths for an HMM} \label{sec:sampling_hmm}
The first step of Algorithm \ref{alg:HMM_knock} consists of sampling from the conditional distribution of the latent variables $Z$ of an HMM, given all the observable variables $X$. This task is closely related to that of finding the most likely a-posteriori sequence of hidden states (i.e.~the Viterbi path) and it can be solved efficiently with a forward-backward sampling algorithm. Earlier examples of this technique are found in \cite{zhu1998} and \cite{Cawley27092003}, in the context of biological sequence alignment and gene splicing, respectively.\footnote{The method described in \cite{Cawley27092003} is slightly different, but essentially equivalent. Instead of proceeding as we suggest, they first compute a collection of ``backward probabilities'', and then sample $Z$ with a forward pass.}\\

For each variable $j \in \{1,\ldots,p\}$, we define the forward probability
\begin{align*} 
  \alpha_j(k) = \P{x_{1:j}, Z_j = k},
\end{align*}
which is the probability of observing the features $X_{1:j}=x_{1:j}$ up to time $j$ and ending up in the hidden state $k$. 
Note that for $j=1$ this is simply
\begin{align*} 
  \alpha_1(k) = q_1(k) f_1(x_1 | k),
\end{align*}
where $q_1(k)$ is the marginal distribution of $Z_1$. The other forward probabilities can be computed recursively as follows:
\begin{align*}
  \alpha_{j+1}(k) = \P{x_{1:(j+1)}, Z_{j+1} = k} 
  & = \sum_l \Pc{x_{j+1}, Z_{j+1} = k}{Z_j = l, x_{1:j}} \alpha_j(l) \\
  & =  \sum_l \Pc{x_{j+1}}{Z_{j+1} = k} \P{Z_{j+1} = k | Z_j = l} \alpha_j(l)\\
  & =  f_{j+1}(x_{j+1}|k) \cdot \left[\sum_l Q_{j+1}(k|l) \, \alpha_j(l)\right]. 
\end{align*}
These equations can be written more compactly in matrix notation:
\[
   \alpha_{j} = (Q_{j} \alpha_{j-1}) \odot \beta_j, \qquad 
   \beta_j(k) = f_j(x_j | k), 
\]
where $\odot$ indicates component-wise multiplication. \\

Having computed the forward probabilities (forward pass), we can now sample from $p(Z|X)$, starting from $Z_p$ and back-tracking along the sequence all the way to $Z_1$. This approach arises naturally from the fact that
\begin{align*}
  \Pc{Z_{1:p}=z_{1:p}}{x_{1:p}}
  = \Pc{Z_{1:(p-1)}=z_{1:(p-1)}}{Z_p=z_p, x_{1:p}} \Pc{Z_p=z_p}{x_{1:p}}.
\end{align*}
This identity suggests that one should start by sampling $z_p$ from the discrete distribution 
\begin{align*}
  \Pc{Z_p=z_p}{x_{1:p}} = \frac{\alpha_p(z_p)}{\sum_k \alpha_p(k)}. 
\end{align*}
Once $z_p$ is chosen, we can think of it as a fixed parameter and turn on to sampling the random variable $Z_{p-1}$. To this end, note that 
\begin{align*}
  \Pc{Z_{1:(p-1)}=z_{1:(p-1)}}{z_p, x_{1:p}} 
  & = \Pc{Z_{1:(p-1)}=z_{1:(p-1)}}{z_p, x_{1:p-1}}\\
  & = \Pc{Z_{1:(p-2)}=z_{1:(p-2)}}{Z_{p-1}=z_{p-1}, x_{1:p}} 
    \underbrace{\Pc{Z_{p-1}=z_{p-1}}{z_p, x_{1:(p-1)}}}_{\propto \, Q_p(z_p | z_{p-1}) \, \alpha_{p-1}(z_{p-1})}. 
\end{align*}
Hence, we sample $z_{p-1}$ from 
\begin{align*}
  \Pc{Z_{p-1}=z_{p-1}}{z_p, x_{1:(p-1)}}
  & = \frac{Q_{p}(z_p | z_{p-1}) \alpha_{p-1}(z_{p-1})}{\sum_k Q_{p}(z_p|k) \alpha_{p-1}(k)}.
\end{align*}
We continue in this fashion and, at step $p-j+1$, we sample $z_j$ from
\begin{align*}
  \Pc{Z_{j}=z_{j}}{z_{j+1}, x_{1:j}}
  & = \frac{Q_{j+1}(z_{j+1} | z_j) \alpha_{j}(z_j)}{\sum_k Q_{j+1}(z_{j+1} | k) \alpha_{j}(k)}.
\end{align*}

To summarize, in the first phase the forward variables are computed with Algorithm \ref{alg:hmm_sampling_forw}. Then, sampling is done with a backward pass as in Algorithm \ref{alg:hmm_sampling_back}.
This process allows one to sample a complete path of latent HMM variables from their conditional law given the corresponding emitted variables $X$. Since the algorithm only involves matrix multiplications and other trivial operations, its computation time is $O(pK^2)$, where $K$ is the size of the state space of the latent Markov chain. This complexity is the same as that of our procedure for generating knockoff copies of a Markov chain. % Therefore, the overall recipe for generating a knockoff copy of an HMM will also be $O(pK^2)$. In the case of $n$ independent HMMs, this trivially becomes $O(npK^2)$.

\begin{algorithm}[!htb]
  \caption{Forward-backward sampling (forward pass)} \label{alg:hmm_sampling_forw}
  \begin{algorithmic}[1] 
    \State{\textbf{initialize} $t = 1$, \; $\alpha_0 = 1$, \; $Q_1(k|l) = q_1(k)$ for all $k,l$}, \; $\beta_j(k) = f_j(x_j | k)$
    \For{$j=1$ \textbf{to} $p-1$}
    \State{\textbf{compute} the forward probabilities $\alpha_j = (Q_{j} \alpha_{j-1}) \odot \beta_j$}
    \EndFor.
  % \Inputs{$x=(x_1, \ldots, x_p)$, $q_1$, $Q = (Q_1, \ldots, Q_{p-1}), f=(f_1, \ldots, f_p)$}
  \end{algorithmic}
\end{algorithm}
\begin{algorithm}[!h]
  \caption{Forward-backward sampling (backward pass)}  \label{alg:hmm_sampling_back}
  \begin{algorithmic}[1] 
    \State{\textbf{initialize} $j = p$, \; $Q_{p+1}(k|l) = 1$ for all $k,l$}
    \For{$j=p$ \textbf{to} $1$ (backward)}
    \State{\textbf{sample} $z_j$ according to $\pi_j(z_j)= \frac{Q_{j+1}(z_{j+1} | z_j) \alpha_{j}(z_j)}{\sum_k Q_{j+1}(z_{j+1} | k)
          \alpha_{j}(k)}
     $
    }
    \EndFor.
  \end{algorithmic}
\end{algorithm}

\section{Hidden Markov models in genome-wide association studies}

Now that we have an algorithm to perform controlled variable selection in problems where the covariates are well described by an HMM, we can discuss its practical applicability to GWAS.

\subsection{Modeling single-nucleotide polymorphisms} \label{sec:HMM_SNP}

In a GWAS, the response $Y$ is the status of a disease or a quantitative trait of interest, while each sample of $X$ consists of the genotype for a set of SNPs. In particular, we consider the case in which $X \in \{0,1,2\}^p$ collects unphased genotypes. For simplicity, in this section we restrict our attention to a single chromosome, since distinct ones are typically assumed to be independent. Several HMMs, with different parametrizations, have been proposed to describe the block-like patterns observed in the distribution of the alleles at adjacent markers. %these variables. 
In this paper, we adopt the specific model implemented in the software \texttt{fastPHASE}, as discussed in \cite{Scheet2006} and outlined below. We opt for this model because we find that it offers both an intuitive interpretation and a remarkable computational efficiency. However, our knockoff construction presented in Section \ref{sec:hmm_knock} is not limited to this choice and could easily be implemented with other alternatives. \\

%While the mechanisms connecting genotypes to phenotypes are extremely complex and mostly unknown \cite{vanderSijde2014}, the structure of the former is better understood \cite{Slatkin2008}. In particular, correlations across different SNPs exhibit block-like patterns that can be described by a hidden Markov model, as discussed in the groundbreaking work of Li and Stephens \cite{Li2003}. This perspective has been broadly adopted in genetics, with successful applications to haplotype phasing \cite{Browning2011} and genotype imputation \cite{Marchini2010}. \\
%The latter problem amounts to inferring missing values in partially observed genotype sequences and is of great importance for genome-wide association studies. Today, most of the best imputation techniques exploit variations of the Li and Stephens model. The software \texttt{Impute} \cite{Marchini2007,howie2009}, \texttt{MaCH} \cite{Li2010}, \texttt{fastPHASE} \cite{Scheet2006}, \texttt{Bimbam}\cite{guam2008} and \texttt{Beagle} \cite{Browning2007, Browning2016} all rely on HMMs and achieve extremely high accuracy. In our work, we describe the distribution of sequences of SNPs by adopting the specific HMM implemented in \texttt{fastPHASE}. This is described in \cite{Scheet2006} and it offers both an intuitive interpretation and remarkable computational efficiency. For the sake of completeness, we summarize it below.\\

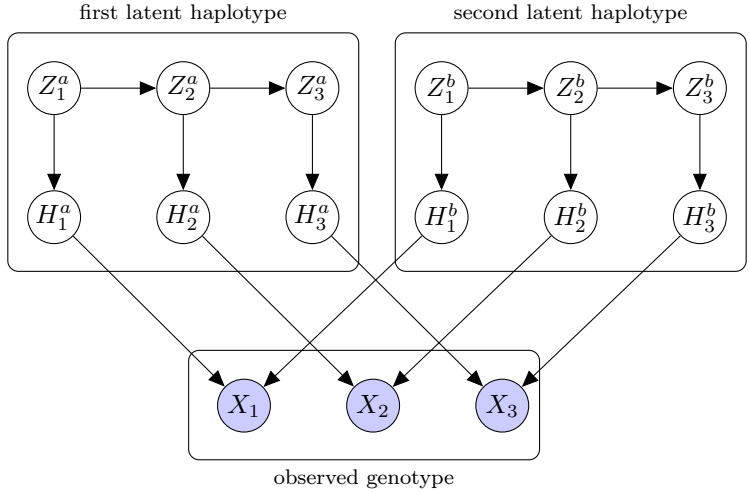
\begin{figure}[!htb]
      \centering
      \begin{tikzpicture}
        \node[latent, fill=white, fill opacity=0.2, text opacity=1] (Z11) {$Z^{a}_{1}$} ; %
        \node[latent, right=of Z11, fill=white, fill opacity=0.2, text opacity=1] (Z12) {$Z^{a}_{2}$} ; %
        \node[latent, right=of Z12, fill=white, fill opacity=0.2, text opacity=1] (Z13) {$Z^{a}_{3}$} ; %
        \node[obs, below=of Z11, fill=white, fill opacity=0.2, text opacity=1] (H11) {$H^{a}_{1}$} ; %
        \node[obs, below=of Z12, fill=white, fill opacity=0.2, text opacity=1] (H12) {$H^{a}_{2}$} ; %
        \node[obs, below=of Z13, fill=white, fill opacity=0.2, text opacity=1] (H13) {$H^{a}_{3}$} ; %
        \node[latent, below right = 2cm and 0.3cm of H12, fill=blue, fill opacity=0.2, text opacity=1] (X1) {$X_{1}$} ; %
        \node[latent, right=of X1, fill=blue, fill opacity=0.2, text opacity=1] (X2) {$X_{2}$} ; %
        \node[latent, right=of X2, fill=blue, fill opacity=0.2, text opacity=1] (X3) {$X_{3}$} ; %
        \node[obs, right=of Z13, fill=white, fill opacity=0.2, text opacity=1] (Z21) {$Z^{b}_{1}$} ; %
        \node[obs, right=of Z21, fill=white, fill opacity=0.2, text opacity=1] (Z22) {$Z^{b}_{2}$} ; %
        \node[obs, right=of Z22, fill=white, fill opacity=0.2, text opacity=1] (Z23) {$Z^{b}_{3}$} ; %
        \node[latent, below=of Z21, fill=white, fill opacity=0.2, text opacity=1] (H21) {$H^{b}_{1}$} ; %
        \node[latent, right=of H21, fill=white, fill opacity=0.2, text opacity=1] (H22) {$H^{b}_{2}$} ; %
        \node[latent, right=of H22, fill=white, fill opacity=0.2, text opacity=1] (H23) {$H^{b}_{3}$} ; %
        \plate[inner sep=0.25cm, xshift=-0.12cm, yshift=0.12cm, label=south:observed genotype] {plateX} {(X1) (X2) (X3)}{}; 
        \plate[inner sep=0.25cm, yshift=0.12cm, label=north:first latent haplotype] {plateX} {(H11) (H12) (H13) (Z11) (Z12) (Z13)} {}; 
        \plate[inner sep=0.25cm, yshift=0.12cm, label=north:second latent haplotype] {plateZt} {(Z21) (Z22) (Z23) (H21) (H22) (H23)} {}; 
        \edge[opacity=1] {Z11} {Z12}; %
        \edge[opacity=1] {Z12} {Z13}; %
        \edge[opacity=1] {Z11} {H11}; %
        \edge[opacity=1] {Z12} {H12}; %
        \edge[opacity=1] {Z13} {H13}; %
        \edge[opacity=1] {Z21} {Z22}; %
        \edge[opacity=1] {Z22} {Z23}; %
        \edge[opacity=1] {Z21} {H21}; %
        \edge[opacity=1] {Z22} {H22}; %
        \edge[opacity=1] {Z23} {H23}; %
        \edge[opacity=1] {H11} {X1}; %
        \edge[opacity=1] {H21} {X1}; %
        \edge[opacity=1] {H12} {X2}; %
        \edge[opacity=1] {H22} {X2}; %
        \edge[opacity=1] {H13} {X3}; %
        \edge[opacity=1] {H23} {X3}; %
      \end{tikzpicture}
      \caption{Sequence of $p=3$ genotype SNPs (blue) as the sum of two i.i.d.~HMM haplotypes (white).}
%\\ Each haplotype sequence is distributed as a HMM with $K$ hidden states, but only the blue nodes can be observed. Equivalently, one can consider the sum of their emissions as a single HMM with $\frac{1}{2}K(K+1)$ hidden states.}
      \label{fig:hap_hmm}
    \end{figure}

Without loss of generality, the unphased genotype of a diploid
individual (e.g.~human) can be seen as the component-wise sum of two
unobserved sequences, called haplotypes $H = (H_1, \ldots, H_p)$. Here
$H_i \in \{0,1\}$ is a binary variable that represents the allele on
the $i$th marker. The main modeling assumption is that the two
haplotypes are i.i.d.~HMMs. This idea is sketched in Figure
\ref{fig:hap_hmm}, for the special case $p=3$. In order to describe
the parametrization of this model, we begin by focusing on a single
sequence $H$. Its distribution is in the same form as the HMM defined
earlier in \eqref{eq:HMM_def}, 
\begin{align*}
\begin{split}
\begin{cases}
  Z  \sim \text{MC}\left(q_1^{\text{hap}}, Q^{\text{hap}}\right) & \text{(latent Markov chain for one haplotype)}, \\
  H_j | Z  \sim H_j | Z_j \overset{\text{ind.}}{\sim} f^{\text{hap}}_j(H_j|Z_j) & \text{(haplotype emission distribution)},
\end{cases}
\end{split}
\end{align*}
with an associated latent Markov chain $Z=(Z_1,\ldots,Z_p)$. Each variable in $Z$ can take one of $K$ possible values, that indicate membership to a specific group of closely related haplotypes. Borrowing from the literature on fuzzy patterns in the DNA sequence, we use the term ``haplotype motifs'' to describe these: each haplotype motif is characterized by specific allele frequencies at the various markers. Intuitively, one can thus see $H$ as a mosaic of segments, each originating from one of $K$ distinct haplotypes motifs, that can be loosely taken as representing the genome of the population founders. It is important to note that while this model provides a good  description of the local patterns of correlation originating from genetic recombination, %but it does not claim to realistically capture the real evolutionary process. 
it is phenomenological in nature and it should not be interpreted as an accurate representation of the real sequence of mutations and recombinations that originate the  haplotypes in the population.

The marginal distribution of the first element of the hidden Markov chain $Z$ is
\begin{align*}
  q_1^{\text{hap}} \left(k\right) = \alpha_{1,k}, \qquad k \in \{1,\ldots,K\},
\end{align*}
while the transition matrices are
\begin{align*}
  Q^{\text{hap}}_j(k'| k)
  & = 
  \begin{cases}
    e^{-r_j} + (1 - e^{-r_j} ) \, \alpha_{j,k'}, & k' = k, \\
    (1 - e^{-r_j} ) \,\alpha_{j,k'}, & k' \neq k.
  \end{cases}
\end{align*}
The parameters $\alpha = (\alpha_{j,k})_{k \in \{1,\ldots,K\}, j \in \{1,\ldots, p\}}$ describe the propensity of different ``haplotype motifs'' to succeed each other. The occurrence of a transition is regulated by the values of $r = (r_1, \ldots, r_p)$, which are intuitively related to the genetic recombination rates.

Once a sequence of ancestral segments is fixed, the allele $H_j$ in position $j$ is sampled from the emission distribution
\begin{align*}
  f^{\text{hap}}_j(h_j; z_j, \theta)
  & = \begin{cases}
    1-\theta_{j,z_j},  &h_j = 0, \\
    \theta_{j,z_j}, & h_j = 1.
  \end{cases}
\end{align*}
The parameters $\theta = (\theta_{j,k})_{k \in \{1,\ldots,K\}, j \in \{1,\ldots, p\}}$ represent the frequency of allele one across all the polymorphisms, in each of the ancestral haplotype motifs. These can be estimated along with $\alpha$ and $r$. \\

Having defined the distribution of $H$, we return our attention to the observed genotype vector. 
%We have assumed that $X$ can be described as the elementwise sum of the two i.i.d.~HMM haplotype sequences defined above.
By definition, the genotype $X$ of an individual is obtained by pairing, marker by marker, the alleles on each of his haplotypes and discarding information on the haplotype of origin (phase). 
Then---under the standard assumptions (i.e.~random mating/Hardy-Weinberg equilibrium)---the population from which the genotype vector of a subject is randomly sampled can be described as the element-wise sum of two i.i.d.~haplotypes with distribution described by the HMM above. 
 Consequently, its distribution is also an HMM. The latent Markov chain has bivariate states, corresponding to unordered pairs of haplotype latent states.
 It is easy to verify that these can take $K_{\text{eff}} = \frac{1}{2}K(K+1)$ possible values. By this construction, it follows that the initial-state probabilities for the genotype model are:
\begin{align} \label{eq:fastphase_1}
  q_1^{\text{gen}} \left(\{k_a, k_b\} \right) 
  & = 
  \begin{cases}
    (\alpha_{1,k_a})^2, & k_a = k_b, \\
    2\alpha_{1,k_a} \alpha_{1,k_b}, & k_a \neq k_b,
  \end{cases}
\end{align}
and the transition matrices are
\begin{align} \label{eq:fastphase_b}
  Q_j^{\text{gen}}(\{k'_{a}, k'_b\}| \{k_{a}, k_b\} )
  & = \begin{cases}
    Q^{\text{hap}}_j(k'_a| k_a) \, Q^{\text{hap}}_j(k'_b| k_b) + Q^{\text{hap}}_j(k'_b| k_a) \, Q^{\text{hap}}_j(k'_a| k_b), & k'_a \neq k'_b, \\
%    Q^{\text{hap}}_j(k'_a, k_a) \, Q^{\text{hap}}_j(k'_b, k_b) + Q^{\text{hap}}_j(k'_b, k_a) \, Q^{\text{hap}}_j(k'_a, k_b) & \text{if } k_a \neq k_b \text{ and } k'_a \neq k'_b, \\
    Q^{\text{hap}}_j(k'_a| k_a) \, Q^{\text{hap}}_j(k'_b| k_b), & \text{otherwise}.
  \end{cases}
\end{align} 
Similarly, the HMM emission probabilities for $X_j$ are:
\begin{align} \label{eq:fastphase_3}
  f_j(x_j; \{k_a, k_b\}, \theta)
  & = \begin{cases}
    (1-\theta_{j,k_a})(1-\theta_{j,k_b}), & x_j = 0, \\
    \theta_{j,k_a}(1-\theta_{j,k_b}) + (1-\theta_{j,k_a})\theta_{j,k_b}, & x_j = 1, \\
    \theta_{j,k_a} \theta_{j,k_b}, & x_j = 2.
  \end{cases}
\end{align}

\subsection{Parameter estimation}
In general, model-free knockoffs are guaranteed to control the FDR when the marginal distribution of $X$ is known exactly. However, exact knowledge is unrealistic in practical applications and some degree of approximation is ultimately unavoidable. Since we have argued that the HMM model in \eqref{eq:fastphase_1}--\eqref{eq:fastphase_3} offers a sensible and tractable description of real genotypes, it makes sense to estimate the $p(2K+1)$ parameters in $(r, \alpha, \theta)$ from the available data. In the usual GWAS setting, one disposes of $n\gg 2K+1$ observations for each of the $p$ sites, so this task is not unreasonable. Moreover, the validity of this approach is empirically verified in our simulations with real genetic covariates, as discussed in the next section. Alternatively, if additional unsupervised observations (i.e.~including only the covariates) from the same population are available, one could consider including them in this phase in order to improve the estimation. \\

In practice, the estimation of the HMM parameters can be efficiently performed through standard EM techniques and it only requires $O(npK^2)$ time, where $n$ is the number of individuals. This procedure is already implemented in the imputation software \texttt{fastPHASE}, which is freely available. The latter fits the model described above, for the original purpose of recovering missing observations, and it conveniently provides us with the estimates $(\hat{r}, \hat{\alpha}, \hat{\theta})$ needed to sample a knockoff copy of the genotype. An important advantage of the HMM representation is that the number of parameters only grows linearly in $p$, thus greatly reducing the risk of overfitting, compared to a multivariate Gaussian approximation. In our case, the model complexity is controlled by the number $K$ of haplotype motifs, which can be chosen by cross-validation (the typical values recommended in \cite{Scheet2006} are around $10$). We have observed that our knockoffs procedure is relatively robust and is not prone to overfitting for a range of different choices of $K$.

\section{Numerical Simulations}

\subsection{Knockoffs for Markov chain variables} \label{sec:sim_dmc}
We begin to demonstrate the use of our procedure by performing numerical experiments in the case of Markov chain variables.

\subsubsection{A toy model}
We consider a vector $X$ of $p=1000$ covariates distributed as a discrete Markov chain taking values in a state space $\mathcal{X} = \{-2,-1,0,+1,+2\}$ of size $K=|\mathcal{X}|=5$. In the notation of \eqref{eq:def_dmc}, this can be written as $X \sim \text{MC}(q_1, Q)$, with an initial distribution $q_1$ assumed to be uniform on $\mathcal{X}$. For each $j \in \{1,\ldots,p-1\}$, we set:
\begin{align*}
  Q_j(k|l)
  & = \begin{cases}
    \frac{1}{K} + \gamma_j \left(1-\frac{1}{K}\right), & k=l, \\
    \left[ 1 - \frac{1}{K} - \gamma_j \left(1-\frac{1}{K}\right) \right] \frac{1}{K-1}, & k \neq l,
  \end{cases}
\end{align*}
where the hyper-parameters $\gamma_j$ are once randomly sampled $\gamma_j \dIid \text{Uniform}\left( [0,0.5] \right)$ and then held constant. \\
Conditional on $X=(X_1,\ldots,X_p)$, the response $Y$ is sampled from
a binomial generalized linear model with a logit link function. The coefficient vector $\beta$ has 60 non-zero elements, which correspond to the set $\mathcal{S}$ of relevant features. In summary,
\begin{align*}
  & Y|X \sim \text{Bernoulli}\left( \text{logit} \left( X^T \beta \right)\right),
  & \text{where} \quad \beta_j= \begin{cases}
    \frac{a}{\sqrt{n}}, & j \in \mathcal{S}, \\
    0, & \text{otherwise.}
  \end{cases}
\end{align*}
Above, the signal amplitude $a$ is a parameter that we can vary in the simulations.

\subsubsection{Effect of signal amplitude}
\begin{figure}[!htb]
  \centering
  \begin{subfigure}[b]{0.45\textwidth}      \captionsetup{justification=centering}
    \centering
    \includegraphics[width=\textwidth]{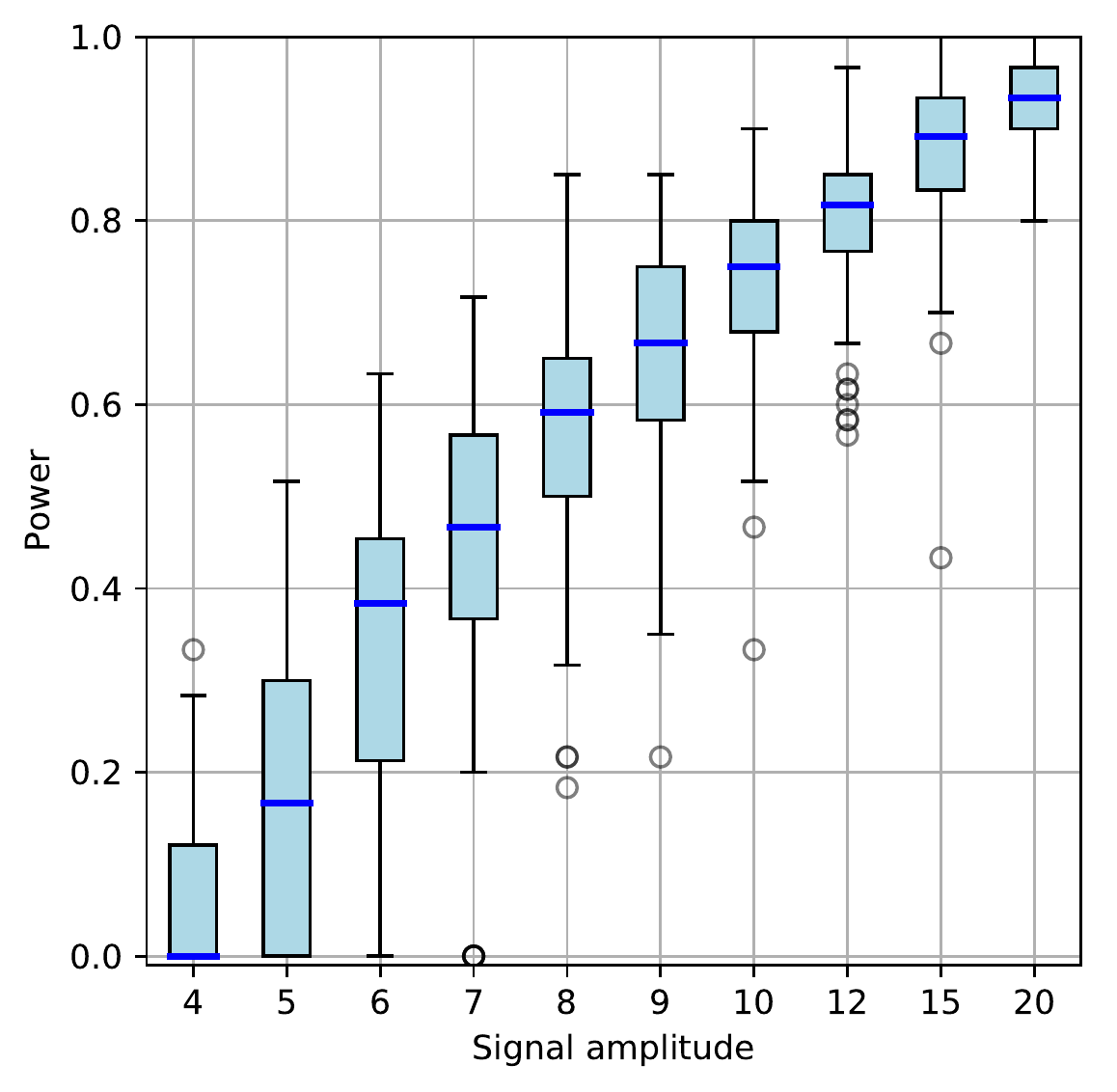}
    \caption{}
    \label{fig:sim_dmc_01_power}
  \end{subfigure}
  \begin{subfigure}[b]{0.45\textwidth}      \captionsetup{justification=centering}
    \centering
    \includegraphics[width=\textwidth]{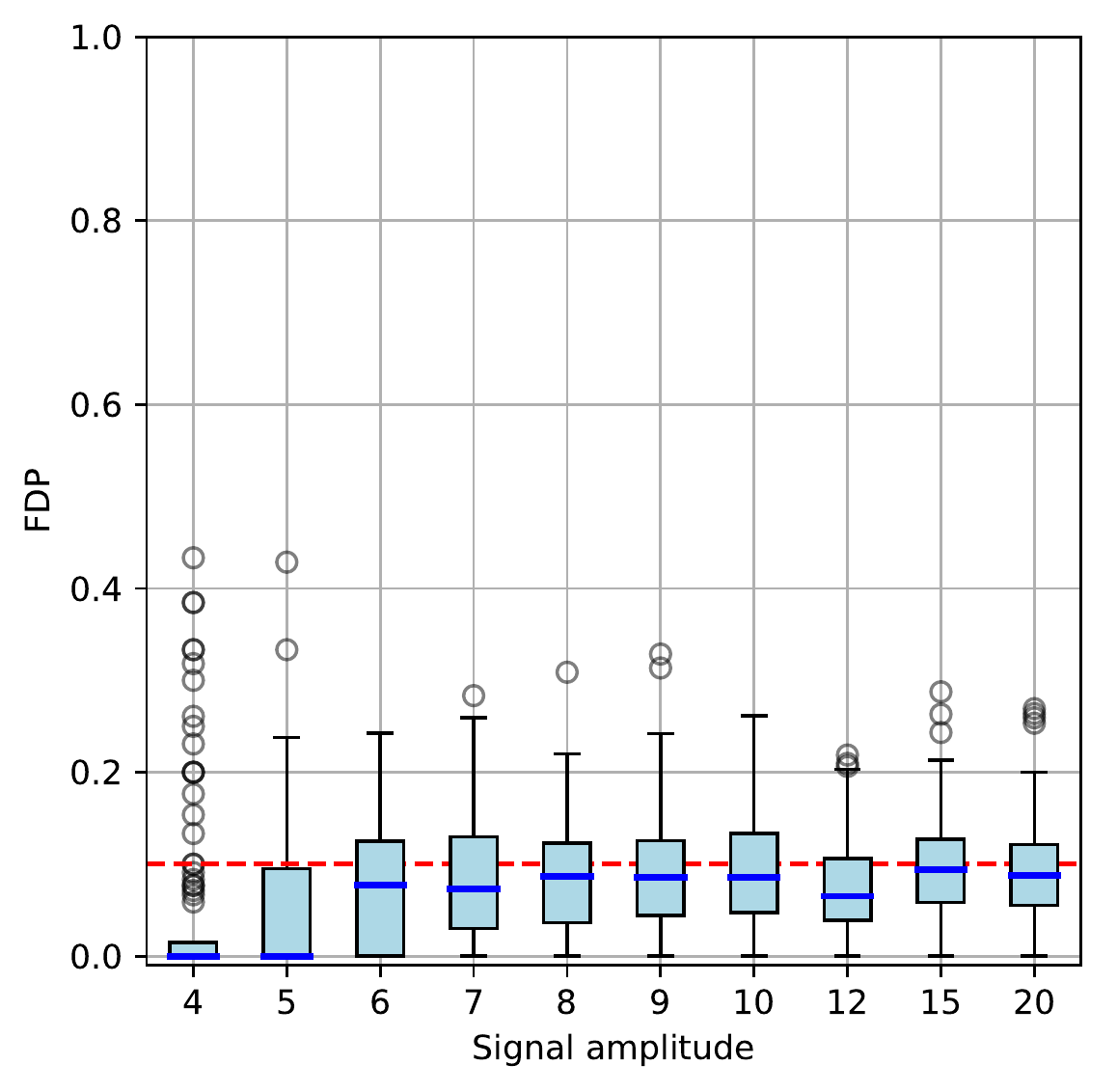}
    \caption{}
  \label{fig:sim_dmc_01_fdr}
  \end{subfigure}
    \caption{Power (a) and FDP (b) of our procedure in a simulation with $n=1000$ and $p=1000$, over 100 independent experiments. Variables are distributed as a discrete Markov chain. The knockoff copies are constructed using the true model parameters. The response $Y|X$ is sampled from a logistic regression model. The dashed red line in (b) indicates the target FDR level $\alpha=0.1$. }
  \label{fig:sim_dmc_01}
\end{figure}

We draw $n=1000$ independent observations of $(X,Y)$ from the model
described above. For different values of the signal amplitude $a$, we
apply the knockoff construction procedure of Section \ref{sec:dmc},
using the true model parameters $(q_1,Q)$. It is interesting to note
that, since $p=n$, the observations are perfectly
separable\footnote{There exists a hyperplane
  in the feature space that perfectly separates the two classes of
  $Y$.} and the maximum likelihood estimate of $\beta$, therefore,
does not exist. This is the reason why it is useful to leverage some sparsity in order to identify the relevant variables. As variable importance measures, we compute $W_j = |\hat{\beta}_j(\lambda^{\text{CV}})|-|\hat{\beta}_{j+p}(\lambda^{\text{CV}})|$, where $\hat{\beta}_j(\lambda^{\text{CV}})$ and $\hat{\beta}_{j+p}(\lambda^{\text{CV}})$ are the logistic regression coefficients for the $j$th variable and its knockoff copy, respectively, regularized with an $\ell_1$-norm penalty chosen by 10-fold cross-validation. Finally, we estimate the set of relevant variables using the $\textit{knockoff+}$ threshold for strict FDR control. The results shown in Figure \ref{fig:sim_dmc_01} and Table \ref{tab:sim_dmc} correspond to 100 independent replications of this experiment. Empirically, our method is confirmed to control the FDR for all values of the signal amplitude. As it should be expected, the actual false discovery proportion (FDP) is not always below the target value, but is quite concentrated around its mean.

% \begin{table}[!htb]
% \begin{centering}
% \begin{longtable}{ |>{\centering\arraybackslash}m{15mm}||c|c||c|c|}
% \hline
% Signal & \multicolumn{2}{c||}{True $F_X$} & \multicolumn{2}{c|}{Estimated $F_X$}  \\ \cline{2-5}
% amplitude & FDR (95\% c.i.) & Power (95\% c.i.) & FDR (95\% c.i.) & Power (95\% c.i.) \\ 
% \hline
% 4& 0.050 $\pm \; 0.020$ & 0.051 $\pm \; 0.018$ &0.054 $\pm \; 0.020$ & 0.064 $\pm \; 0.020$ \\ 
% 5& 0.057 $\pm \; 0.017$ & 0.154 $\pm \; 0.031$ &0.062 $\pm \; 0.019$ & 0.155 $\pm \; 0.031$ \\ 
% 6& 0.083 $\pm \; 0.014$ & 0.329 $\pm \; 0.034$ &0.078 $\pm \; 0.015$ & 0.312 $\pm \; 0.035$ \\ 
% 7& 0.084 $\pm \; 0.014$ & 0.446 $\pm \; 0.031$ &0.091 $\pm \; 0.015$ & 0.449 $\pm \; 0.031$ \\ 
% 8& 0.086 $\pm \; 0.012$ & 0.566 $\pm \; 0.025$ &0.089 $\pm \; 0.013$ & 0.560 $\pm \; 0.029$ \\ 
% 9& 0.092 $\pm \; 0.013$ & 0.658 $\pm \; 0.024$ &0.088 $\pm \; 0.013$ & 0.653 $\pm \; 0.023$ \\ 
% 10& 0.093 $\pm \; 0.011$ & 0.730 $\pm \; 0.020$ &0.096 $\pm \; 0.011$ & 0.741 $\pm \; 0.017$ \\ 
% 12& 0.076 $\pm \; 0.010$ & 0.798 $\pm \; 0.016$ &0.078 $\pm \; 0.011$ & 0.795 $\pm \; 0.016$ \\ 
% 15& 0.096 $\pm \; 0.011$ & 0.874 $\pm \; 0.016$ &0.092 $\pm \; 0.012$ & 0.878 $\pm \; 0.014$ \\ 
% 20& 0.094 $\pm \; 0.011$ & 0.930 $\pm \; 0.009$ &0.098 $\pm \; 0.011$ & 0.933 $\pm \; 0.009$ \\ 
% \hline
% \end{longtable}
% \end{centering}
% \end{table}

\subsubsection{Robustness to overfitting}
\begin{figure}[!htb]
  \centering
  \begin{subfigure}[b]{0.45\textwidth}      \captionsetup{justification=centering}
    \centering
    \includegraphics[width=\textwidth]{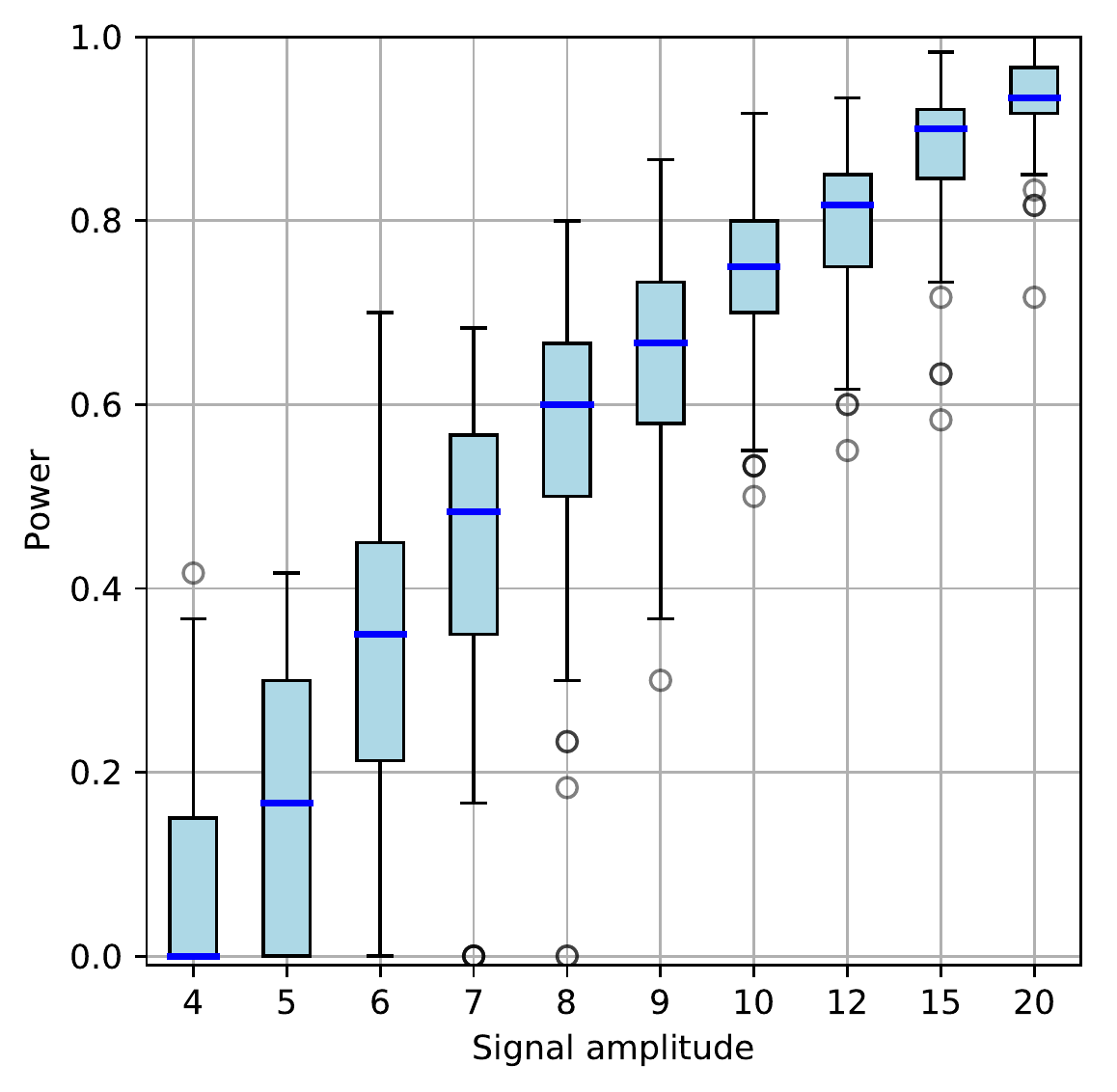}
    \caption{}
    \label{fig:sim_dmc_03_power}
  \end{subfigure}
  \begin{subfigure}[b]{0.45\textwidth}      \captionsetup{justification=centering}
    \centering
    \includegraphics[width=\textwidth]{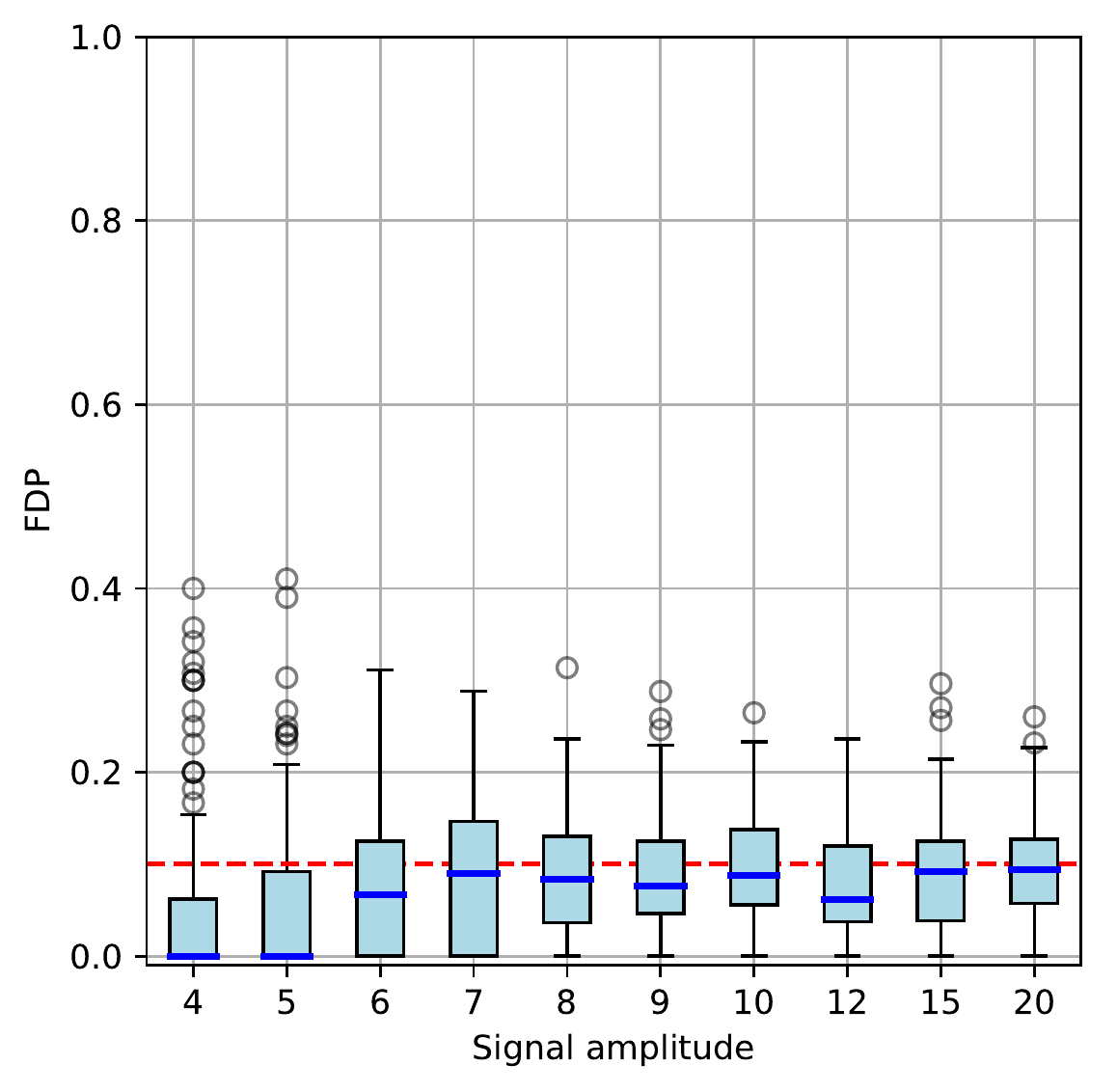}
    \caption{}
  \label{fig:sim_dmc_03_fdr}
  \end{subfigure}
    \caption{Power (a) and FDP (b) of our procedure with simulated Markov chain covariates, with knockoffs sampled using estimates of the transition matrices obtained from the same dataset. The setup is otherwise the same as that in Figure \ref{fig:sim_dmc_01}.}
  \label{fig:sim_dmc_03}
\end{figure}

In the previous example, we generated the knockoff variables using the real distribution of $X$. However, in most practical applications this is not known exactly and it must be estimated from the available data. In a more realistic situation one may have some prior knowledge that a Markov chain is a good model for the covariates, but ignore the exact form of the transition matrices. Therefore, we repeat the previous experiment, generating instead the knockoff copies $\tilde{X}$ from the fitted values of the Markov chain parameters. The estimates $(\hat{q}_1, \hat{Q})$ are obtained by maximum-likelihood with Laplace smoothing\footnote{This is a well-known technique that can be used to improve the estimation of the transition matrices. In order to avoid estimating any transition probabilities as zero, we simply add one to all transition counts.} on all the available observations of $X$. The results shown in Figure \ref{fig:sim_dmc_03} and Table \ref{tab:sim_dmc} are very similar to those of Figure \ref{fig:sim_dmc_01}. This shows that the FDR is still controlled, and it also suggests that our procedure is robust to fitting the feature distribution. \\

\begin{table}[!htbp]
\centering
\begin{tabular}{ |>{\centering\arraybackslash}m{15mm}||c|c||c|c|}\hline
Signal & \multicolumn{2}{c||}{True $F_X$} & \multicolumn{2}{c|}{Estimated $F_X$}  \\ \cline{2-5}amplitude & FDR (95\% c.i.) & Power (95\% c.i.) & FDR (95\% c.i.) & Power (95\% c.i.) \\ 
\hline
4& 0.050 $\pm \; 0.020$ & 0.051 $\pm \; 0.018$ &0.054 $\pm \; 0.020$ & 0.064 $\pm \; 0.020$ \\ 
5& 0.057 $\pm \; 0.017$ & 0.154 $\pm \; 0.031$ &0.062 $\pm \; 0.019$ & 0.155 $\pm \; 0.031$ \\ 
6& 0.083 $\pm \; 0.014$ & 0.329 $\pm \; 0.034$ &0.078 $\pm \; 0.015$ & 0.312 $\pm \; 0.035$ \\ 
7& 0.084 $\pm \; 0.014$ & 0.446 $\pm \; 0.031$ &0.091 $\pm \; 0.015$ & 0.449 $\pm \; 0.031$ \\ 
8& 0.086 $\pm \; 0.012$ & 0.566 $\pm \; 0.025$ &0.089 $\pm \; 0.013$ & 0.560 $\pm \; 0.029$ \\ 
9& 0.092 $\pm \; 0.013$ & 0.658 $\pm \; 0.024$ &0.088 $\pm \; 0.013$ & 0.653 $\pm \; 0.023$ \\ 
10& 0.093 $\pm \; 0.011$ & 0.730 $\pm \; 0.020$ &0.096 $\pm \; 0.011$ & 0.741 $\pm \; 0.017$ \\ 
15& 0.096 $\pm \; 0.011$ & 0.874 $\pm \; 0.016$ &0.092 $\pm \; 0.012$ & 0.878 $\pm \; 0.014$ \\ 
20& 0.094 $\pm \; 0.011$ & 0.930 $\pm \; 0.009$ &0.098 $\pm \; 0.011$ & 0.933 $\pm \; 0.009$ \\ 
\hline
\end{tabular}
\caption{FDR and average power in the numerical experiments of Figure \ref{fig:sim_dmc_01} and \ref{fig:sim_dmc_03}. We compare the results obtained with knockoff variables created using the exact (left) and estimated (right) Markov chain model parameters.} \label{tab:sim_dmc}
\end{table}

\begin{figure}[!htb]
  \centering
  \begin{subfigure}[b]{0.45\textwidth}      \captionsetup{justification=centering}
    \centering
    \includegraphics[width=\textwidth]{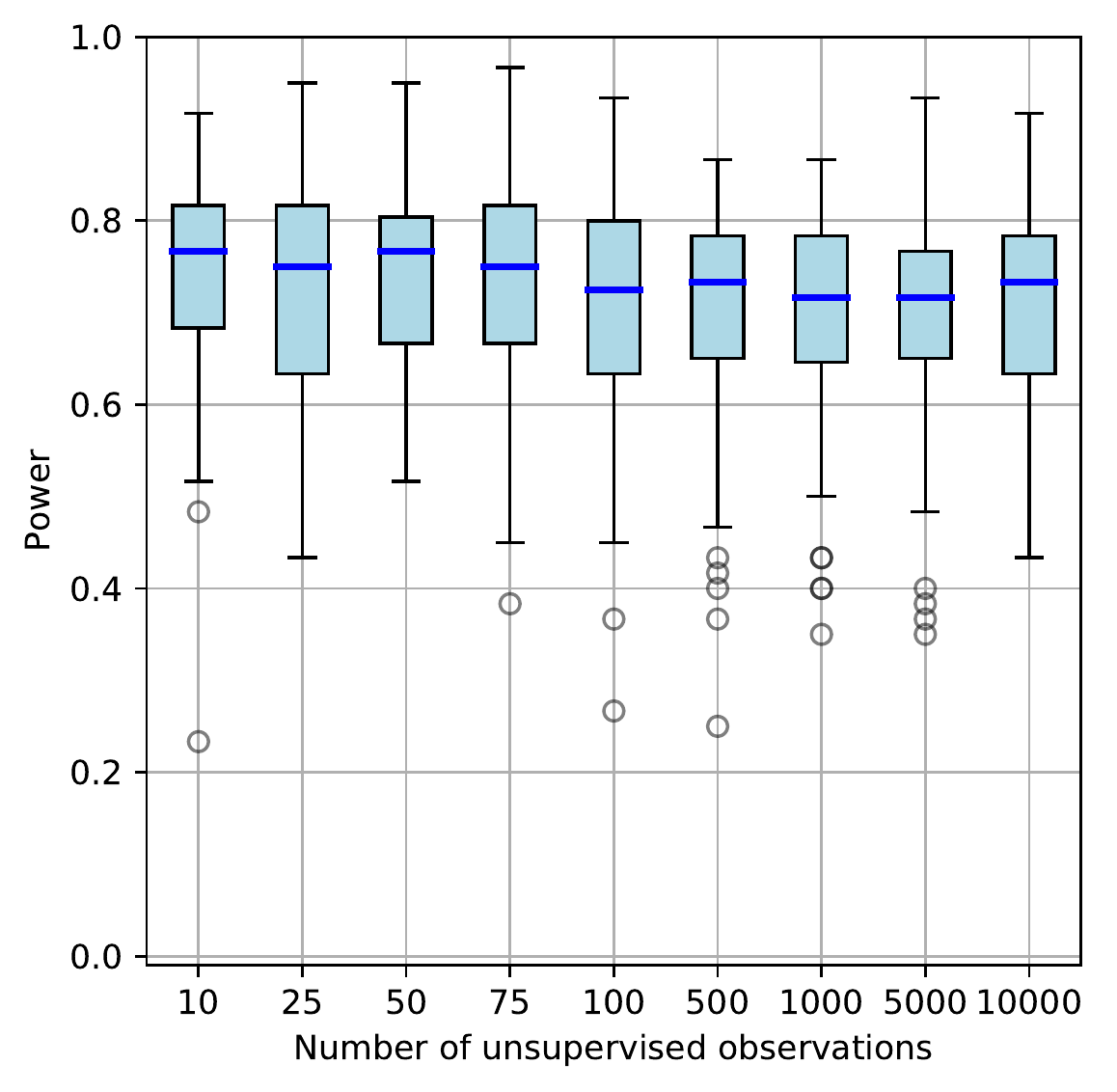}
    \caption{}
    \label{fig:sim_dmc_02_power}
  \end{subfigure}
  \begin{subfigure}[b]{0.45\textwidth}      \captionsetup{justification=centering}
    \centering
    \includegraphics[width=\textwidth]{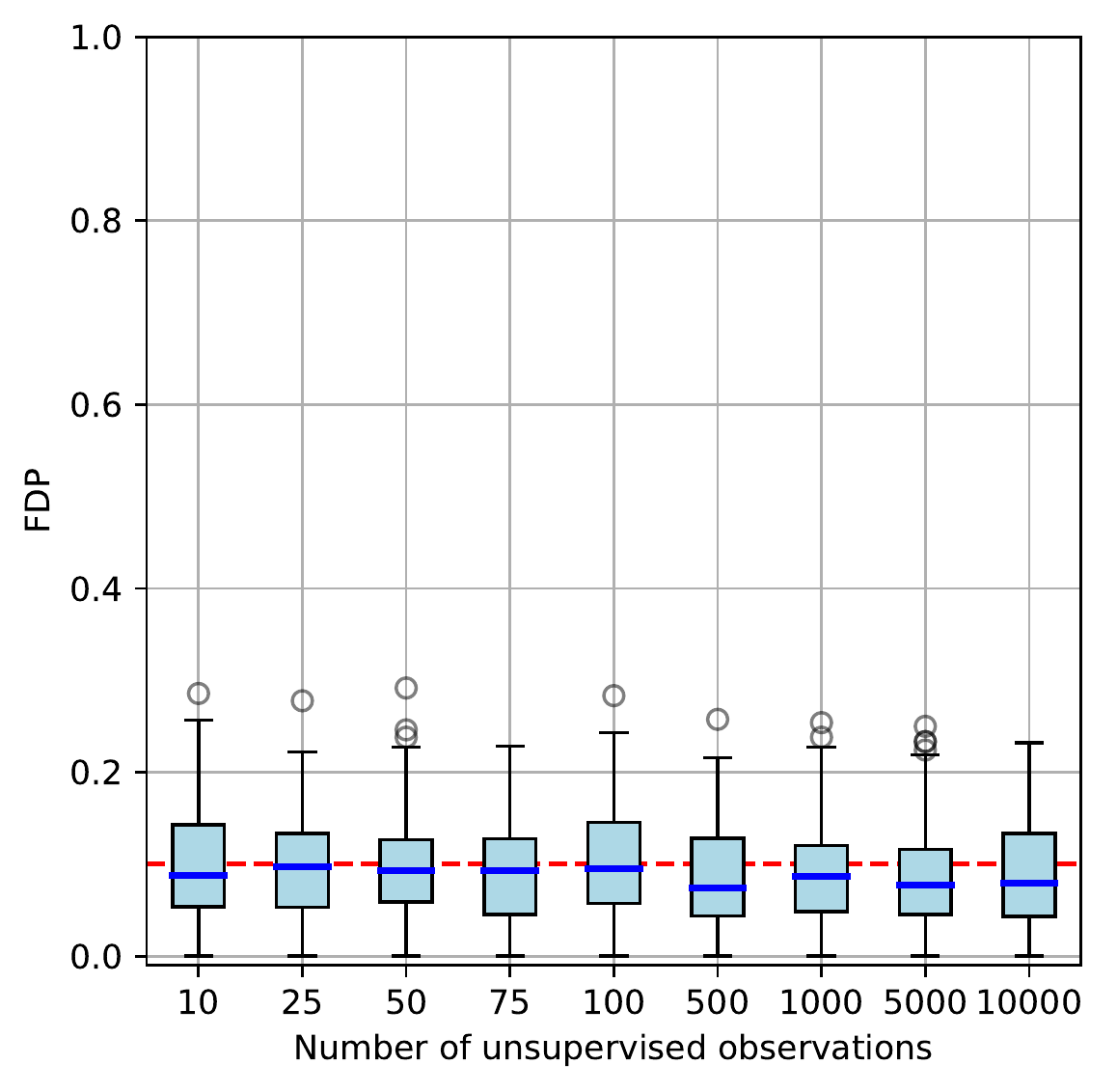}
    \caption{}
  \label{fig:sim_dmc_02_fdr}
  \end{subfigure}
    \caption{Power (a) and FDP (b) of our procedure with simulated Markov chain covariates. Knockoffs are sampled using estimates of the transition matrices obtained from an independent dataset of $n_u$ unsupervised observations of $X$, for different values of $n_u$. The signal amplitude is $a=10$. The setup is otherwise the same as that in Figure \ref{fig:sim_dmc_01}.}
  \label{fig:sim_dmc_02}
\end{figure}

Alternatively, if additional unsupervised samples are available, one can use them to improve the estimation of the covariate distribution. We illustrate this idea by generating unlabeled datasets of varying size $n_\text{u}$, from the same population. In principle, one could use both the supervised and the unsupervised observations of $X$ to estimate the parameters of $F_X$. However, we choose to fit the parameters only on the latter, in order to better observe the effect of overfitting. For a range of values of $n_\text{u}$, we compute $(\hat{q}_1, \hat{Q})$ and proceed as in the previous examples, repeating the experiment 100 times. The results are shown in Figure \ref{fig:sim_dmc_02}. We observe that our procedure is robust to overfitting. Even in the extreme cases in which $n_u$ is very small (i.e.~$n_u \leq 50$), the empirical FDR is below the nominal value, while for larger values of $n_u$ the validity of the FDR control is clear.
%This is particularly reassuring because these worst cases would not even be particularly relevant from a practical perspective. In fact, the results shown in Figure \ref{fig:sim_dmc_03} suggest that one can safely use all observations for fitting the distribution of $X$, so in this example one should be mostly interested in the regime of $n_u \geq n = 1000$.

\FloatBarrier

\subsection{Knockoffs for HMM variables} \label{sec:sim_hmm}
We continue our numerical experiments by generating knockoff copies of an HMM.

\subsubsection{A toy model}
We consider a vector $X$ of $p=1000$ covariates distributed as the HMM
defined below. The parametrization that we adopt is loosely inspired
by the \textit{left-right} models used for speech recognition
\cite{juang1991}, but we do not aim to realistically simulate any
specific application. Instead, we prefer to keep the model extremely
simple for the sake of exposition. Here, the latent Markov chain
$Z \sim \text{MC}(q_1, Q)$ takes on values in $\{0,1,\ldots,K-1\}^p$ and
its states evolve ``clockwise'' according to
\begin{align*}
   & q_1(k) = \begin{cases}
     1,  & k =1, \\
     0,  & \text{otherwise},
   \end{cases}
        & Q_j(k|l) = \begin{cases}
          0.9, &  k=l, \\
          0.1, & k= l+1 \mod K, \\ % \text{ or } (k,l)=(1,K)\\
          0, & \text{otherwise},
        \end{cases}
        && j \in \{2,\ldots,p\},
\end{align*}
for $k,l \in \{0,1,\ldots,K-1\}$.
Concretely, we let $K=9$ and we assume for simplicity that all
observed variables $X_j$ take on values in a set $\mathcal{X} = \{-4,-3,\ldots,+3,+4\}$, also of size $K$. The emission probabilities $f_j(x|z)$ are defined, for some $\gamma \in (0,1)$, as
\begin{align*}
  f_j(x|z)
  & = \begin{cases}
    \frac{\gamma}{2}, & (x+4) = z \text{ or } (x+4) = z+1, \\
    \frac{\gamma}{2}, & (x+4) = 0 \text{ and } z = K-1, \\
    \frac{1-\gamma}{K-2}, & \text{otherwise}.
  \end{cases}
\end{align*}
In this example, we set $\gamma=0.35$ because we have observed empirically that it yields an interesting structure with moderately strong correlations.\\

Conditional on $X=(X_1,\ldots,X_p)$, the response $Y$ is sampled from the same binomial generalized linear model of Section \ref{sec:sim_dmc}. Again, we vary the signal amplitude in the simulations.

\subsubsection{Effect of signal amplitude}
\begin{figure}[!htb]
  \centering
  \begin{subfigure}[b]{0.45\textwidth}      \captionsetup{justification=centering}
    \centering
    \includegraphics[width=\textwidth]{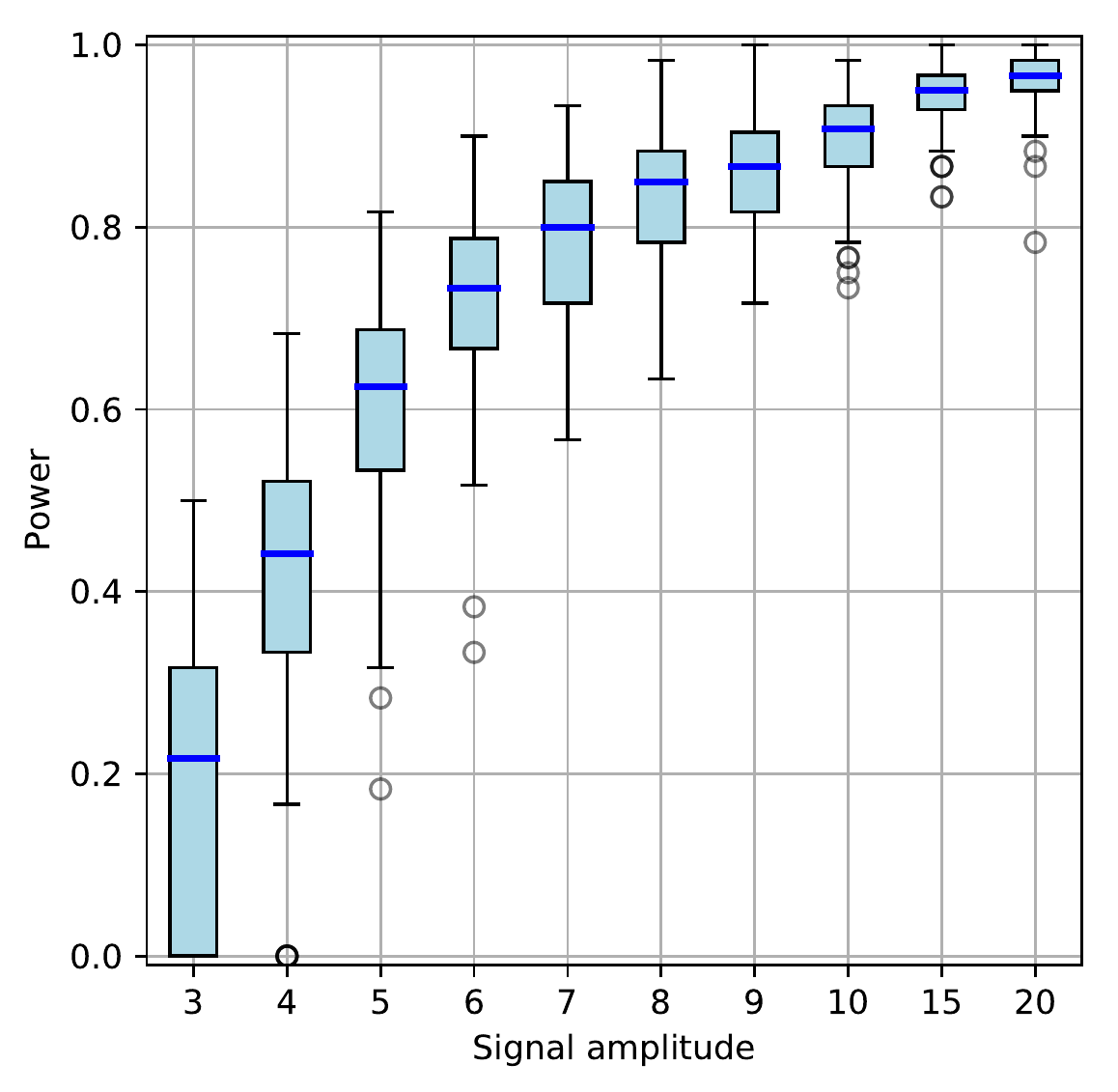}
    \caption{}
    \label{fig:sim_hmm_01_power}
  \end{subfigure}
  \begin{subfigure}[b]{0.45\textwidth}      \captionsetup{justification=centering}
    \centering
    \includegraphics[width=\textwidth]{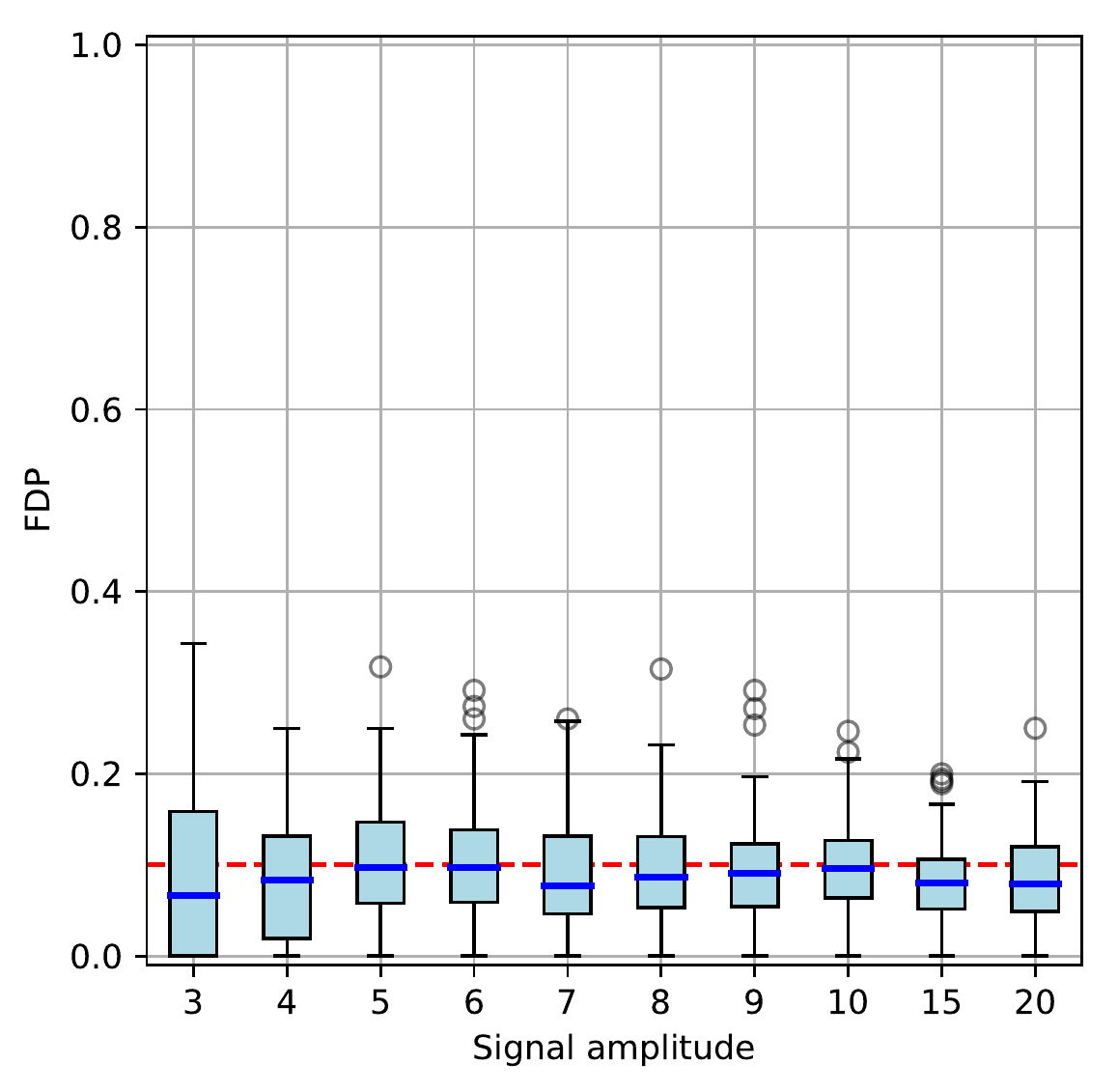}
    \caption{}
  \label{fig:sim_hmm_01_fdr}
  \end{subfigure}
    \caption{Power (a) and FDP (b) of our procedure in a simulation with $n=1000$ and $p=1000$, over 100 independent experiments. Variables are distributed as a discrete hidden Markov model. The knockoff copies are constructed using the true model parameters. The response $Y|X$ is sampled from a logistic regression model. The dashed red line in (b) indicates the target FDR level $\alpha=0.1$. }
  \label{fig:sim_hmm_01}
\end{figure}

We simulate $n=1000$ independent observations of $(X,Y)$ from the
model described above. For different values of the signal amplitude
$a$, we apply our method to construct knockoff copies of the HMM,
using the exact model parameters. We select relevant variables after
computing the same importance measures as in Section
\ref{sec:sim_dmc}, and applying the filter with a $\textit{knockoff+}$
threshold (target $\alpha=0.1$). The power and FDP shown in Figure
\ref{fig:sim_hmm_01} and Table \ref{tab:sim_hmm} correspond to 100
independent replications of this experiment. The results confirms that
our procedure accurately controls the FDR for all values of the signal amplitude.

\subsubsection{Robustness to overfitting}
\begin{figure}[!htb]
  \centering
  \begin{subfigure}[b]{0.45\textwidth}      \captionsetup{justification=centering}
    \centering
    \includegraphics[width=\textwidth]{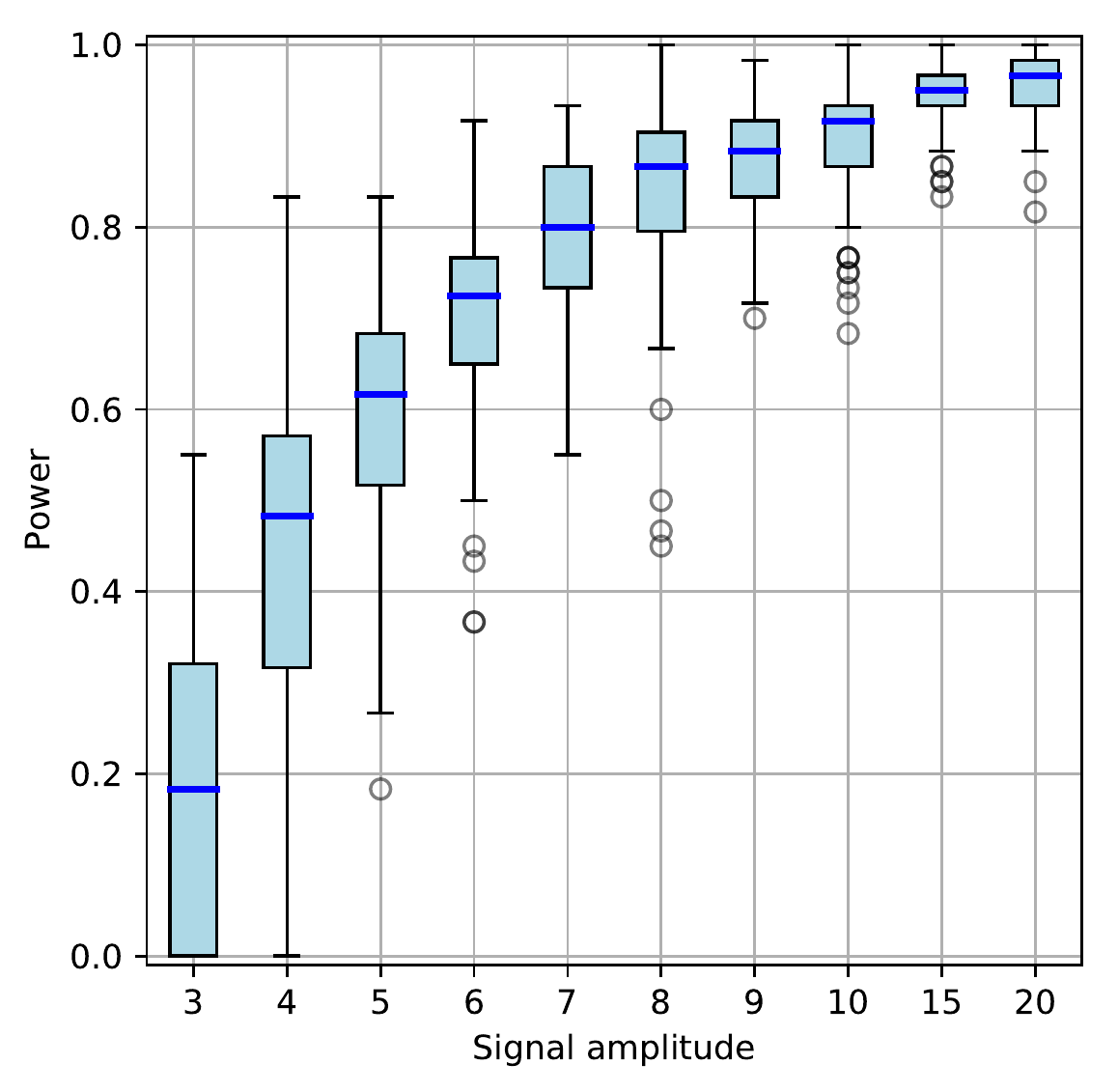}
    \caption{}
    \label{fig:sim_hmm_03_power}
  \end{subfigure}
  \begin{subfigure}[b]{0.45\textwidth}      \captionsetup{justification=centering}
    \centering
    \includegraphics[width=\textwidth]{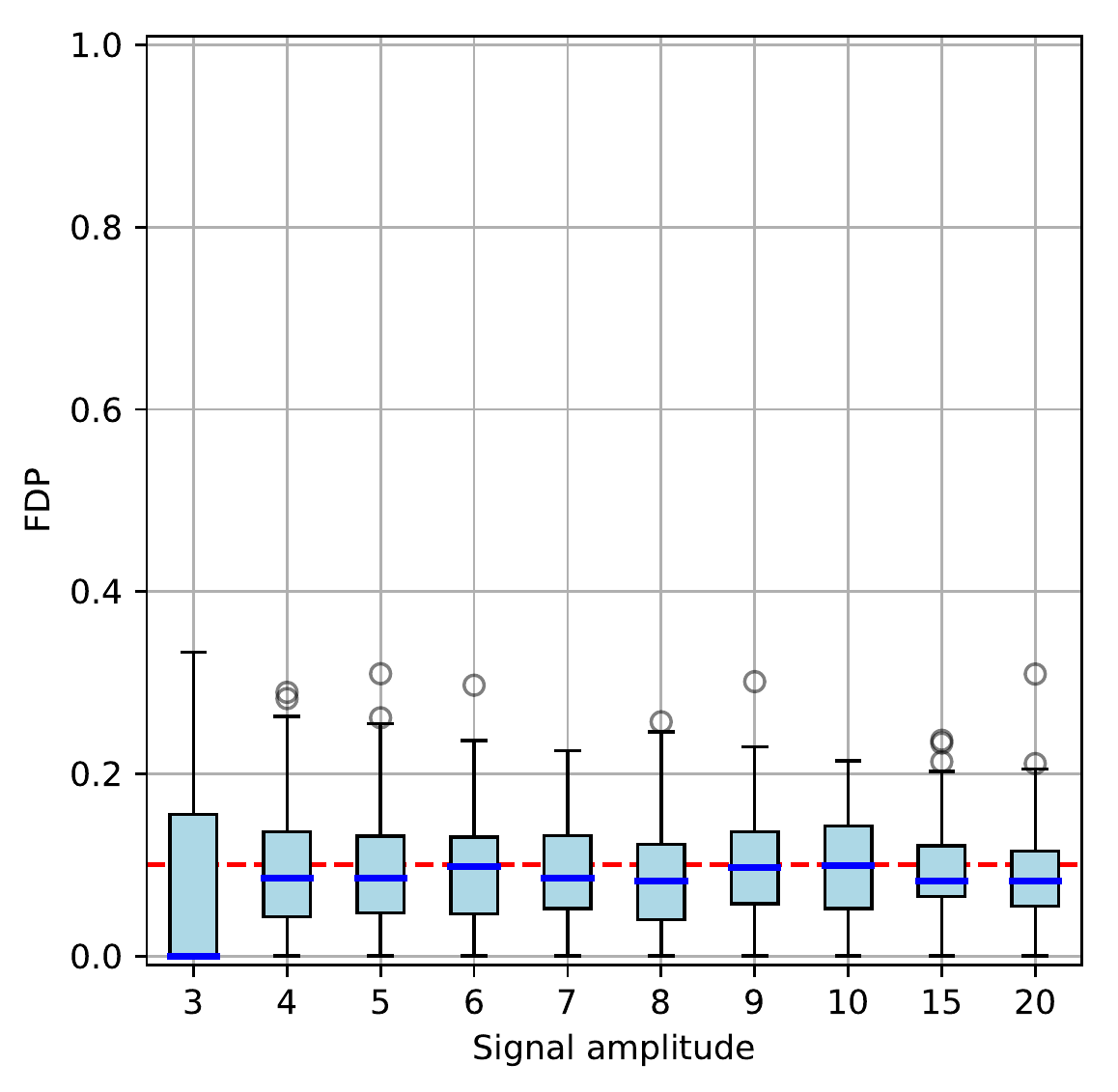}
    \caption{}
  \label{fig:sim_hmm_03_fdr}
  \end{subfigure}
    \caption{Power (a) and FDP (b) of our procedure with simulated HMM covariates, with knockoffs sampled using the HMM parameters fitted by EM on the same dataset. The setup is otherwise the same as that in Figure \ref{fig:sim_hmm_01}.}
  \label{fig:sim_hmm_03}
\end{figure}

In the previous example, we have sampled the knockoff variables by exploiting our knowledge of the true distribution of $X$. Now, we continue as in Section \ref{sec:sim_dmc} to verify the robustness of our procedure to the estimation of $F_X$. Instead of using the exact values of $(q_1, Q, f)$, we fit them on the available data using the Baum-Welch algorithm \cite{rabiner1989}. The power and FDR shown in Figure \ref{fig:sim_hmm_03} and Table \ref{tab:sim_hmm} are estimated over 100 replications, for different values of the signal amplitude. Similarly to the earlier example with Markov chain covariates, our technique behaves robustly and maintains control as expected. \\

\begin{table}[!htbp]
\centering 
\begin{tabular}{ |>{\centering\arraybackslash}m{15mm}||c|c||c|c|}\hline
Signal & \multicolumn{2}{c||}{True $F_X$} & \multicolumn{2}{c|}{Estimated $F_X$}  \\ \cline{2-5}amplitude & FDR (95\% c.i.) & Power (95\% c.i.) & FDR (95\% c.i.) & Power (95\% c.i.) \\ 
\hline
2& 0.037 $\pm \; 0.019$ & 0.030 $\pm \; 0.014$ &0.049 $\pm \; 0.025$ & 0.029 $\pm \; 0.013$ \\ 
3& 0.091 $\pm \; 0.019$ & 0.196 $\pm \; 0.028$ &0.078 $\pm \; 0.019$ & 0.189 $\pm \; 0.033$ \\ 
4& 0.082 $\pm \; 0.013$ & 0.414 $\pm \; 0.030$ &0.094 $\pm \; 0.014$ & 0.432 $\pm \; 0.040$ \\ 
5& 0.102 $\pm \; 0.013$ & 0.610 $\pm \; 0.023$ &0.094 $\pm \; 0.013$ & 0.592 $\pm \; 0.026$ \\ 
6& 0.105 $\pm \; 0.012$ & 0.726 $\pm \; 0.020$ &0.093 $\pm \; 0.011$ & 0.708 $\pm \; 0.022$ \\ 
7& 0.090 $\pm \; 0.012$ & 0.781 $\pm \; 0.017$ &0.093 $\pm \; 0.011$ & 0.790 $\pm \; 0.018$ \\ 
8& 0.093 $\pm \; 0.012$ & 0.830 $\pm \; 0.015$ &0.086 $\pm \; 0.011$ & 0.839 $\pm \; 0.020$ \\ 
9& 0.093 $\pm \; 0.011$ & 0.865 $\pm \; 0.013$ &0.099 $\pm \; 0.010$ & 0.877 $\pm \; 0.012$ \\ 
10& 0.097 $\pm \; 0.009$ & 0.896 $\pm \; 0.011$ &0.099 $\pm \; 0.011$ & 0.898 $\pm \; 0.012$ \\ 
15& 0.083 $\pm \; 0.009$ & 0.945 $\pm \; 0.007$ &0.093 $\pm \; 0.010$ & 0.950 $\pm \; 0.007$ \\ 
20& 0.086 $\pm \; 0.009$ & 0.965 $\pm \; 0.006$ &0.092 $\pm \; 0.010$ & 0.954 $\pm \; 0.007$ \\ 
\hline
\end{tabular}
\caption{FDR and average power in the numerical experiments of Figure \ref{fig:sim_hmm_01} and \ref{fig:sim_hmm_03}. We compare the results obtained with knockoff variables created using the exact (left) and estimated (right) hidden Markov model parameters.} \label{tab:sim_hmm}
\end{table}

\begin{figure}[!htb]
  \centering
  \begin{subfigure}[b]{0.45\textwidth}      \captionsetup{justification=centering}
    \centering
    \includegraphics[width=\textwidth]{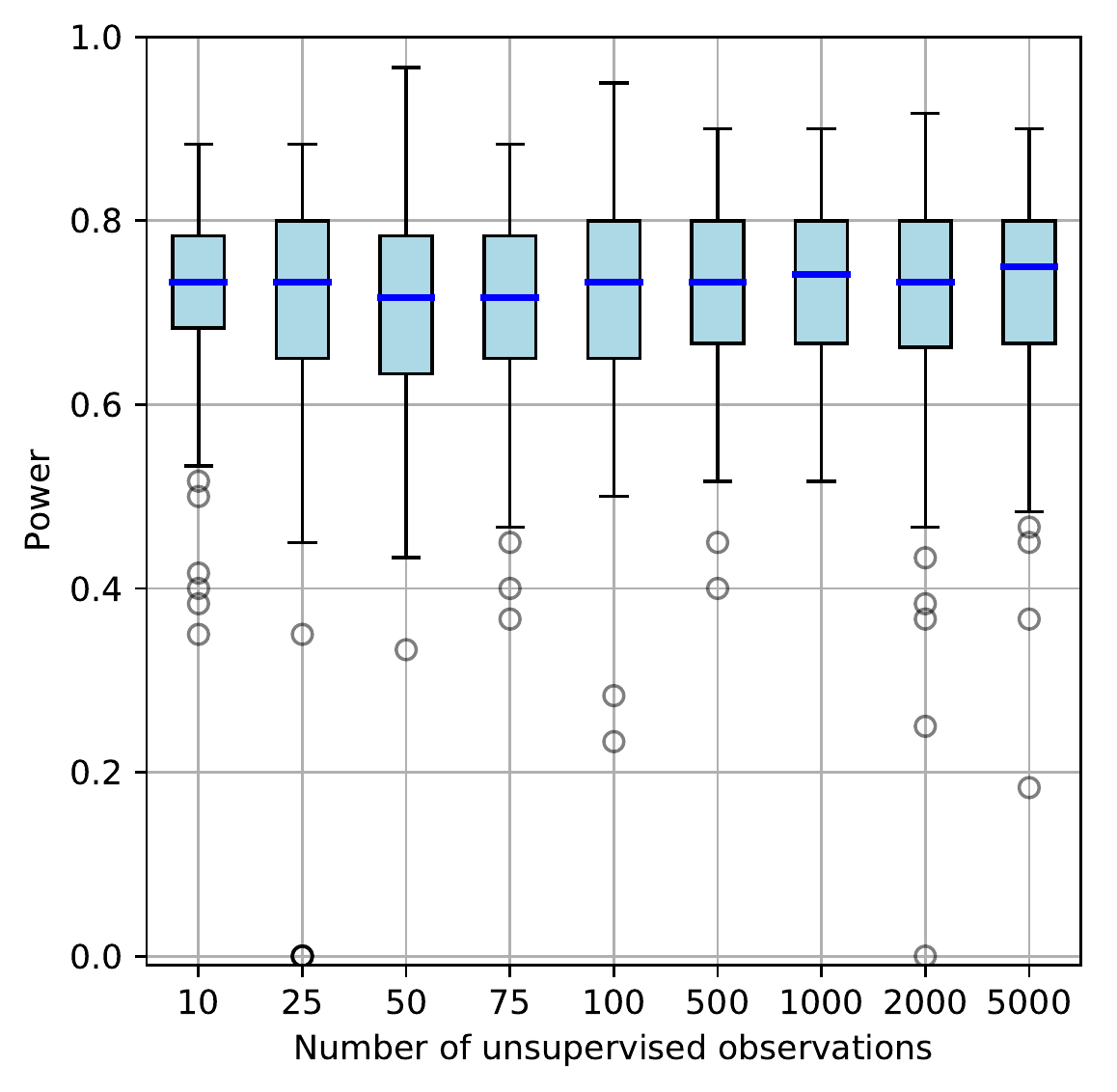}
    \caption{}
    \label{fig:sim_hmm_02_power}
  \end{subfigure}
  \begin{subfigure}[b]{0.45\textwidth}      \captionsetup{justification=centering}
    \centering
    \includegraphics[width=\textwidth]{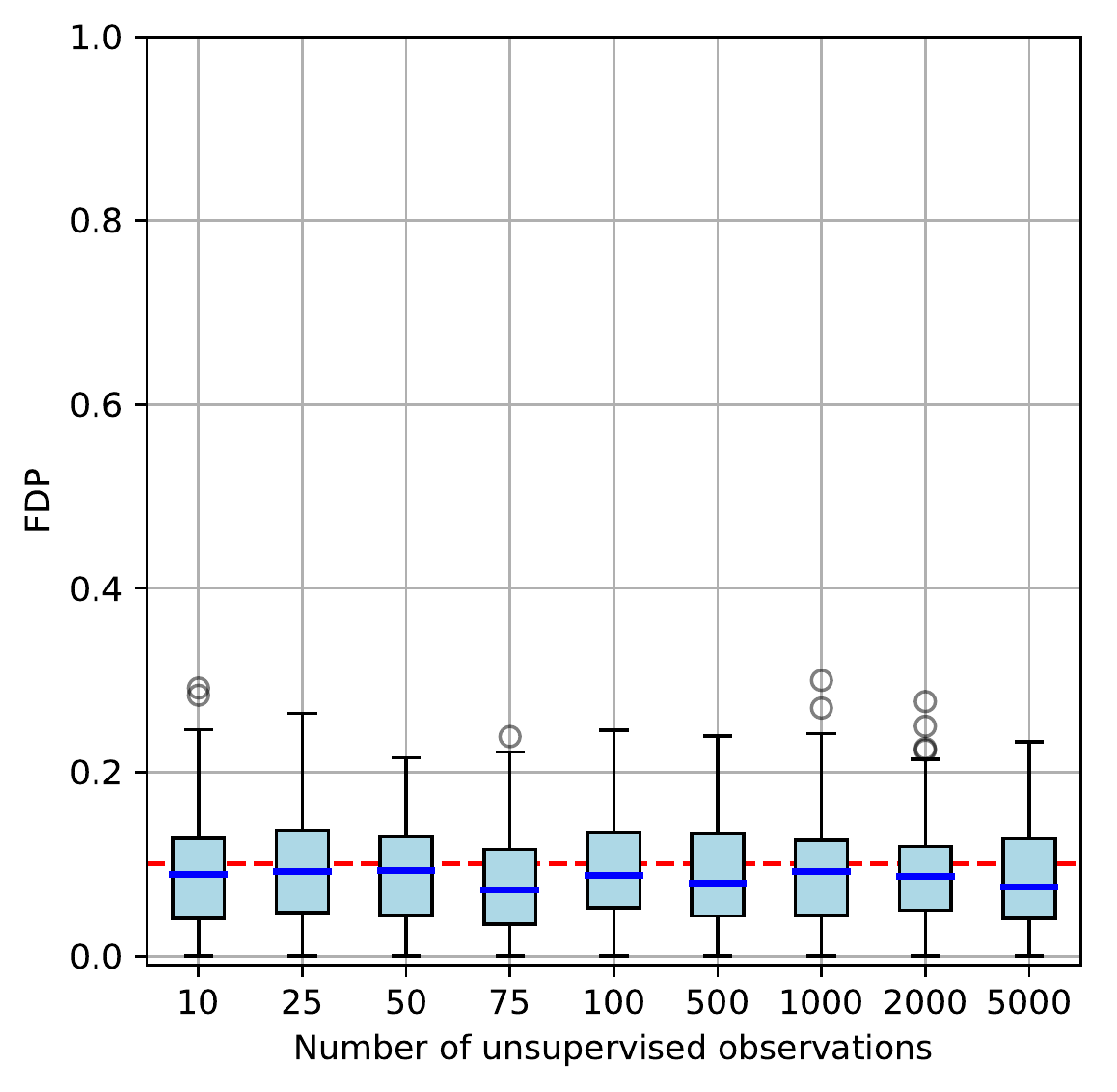}
    \caption{}
  \label{fig:sim_hmm_02_fdr}
  \end{subfigure}
    \caption{Power (a) and FDP (b) of our procedure with simulated HMM covariates. Knockoffs are sampled using parameter estimates obtained with EM from an independent dataset of $n_u$ unlabeled observations of $X$, for different values of $n_u$. The signal amplitude is $a=6$. The setup is otherwise the same as in Figure \ref{fig:sim_hmm_01}.}
  \label{fig:sim_hmm_02}
\end{figure}

Finally, we repeat the experiment by fitting the HMM parameters on an independent and unsupervised dataset of size $n_\text{u}$, for different values of $n_u$. The results are shown in Figure \ref{fig:sim_hmm_02} and they correspond to a range of values for $n_u$ and fixed signal amplitude $a=6$. Again, the FDR is consistently controlled. It should not be suprising that this works even when $n_u$ is as small as 10. Unlike the numerical experiments with the Markov chain variables considered earlier, the transition matrices and emission probabilities for this HMM are homogeneous for all covariates (i.e.~$Q_j = Q_{j+1}$, $\forall j$). This simple model results in fewer parameters to be estimated, thus contributing to the overall robustness.

\FloatBarrier

\subsection{Numerical simulation with real genetic covariates} \label{sec:sim_gwas}
The results in Section \ref{sec:sim_hmm} suggest that our procedure is robust when the HMM parameters of the covariate distribution are estimated from the available data. However, in those cases the true underlying distribution was indeed decided by us to be an HMM. In this section, we verify that the same robustness holds when the covariates consist of real SNPs data collected in the context of a GWAS. \\

We consider 29,258 SNPs on chromosome one, genotyped in  14,708 individuals  by the Wellcome Trust Case Control Consortium \cite{WTCCC2007}. This is the same set of covariates  analyzed in Section 6.1 of \cite{candes2016} and we apply the pre-processing steps described there.
We simulate the response according to a conditional logistic regression model of $Y|X$ with $60$ randomly chosen non-zero coefficients.
Before proceeding to the data analysis with the knockoff framework, we need to prune the SNPs to make sure that there are no pairs of extremely highly correlated variables among the regressors.\footnote{We allow the largest correlation between any two variables to be at most equal to 0.5.} This is needed in order for any model selection method to carry out meaningful distinctions between variables. We  use the approach described in \cite{candes2016}, where a representative is chosen for any cluster of highly correlated SNPs, by selecting the variant among these that is most strongly associated to the phenotype in a hold-out set of 1000 observations (see Section \ref{DataAnalysis} for details).
This leaves us with a total of 5260 variants.
%In order to make this experiment more consistent with our real data analysis, we cluster together highly correlated SNPs and hold out 1000 observations to select group representatives. We thus reduce the effective number of variables to 5260 and remove extremely high correlations. For this, we use the same method of \cite{candes2016}, as summarized with more details in the next section.
Then, we split the rows of $X$ into 10 folds and separately fit the HMM of Section \ref{sec:HMM_SNP} with \texttt{fastPHASE}, using the default configuration and assuming the presence of $K=12$ latent haplotype clusters.
 Once the parameter estimates are obtained, we construct our knockoff variables according to Algorithm \ref{alg:HMM_knock}.\footnote{The 1000 observations used to select the cluster representatives are partially reused in each of the 10 folds, according to the same method described in the next section, without violating the knockoff echangeability property required for FDR control.} 
 With our software implementation, this last step takes approximatively $0.1$ seconds on a single core of an Intel Xeon CPU (2.60GHz) for each individual.\footnote{This gives an idea of the real computational cost of our algorithm to create a knockoff copy of an HMM when $n=1$, $p=5260$ and the effective number of possible latent states is $K_{\text{eff}}=\frac{1}{2} K(K+1) = 78$. The latter expression follows from the parametrization described in Section \ref{sec:HMM_SNP}, which assumes that a genotype is given by the sum of two haplotypes.} We run the knockoffs procedure on each fold by computing the same feature importance measures $W_j = |\hat{\beta}_j(\lambda^{\text{CV}})|-|\hat{\beta}_{j+p}(\lambda^{\text{CV}})|$ as in Section \ref{sec:sim_dmc}, based on regularized logistic regression with $\ell_1$-norm penalty tuned by cross-validation. The selection threshold is chosen as to enforce strict FDR control at level $\alpha=0.1$. 

\begin{figure}[!htb]
  \centering
  \begin{subfigure}[b]{0.45\textwidth}      \captionsetup{justification=centering}
    \centering
    \includegraphics[width=\textwidth]{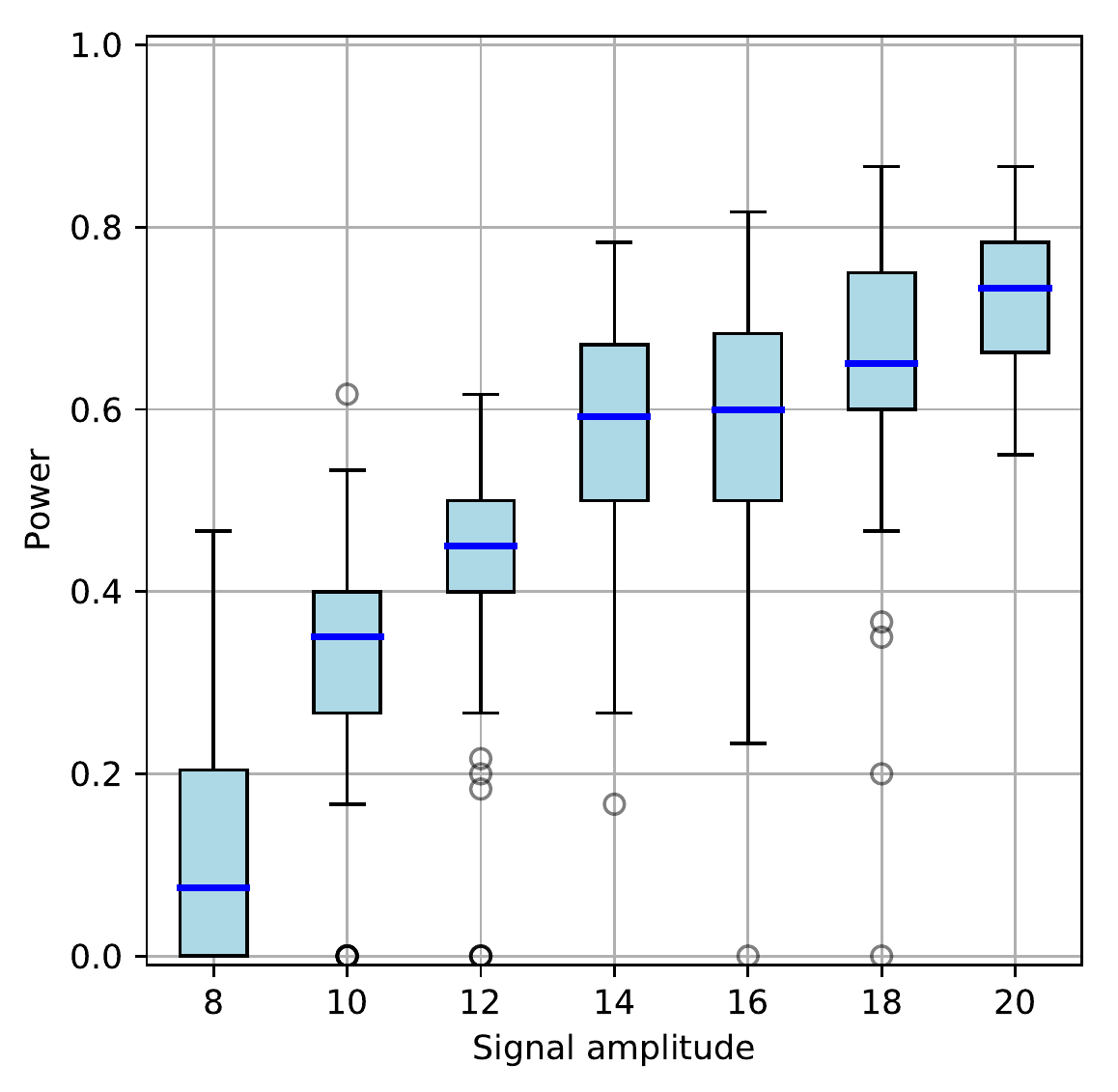}
    \caption{}
    \label{fig:sim_gwas_01_power}
  \end{subfigure}
  \begin{subfigure}[b]{0.45\textwidth}      \captionsetup{justification=centering}
    \centering
    \includegraphics[width=\textwidth]{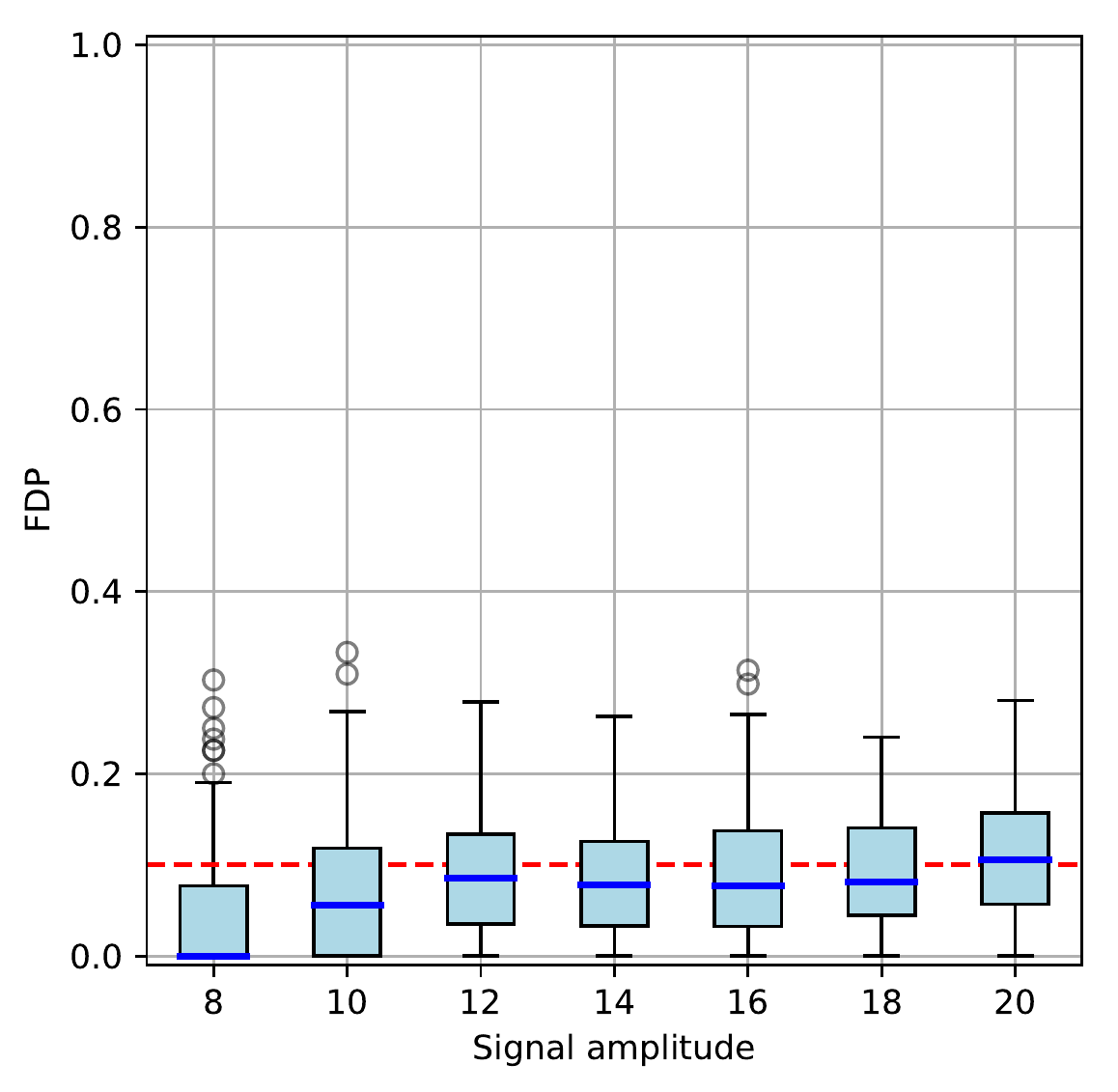}
    \caption{}
  \label{fig:sim_gwas_01_fdr}
  \end{subfigure}
    \caption{Power (a) and FDP (b) of our procedure with real genetic variables. 
One boxplot represents 100 experiments with a total of 10 different logistic regression models for $Y|X$. Apart from the construction of the knockoff variables, this setup is analogous to that of Figure 9 in \cite{candes2016}. The dashed red line in (b) indicates the target FDR level $\alpha=0.1$. }
  \label{fig:sim_gwas_01}
\end{figure}

The power and FDP are estimated by comparing our selections to the exact coefficients in the logistic model. For this purpose, a discovery is considered true if and only if any of the highly correlated SNPs in the selected cluster has a non-zero coefficient. The entire experiment is repeated 10 times, starting with the choice of the logistic regression model. This yields a total of 100 point estimates for the power and FDR of our procedure in the unconditional model. The empirical distribution of these two quantities is shown in Figure \ref{fig:sim_gwas_01} and Table \ref{tab:sim_gwas}, for different values of the signal amplitude. We observe that the FDR is consistently controlled and the FDP is reasonably concentrated. \\

The results of this experiment suggest that we can safely proceed with the analysis of GWAS data. Our confidence partially derives from the fact that our procedure enjoys the rigorous robustness of model-free knockoffs for any conditional distribution of the phenotype. As far as type-I error control is concerned, it does not seem consequential that in this experiment we have chosen to simulate the response from a generalized linear model. In fact, the FDR is provably controlled for any $F_{Y|X}$, provided that $F_X$ is well-specified. Since we have not artificially simulated the covariates, but used instead real genotypes, we can see no reason why our procedure should not similarly control the FDR once applied to GWAS data.

\begin{table}[!htbp]
\centering 
\begin{tabular}{ |c||c|c|}\hline
Signal amplitude & FDR (95\% c.i.) & Power (95\% c.i.) \\ 
\hline
8& 0.040 $\pm \; 0.014$ & 0.121 $\pm \; 0.026$ \\ 
10& 0.075 $\pm \; 0.015$ & 0.321 $\pm \; 0.026$ \\ 
12& 0.089 $\pm \; 0.014$ & 0.425 $\pm \; 0.024$ \\ 
14& 0.086 $\pm \; 0.012$ & 0.571 $\pm \; 0.024$ \\ 
16& 0.093 $\pm \; 0.015$ & 0.586 $\pm \; 0.028$ \\ 
18& 0.091 $\pm \; 0.012$ & 0.651 $\pm \; 0.026$ \\ 
20& 0.112 $\pm \; 0.013$ & 0.718 $\pm \; 0.015$ \\ 
\hline
\end{tabular}
\caption{FDR and average power in the numerical experiments of Figure \ref{fig:sim_gwas_01} with real genetic variables.} \label{tab:sim_gwas}
\end{table}

\section{Applications to GWAS data}

We apply our procedure to data from two GWAS: the Northern Finland 1996 Birth Cohort   study of metabolic syndrome (NFBC) \cite{Sabatti2009} and the Wellcome Trust Case Control Consortium (WTCCC) \cite{WTCCC2007}.

\subsection{Analysis of GWAS data}\label{DataAnalysis}

%\subsection{Genetic analysis of Crohn's disease}
\textbf{Datasets.}  NFBC (dbGaP accession number phs000276.v2.p1) comprises observations on  5402 individuals from northern Finland, including genotypes for $\approx 300,000$ SNPs and nine phenotypes.  We focus on measurements of cholesterol (HDL and LDL), triglyceride levels (TG) and height (HT), as there is a rich literature on their genetic bases that we can rely upon for comparison.
Since not all outcome measurements are available for every subject, the effective values of $n$ are different for each phenotype and a little lower than 5402. 

We analyze the control ($n$=2996) and  Crohn's disease (CD) ($n$= 1917) samples from the WTCCC: all of these are typed at   $p=377,749$ SNPs. \\

\noindent \textbf{Data pre-processing.} We follow the pre-processing steps of \cite{Sabatti2009} and \cite{barber2016} for the NFBC data. This reduces the total number of SNPs to $p=328,934$. Cholesterol and triglycerides levels are log-transformed prior to analysis, and all response variables are regressed on 
 the top five principal components of the genotype matrix to correct for population stratification \cite{Price2006}. The residuals from these regressions define the phenotypes we actually analyze. 
%In the NFBC dataset, the measurements of cholesterol and triglycerides levels are log-transformed. We follow the same pre-processing steps of \cite{Sabatti2009}, and thus reduce the number of SNPs to $p=328,934$. Moreover, we exclude some subjects according to the same criterion as in \cite{barber2016}. Then, we  follow the approach of \cite{barber2016} to correct for population stratification. Their approach is originally inspired by \cite{Price2006} and it consists of regressing the response on the top five principal components of the design matrix.
The WTCCC data does not require additional pre-processing \cite{WTCCC2007}. A summary of both datasets is shown in Table \ref{tab:analysis_dimensions}. \\
% This removes any features that either
% \begin{itemize}
%   \item are not SNPs (e.g. some copy number variations);
%   \item have more than 5\% of the observations missing;
%   \item are constant across all samples;
%   \item correspond to locations that cannot be aligned to the genome;
%   \item are rejected by a $\chi^2$ test for Hardy-Weiberg equilibrium at level 0.01\%;
%   \item are located on chromosome 23.
% \end{itemize}

\noindent \textbf{SNP pruning.}
The presence of very high correlations between neighboring SNPs is a well-known issue in genotype association studies and it can be clearly observed in our data.
 %the design matrices from the two sources that we consider.
  Since many SNPs are very similar to each other, the most compelling scientific question lies in the identification of relevant clusters of tightly linked sites, rather than individual markers. 
  Indeed, the results of the standard GWAS analysis are interpreted as identifying loci (positions in the genome) rather than individual variants---effectively clustering the rejected hypotheses.
%  For this reason, in the canonical approach based on marginal testing, the rejected hypotheses are a posteriori aggregated into groups. However, we have reasons to proceed differently. First, it has been argued in \cite{Brzyski2017} that this approach may be misleading, since FDR control at the individual level does not imply control at the group level. 
However, a na\"ive a-posteriori aggregation of results can inflate the FDR, as the counting of discoveries must be redefined. This issue has been addressed before in special cases \cite{Pacifico2004,Siegmund2011}, but the problem remains that a-posteriori aggregation is intrinsically ill-suited for high-dimensional problems in which the small sample size imposes a limited resolution and makes it fundamentally impossible to distinguish between highly correlated variables. 
% is intrinsically ill-suited for multivariate variable selection techniques, such as ours, while it can easily inflate the proportion of false discoveries in result in inflation. \\
%  . This last point can be easily seen by imagining the extreme case in which two SNPs are identical. In the traditional framework, their marginal regression p-values would also be identical and both variables would be selected and eventually aggregated. However, a multivariate method would find itself in the uncomfortable position of having to arbitrarily choose one of them, and may consequently suffer in terms of power. 
A more natural solution consists of grouping the SNPs a priori, before performing variable selection. By following the steps of \cite{candes2016}, we implement an additional pre-processing phase of single-linkage hierarchical clustering, using the empirical correlations as a similarity measure. The SNP clusters are identified by finding the lowest possible cutoff in the dendrogram such that the highest correlation does not exceed 0.5 within any group.  Then, we spend a randomly selected subset of the observations (i.e.~20\% of the total $n$) to perform marginal t-tests between each variable and the response. The SNP with the smallest p-value in a cluster is chosen as its representative, to be later used with the knockoffs procedure. In both datasets, this process decreases the effective number of variables by a little over $80\%$, as summarized in Table \ref{tab:analysis_dimensions}. \\

It must be remarked that the samples used to identify the cluster
representatives are not wasted as they % The choice of promising
%representatives is beneficial to the power of our main variable
% selection analysis. 
can be partially reused without compromising the rigorous FDR-control
guarantees. As shown in \cite{barber2016,candes2016}, they can be
exploited without violating the exchangeability property
\eqref{eq:knock_cond_2}, provided that the corresponding knockoff
copies are created identical to the original variables. Alone, these
identical knockoffs would not provide any information to distinguish
the relevant variables from the nulls. However, they are useful in
improving the accuracy of the importance measures in the knockoff
statistics computed for the remaining 80\% of the data.
\begin{table}[!htb]
\begin{centering}
\begin{tabular}{ |c|c|c |c |c |}
\hline
Data source & Response  & $n$ & $p$ (pre-clustering) & $p$ (post-clustering) \\
\hline
  NFBC & HDL (quantitative) & 4700 & 328,934 & 59,005 \\  
  NFBC & LDL (quantitative)& 4682 & 328,934 & 59,005 \\
  NFBC & TG (quantitative) & 4644 & 328,934 & 59,005 \\
  NFBC & HT (quantitative) & 5302 & 328,934 & 59,005 \\
  WTCCC & CD (binary) & 4913 & 377,749 & 71,145 \\ 
\hline
\end{tabular}
\caption{Summary of the datasets considered in our analysis. The value of $n$ indicates the number of samples for each response, while the last two columns show the corresponding number of variables before and after clustering. Since clustering was performed on the same empirical correlation matrix for all NFBC traits, the same number of clusters are found. However, the cluster representives may be different because they are selected based on the response.} \label{tab:analysis_dimensions}
\end{centering}
\end{table}\newline

\noindent \textbf{Knockoff construction.} In order to apply Algorithm
\ref{alg:HMM_knock} to construct the knockoff variables, we estimate
the HMM parameters $(\hat{r}, \hat{\alpha}, \hat{\theta})$ of Section
\ref{sec:HMM_SNP} using \texttt{fastPHASE}. We perform this separately
for each of the first 22 chromosomes in the WTCCC and the NFBC
data. Since the estimation of the covariate distribution does not make
use of the response, we only compute one set of estimates for the NFBC
using all of the corresponding SNP sequences. In both cases, we run
\texttt{fastPHASE} with a pre-specified number of latent haplotype
clusters $K=12$. In its default configuration, the imputation software
estimates $\hat{\alpha}$ with the additional constraint that
$\alpha_{j,k}$ can only depend on the first index $j$. For simplicity, we do not modify this setting. \\

\noindent \textbf{Knockoff statistics and filter.}
We compute the variable importance measures as in Section \ref{sec:sim_gwas}, by performing a Lasso regression of $Y$ on the (standardized) knockoff-augmented matrix of covariates $[X, \tilde{X}] \in \{0,1,2\}^{n \times 2p}$, with a regularization parameter $\lambda$ chosen through 10-fold cross-validation. In the case of the Crohn's disease study, in which the response is binary, the Lasso is replaced by logistic regression with an $\ell_1$-norm penalty.
%The symmetrized knockoff statistics are then simply obtained by taking the difference of the absolute values of these coefficients. 
%For $j \in \{1,\ldots,p\}$, they are formally defined as 
%$W_j = |\hat{\beta}_j(\hat{\lambda}^{\text{cv}})| - |\hat{\beta}_{j+p}(\hat{\lambda}^{\text{cv}})|$.
Then, relevant SNPs are selected by applying the knockoff filter with the typical \textit{knockoff} threshold for the target FDR $\alpha=0.1$. \\

\subsection{Results}
\textbf{Selections.} We carried the analysis described above on the
four datasets of Table \ref{tab:analysis_dimensions}. Since the model-free knockoffs method is based on a random sample of $\tilde{X}$, in each case our selections depend on its specific realization. Repeating our procedure multiple times and choosing one $\tilde{X}$ after looking at the results would obviously violate the exchangeability conditions required for FDR control. Therefore, we choose instead to report all findings that are selected at least 10 times over 100 independent repeats of the knockoffs procedure. 
This allows us to provide the reader with both an impression of the variability  and an informal measure of confidence for the selections.  Our findings are summarized in the Appendix. \\

% \ejc{What findings should we truly report? Should we aggregate in some way? Also, why at least 10 times out of 100? 
% \color{red} CHIARA: as far as I know, we do not know how to aggregate. Looking at the number 10\% seems small, but it is by looking at the things that are not identified too often that you see more difference from the marginal tests. \color{black} \ms{I agree with Chiara here. I do not intepret what we are doing as ``reporting a finding'' in the usual sense. We are just showing what is likely to be discovered by someone who runs our procedure only once, since this is all one is really supposed to do.} \\

\noindent\textbf{Evaluation of findings.}
Unfortunately, we do not have enough experimental evidence to assess which of our findings are true or false discoveries. However, we can compare our results to those of studies carried out on much larger samples and consider these as the only available approximation of the truth. For lipids we will rely on \cite{WillerConsortium2013} ($n$=188,577), for height on \cite{Wood2014,Marouli2017}  ($n$= 253,288 and 711,428), and for Crohn's disease on \cite{Franke2010} (22,000 cases and 29,000 controls). Since each of these studies includes a slightly different set of SNPs and our features represent clusters of highly correlated SNPs, some care has to be taken in deciding when the same finding appears in two studies. Each of our SNP clusters spans a genomic locus that can be described by the positions of the first and last SNP. We consider one of our findings to be replicated in the larger study if the latter reported as significant a SNP whose position is within the genomic locus spanned by the cluster of SNPs discovered by our method. Additionally, we highlight clusters that, while not satisfying the definition of ``replicated'' given above, % are not confirmed but whose 
are less then 0.5 Mb away from a SNP reported in the meta-analyses. These are marked by an asterisk in the supplementary tables contained in the Appendix, to indicate that some independent supporting evidence is available. \\

\noindent \textbf{Lipids.} The results for HDL and LDL cholesterol are shown in Supplementary Table \ref{table:res_HDL} and \ref{table:res_LDL}, respectively. In addition to the results in \cite{WillerConsortium2013}, we compare our findings to those in  Sabatti et al. \cite{Sabatti2009}, an analysis of our same data based on marginal tests with a level of $5 \cdot 10^{-7}$.\footnote{The significance threshold adopted in \cite{Sabatti2009} is different from the canonical $5 \cdot 10^{-8}$ of GWAS. It was chosen a-posteriori to approximate the threshold obtained by applying the Benjiamini-Hochberg procedure for FDR control at level $\alpha=0.05$.} On average, our method makes 8 discoveries for HDL and 9.8 for LDL. These numbers can be compared to the 5 and 6 discoveries\footnote{In \cite{Sabatti2009}, several SNPs belonging to the same autosomal locus on chromosome 11 are reported as significant for LDL and a similar issue also occurs with HDL. For the purpose of this comparison, we consider them as one since in our analysis they all belong to the same highly correlated cluster. In contrast, our procedure rarely selects clusters with overlapping physical positions and we do not further aggregate our findings, because we have already pruned the variables so that SNPs in different clusters have correlation smaller than 0.5.} respectively reported in \cite{Sabatti2009}.\footnote{An additional association for LDL is also found in \cite{Sabatti2009} on the X chromosome, which we have not analyzed.}
%Similarly, we look at the results of Barber and Candes \cite{barber2016}, which use the same data and pre-processing steps. They apply an earlier version of the knockoff method that assumes a fixed design matrix and a linear regression model for the response. Despite the fundamentally different assumptions, they set the same nominal level $\alpha=0.1$ for the FDR.\\
%In the case of HDL cholesterol, we find on average 8 SNPs over 100 knockoff resamples. The power thus appears to be twice as large as in Barber and Candes. Of course, one should expect the presence of a few false positives, but the assumptions on which our FDR guarantee rely are not as questionable as a linear regression model. \\
Among our new findings, some SNPs have been confirmed by the meta-analysis in \cite{WillerConsortium2013}, while others can be found in the works of different authors. However, we prefer to avoid an extensive search over the entire  existing literature to avoid selection bias.
%and only compare our results with the two pre-specified papers. 

We discover on average 2.8 SNPs associated to triglycerides. This is less than the 4 variants identified in \cite{Sabatti2009}, but some of our findings are different and one of the additional ones is confirmed by the meta-analysis. \\

%\noindent \textbf{Triglycerides.} 
%%The other lipid trait analyzed in this paper is the level of triglycerides. Similarly to cholesterol, we compare our findings in Table \ref{table:res_TRIG} with the work of \cite{Sabatti2009} on the NFBC and the large meta-analysis of \cite{WillerConsortium2013}. 
%On average, we make 2.8 discoveries. Our power would thus appear slightly lower than that of the latter (they make 4 discoveries), but some of the SNPs that we select are different. \\

\noindent \textbf{Height.} Height is the last trait from the NFBC that we consider. This is known to be a highly polygenic phenotype, with over 700 known variants. However, the effect of each of these variants 
 is very weak and one should not expect to make many discoveries with a dataset as small as ours. 
 We obtain some validation by comparing our findings to the meta-analyses in \cite{Wood2014,Marouli2017}, as shown in Supplementary Table \ref{table:res_height}. % Both are very large meta-analysis based on samples collected from 253,288 and 711,428 individuals, respectively.
% We compare our findings in Supplementary Table \ref{table:res_height}. There, we see that 
 Our method discovers 2 relevant SNP clusters, on average. Since this may appear low at first sight, it should be remarked that to the best of our knowledge no other study has found associations for height using only the NFBC data.\footnote{The longitudinal study in \cite{Sovio2009} has looked for genetic variants associated with height using exclusively the NFBC data. However, none of their reported findings achieves the GWAS significance threshold.} Of the 4 sites that we select at least 10\% of the times, 3 are validated by meta-analysis. The remaining one only appears with frequency equal to 12\% and could not be confirmed. \\

\noindent \textbf{Crohn's disease.} Our findings on the Crohn's disease data are summarized in Supplementary Table \ref{table:res_chrons}, where we compare them to the meta-analysis in \cite{Franke2010} and the original work of the WTCCC \cite{WTCCC2007}. %The latter uses our same data, while the former enjoys about 10 times as much (22,000 cases and 29,000 controls). 
Moreover, we also consider the results of Candes et al. \cite{candes2016}. Their work is the most similar to ours because it uses the same data, pre-processing and clustering method, as well as the overall knockoff methodology. The important distinction is that they construct their knockoff variables differently. Instead of fitting an HMM to the SNP sequences, they assume that the values of the SNPs follow a multivariate normal distribution. Their nominal FDR target $\alpha=0.1$ is the same as ours, and the WTCCC also aims at controlling the Bayesian FDR at approximately the same level. Our method makes 22.8 discoveries on average, versus 18 in \cite{candes2016} and the 9 of the WTCCC. In addition to an apparently higher power in this case, our procedure can in general be expected to enjoy a more principled and safer FDR guarantee. Nowhere have we made the unrealistic assumptions of the WTCCC on the conditional model for the response nor those of \cite{candes2016} on the model for the covariates.

Several of the additional findings that we make have been confirmed in \cite{Franke2010}, as shown in Supplementary Table \ref{table:res_chrons}. Some of the other selected SNPs may be new discoveries. In this sense, it is encouraging to observe that rs11627513, rs4263839 are reported in the meta-analysis of \cite{Liu2015}. The same work also links rs7807268 to the related inflamatory bowel disease, using data from a cohort of 86,682 individuals. \\

\begin{figure}[!htb]
  \centering
  \includegraphics[width=0.7\textwidth]{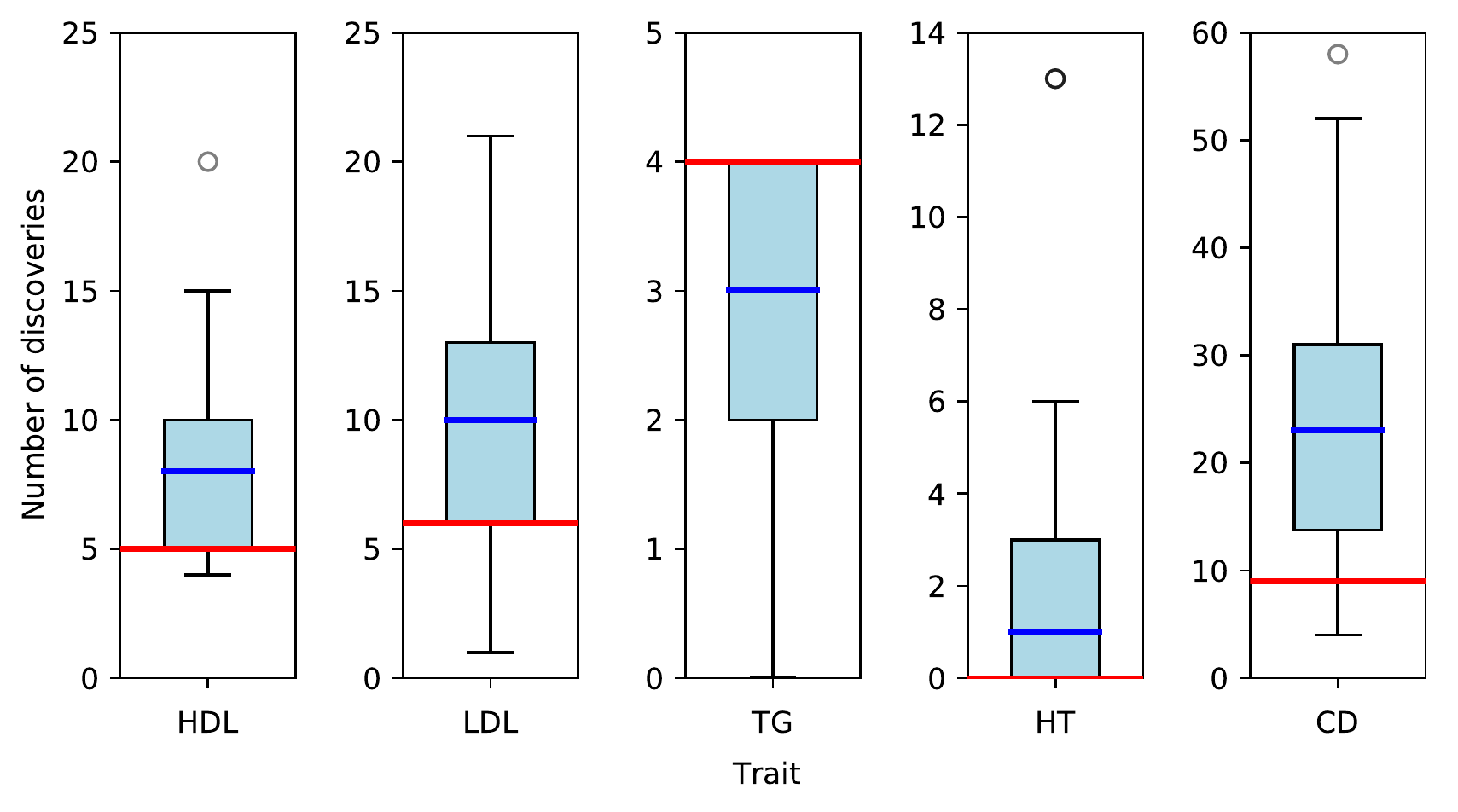}
  \caption{Number of discoveries made on different GWAS datasets. The boxplots refer to our method, for 100 independent realizations of the knockoff variables. The thick red lines indicate the number of discoveries made by the standard genomic analysis of \cite{Sabatti2009} (for HDL, LDL, TG) and \cite{WTCCC2007} (for CD), with the same data.}
  \label{fig:data_results_num}
\end{figure}

\begin{figure}[!htb]
  \centering
  \includegraphics[width=0.7\textwidth]{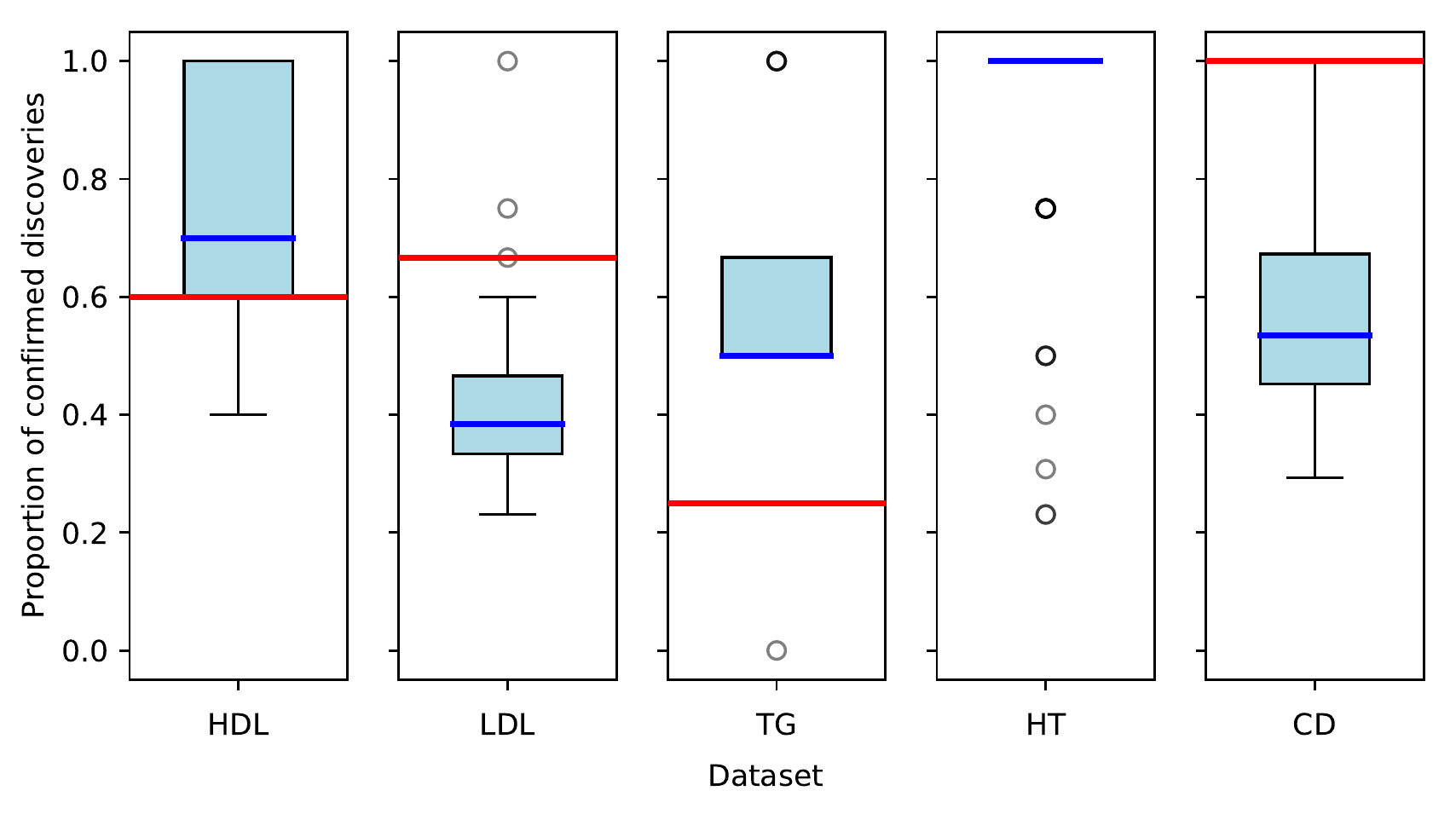}
  \caption{Proportion of the discoveries made with our method that are confirmed by a meta-analysis of \cite{WillerConsortium2013} (HDL, LDL, TG), \cite{Wood2014, Marouli2017} (HT) and \cite{Franke2010} (CD). The boxplots refer to 100 independent realizations of our knockoff variables. The thick red lines corresponds to the results published in the papers that first analyzed our datasets: \cite{Sabatti2009} (HDL, LDL, TG) and \cite{WTCCC2007} (CD).}
  \label{fig:data_results_prop}
\end{figure}

% \ejc{Food for thought: We criticize marginal methods but then compare our findings with these methods. Also, we tend to find the same things.}
% \color{red}
% CHIARA: I think you need a summary of the results over all. I understand you do not want to estimate FDR considering the meta-analysis as ground truth, but you need a take home message. 
% I do not think it is a problem to compare the findings with large meta-analysis: these are based on "silly marginal tests" but have such larger sample sizes that really can be considered closer to the truth than what we can get on the small datasets.
% One think we might want to emphatize is that we are biased towards discovering additive effects, since we analyzed the data with an additive model and we compare the results to those of methods that really can pick up well additive effects only.
% \color{black}
% \ms{Added summary below. Hopefully it addresses the main concerns that have been raised. I have not computed and compared the proportion of variance explained. I'd rather not do that unless you think it's really necessary, because I think it would take some time to do it in such a way that it does not add additional confusion.}\\ 
\noindent\textbf{Summary.} The results of our data analysis show that our procedure identifies a larger number of potentially significant loci than the traditional methods based on marginal testing (except in the case of triglycerides, for which very few findings are obtained with either approach). In Figure \ref{fig:data_results_num}, the distribution of the number of discoveries over 100 independent realizations of our knockoff variables is compared to the corresponding fixed quantity from the standard genomic analysis on the same dataset. We can thus verify that, while model-free knockoffs are intrinsically random, we consistently select more variables. We can expect that many of our new findings are valid, but it is impossible to compute the statistical power or the FDR in a GWAS without having access to the ground truth. We find it nonetheless tempting to look at the proportion of our discoveries that is confirmed by the corresponding meta-analyses. Its distribution is shown in Figure \ref{fig:data_results_prop}, separately for each dataset, and without counting those loci that are only partially confirmed (i.e.~marked by an asterisk in Appendix \ref{sec:tables}). If we were to try to na\"ively estimate the FDR from these plots, we would obtain a value much larger than the target level $\alpha=0.1$. However, such an estimate would be heavily biased and not very meaningful, since none of the meta-analyses is believed to have correctly identified all revelant associations. Instead, some perspective can be gained by comparing our proportion of confirmed discoveries to that obtained with marginal testing on the same data. In the case of HDL cholesterol and triglycerides, we note that our confirmed proportion is appreciably higher, even though one may have intuitively expected a better aggreement between studies relying on the same testing framework.

In general, it should not be surprising that our results are at least partially consistent with those of previous studies. In spite of the fact that our methodology relies on fundamentally different principles, we have selected relevant variables after computing importance measures based on generalized linear regression. The robustness of our type-I error control is completely unaffected by the validity of such model, but a bias towards the discovery of additive linear effects naturally arises. In future studies, one could discover additional associations by easily deploying our procedure with more complex non-linear measures of feature importance. \\

% \subfile{sections/applications/nfbc}

% %\subsubsection{Genetic Analysis of Chron's disease}

% %\subsubsection{Simulations with Genetic Design Matrix}

\section{Discussion}

In this paper, we have shown that one can efficiently generate exact
knockoff copies of a hidden Markov model. This result extends the
applicability of model-free knockoffs beyond the special case of
variables following a multivariate normal distribution. Our
experiments on real and simulated data provide empirical confirmation
of the validity of our entire approach to controlling the selection of
relevant variables.
At this point, we must note that some important questions still remain unanswered, while we bring about new directions for future research.

\begin{itemize}[leftmargin=*]
  \item \textbf{Randomness.} Methods based on model-free knockoffs are intrinsically random. Conditionally on the observed $X$ and $Y$, the selection set depends on the specific realization of the knockoff variables $\tilde{X}$. In the applications described earlier, we have observed that different repetitions of our procedure provide reasonably consistent but different answers on the same data. At this point, it is not clear how to best aggregate the different results.
  \item \textbf{Group selections.} In the presence of extremely high correlatations among the covariates, it is often interesting to ask whether the response depends on a particular group of variables, rather than on each individual one. In our analysis of genetic data, we addressed this point by clustering the variables during the pre-processing phase and restricting the inference to the representatives for each group. Alternatively, one could try to adapt the idea of group-knockoffs in \cite{Dai2016} for our method.
  \item \textbf{HMM parametrization.} We have already mentioned that there exist other forms of HMM that could be adopted for the analysis of genetic data, in addition to that discussed in this paper. Different parametrizations have been developed within the genotype imputation community, and they can be easily exploited by our procedure. For example, if a collection of known haplotypes is available, it is possible to include them in the description of $F_X$ used to generate the knockoff copies. It would be interesting to investigate from an applied perspective the relative advantages of one choice over the other.
\item \textbf{Feature importance measures.} In the simulations and data analysis of this paper we have computed the knockoff statistics using importance measures based on the cross-validated (logistic) Lasso. Therefore, even though our FDR control does not rely on any assumptions of linearity, the power may be negatively affected if the true likelihood is far from linear. In order to fully exploit the flexibility and robustness of model-free knockoffs, it would be interesting to explore the use of alternative statistics that can better capture interactions and non-linearities (e.g.~importance measures based on trees and ensemble methods).
  \item \textbf{Beyond HMMs.} At this point, we know how to perform controlled variable selection with model-free knockoffs in the special cases where the variables can be described by either an HMM or a multivariate normal distribution. Can this be extended to other classes of covariates? For instance, one may want to consider more general graphical models with a higher-dimensional structure.
\end{itemize}

In conclusion, we believe that this work offers a significative development within the model-free knockoff framework and it provides a useful statistical contribution to research in genomics. We have argued that our procedure offers a new powerful and natural way of performing variable selection in GWAS, with rigorous finite-sample control of type-I errors relying solely on mild and principled assumptions. Our numerical examples and the data analysis demonstrate its remarkable advantages over marginal testing, which can only be expected to increase as the sample size of the available datasets grows. In fact, with more data at our disposal, we will be able to more accurately estimate the genotype model parameters used to generate the knockoff copies. Moreover, the higher resolution that comes with more observations will allow us to detect important variables that contribute to the response through non-linearities and interactions, as complex and non-parametric measures of variable importance can be easily included in our procedure.

\subsection*{Acknowledgements}

E.~C.~was partially supported by the Office of Naval Research under
grant N00014-16-1-2712, and by the Math + X Award from the Simons
Foundation. C.~S.~was partially supported by HG006695 and MH101782 and
the Simons Foundation through the Math + X program. We thank Lucas
Janson for inspiring discussions and for sharing his computer code.

\printbibliography

% %% APPENDIX
\section*{Appendices}
\addcontentsline{toc}{section}{Appendices}
\renewcommand{\thesubsection}{\Alph{subsection}}

% TABLE OF RESULTS

\subsection{Model-Free Knockoffs for Discrete Markov Chains}
\begin{proof}[Proof of Proposition \ref{prop:MC_knock}]
  Let us define $Q_{p+1}(k|l)=1$ for all $k,l \in \{1,\ldots,K\}$. We
  proceed by induction, assuming the induction hypothesis that, for some fixed $j \in \{1,\ldots,p-1\}$, the SCIP algorithm samples all knockoff copies $\tilde{X}_{1:j}$ according to \eqref{eq:MC_knock_P}. The main step is to show that $\tilde{X}_{j+1}$ is also sampled according to \eqref{eq:MC_knock_P}.
By construction, the SCIP samples $\tilde{X}_{j+1}$ from
\begin{align*}
  \mathbb{P} \Big[ & X_{j+1} =\tilde{x}_{j+1} \Big| X_{-(j+1)} = x_{-(j+1)}, \tilde{X}_{1:j}=\tilde{x}_{1:j} \Big]
    \\
&    \propto \P{X_{j+1}=\tilde{x}_{j+1}, X_{-(j+1)} = x_{-(j+1)}, \tilde{X}_{1:j}=\tilde{x}_{1:j}} \\
  & \propto \P{X_{j+1}=\tilde{x}_{j+1}, X_{-(j+1)} = x_{-(j+1)}, \tilde{X}_{1:(j-1)}=\tilde{x}_{1:(j-1)}} \\
    & \quad \times \Pc{\tilde{X}_{j}=\tilde{x}_{j}}{X_{j+1}=\tilde{x}_{j+1}, X_{-(j+1)} = x_{-(j+1)}, \tilde{X}_{1:(j-1)}=\tilde{x}_{1:(j-1)}} \\
  & \propto \P{X_{j+1}=\tilde{x}_{j+1}, X_{-(j+1)} = x_{-(j+1)}} \Pc{\tilde{X}_{1:(j-1)}=\tilde{x}_{1:(j-1)}}{X_{j+1}=\tilde{x}_{j+1}, X_{-(j+1)} = x_{-(j+1)}}  \\
     &\qquad \times \Pc{\tilde{X}_{j}=\tilde{x}_{j}}{X_{j+1}=\tilde{x}_{j+1}, X_{-(j+1)} = x_{-(j+1)}, \tilde{X}_{1:(j-1)}=\tilde{x}_{1:(j-1)}}.
\end{align*}
Since we are only interested in the dependence on $\tilde{x}_{j+1}$, the first term above can be simplified as:
\begin{align*}
  \P{X_{j+1}=\tilde{x}_{j+1}, X_{-(j+1)} = x_{-(j+1)}}
  & \propto Q_{j+1}(\tilde{x}_{j+1}|x_{j}) \, Q_{j+2}(x_{j+2}|\tilde{x}_{j+1}).
\end{align*}
From the induction hypothesis it follows that the second term is constant with respect to $\tilde{x}_{j+1}$. This is the case because, according to \eqref{eq:MC_knock_P}, the distribution of $\tilde{X}_i$ only depends on $X_{i-1}, X_{i+1}$ and $\tilde{X}_{i-1}$, for all $i \leq j$. Therefore, the conditional distribution of all $\tilde{X}_{1:(j-1)}$ only depends on $X_{1:j}$.

At this point, we can focus on the third term:
%, while the third one becomes:
\begin{align*}
  \mathbb{P} \Big[ \tilde{X}_{j}=\tilde{x}_{j} & \Big| X_{j+1}=\tilde{x}_{j+1}, X_{-(j+1)} = x_{-(j+1)}, \tilde{X}_{1:(j-1)}=\tilde{x}_{1:(j-1)} \Big] \\
  & = \frac{Q_{j}(\tilde{x}_{j}|x_{j-1}) \, Q_{j}(\tilde{x}_{j}|\tilde{x}_{j-1}) \, Q_{j+1}(\tilde{x}_{j+1}|\tilde{x}_{j})}{\mathcal{N}_{j-1}(\tilde{x}_{j}) \, \mathcal{N}_{j}(\tilde{x}_{j+1})}
  \propto \frac{Q_{j+1}(\tilde{x}_{j+1}|\tilde{x}_{j})}{\mathcal{N}_{j}(\tilde{x}_{j+1})}.
\end{align*}
The equality above follows from the fact that the SCIP algorithm
samples $\tilde{X}_{j}$ independently of $X_{j}$ as in \eqref{eq:MC_knock_P}.
Thus we can conclude that
\begin{align*}
  \Pc{X_{j+1}=\tilde{x}_{j+1}}{X_{-(j+1)} = x_{-(j+1)}, \tilde{X}_{1:j}=\tilde{x}_{1:j}}
  & \propto Q_{j+1}(\tilde{x}_{j+1}|x_{j}) \, Q_{j+2}(x_{j+2}|\tilde{x}_{j+1}) \frac{Q_{j+1}(\tilde{x}_{j+1}|\tilde{x}_{j})}{\mathcal{N}_{j}(\tilde{x}_{j+1})}.
\end{align*}
This proves that the induction hypothesis also holds for $j+1$. The special case $j=1$ remains to be considered. However, this is straightforward since the SCIP algorithm samples $\tilde{X}_1$, independently of $X_1$, from
  \begin{align*}
    \Pc{X_1=\tilde{x}_1}{X_{-1} = x_{-1}}
    & = \Pc{X_1=\tilde{x}_1}{X_{2} = x_{2}} 
    \propto \P{X_1=\tilde{x}_1, X_{2} = x_{2}} 
     = q_1(\tilde{x}_1) Q_2(x_2|\tilde{x}_1).
  \end{align*}

\end{proof}

%\subfile{sections/proofs/app_hmm}

\subsection{Results of data analysis} \label{sec:tables}
We report below the findings of our data analysis performed on the five phenotypes considered in this paper. 
 An asterisk indicates the presence of a confirmed SNP association
 within 0.5 Mb of our discovered cluster. We also compute marginal p-values with the
 standard univariate analysis for all selected SNPs and 
 show the smallest one in each cluster. It must be remarked that our p-values are not identical 
 to those in the original studies, since
 we have made slightly different methodological choices in the pre-processing and pruning
 phases, as detailed in Section \ref{DataAnalysis}. It interesting to look at these p-values because they highlight that many of the marginal signals are weak and could not have been detected by a traditional procedure.

\subsubsection{HDL cholesterol}
\begin{center}
\begin{longtable}{
  |>{\centering\arraybackslash}m{14mm}|
  >{\centering\arraybackslash}m{25mm}|
  >{\centering\arraybackslash}m{6mm}|
  >{\centering\arraybackslash}m{22mm}|
  >{\centering\arraybackslash}m{20mm}|
  >{\centering\arraybackslash}m{20mm}|
  >{\centering\arraybackslash}m{18mm}|}
%\begin{longtable}{ |c|c|c|c|c|c|}
\hline
\shortstack{Selection \\ frequency} & \shortstack{SNP \\ (cluster size)} & Chr. 
  & Position range (Mb) & Confirmed in Willer et al. \cite{WillerConsortium2013} 
  & Found in Sabatti et al. \cite{Sabatti2009} & Marginal p-value \\
%Sel. Frequency & SNP (cluster size) & Chrom. & Pos. Range & Sel. Freq. (C.) & Confirmed \\ 
\hline
100\% & rs1532085 (4) & 15 & 58.68--58.7 & rs1532085 & rs1532085 & $1.33 \cdot 10^{-12}$ \\
\hline
100\% & rs7499892 (1) & 16 & 57.01--57.01 & rs3764261 & rs3764261 & $9.55 \cdot 10^{-17}$  \\
\hline
100\% & rs1800961 (1) & 20 & 43.04--43.04 & rs1800961 &  & $2.84 \cdot 10^{-8}$ \\
\hline
99\% & rs1532624 (2) & 16 & 56.99--57.01 & rs3764261 & rs3764261  & $3.08 \cdot 10^{-34}$ \\
\hline
95\% & rs255049 (142) & 16 & 66.41--69.41 & rs16942887 & rs255049 & $1.76 \cdot 10^{-08}$ \\
\hline
57\% & rs10096633 (19) & 8 & 19.73--19.94 &  &   & $5.33 \cdot 10^{-06}$\\
\hline
55\% & rs9898058 (1) & 17 & 47.82--47.82 &  &   & $1.43 \cdot 10^{-06}$ \\
\hline
51\% & rs17075255 (59) & 5 & 164.28--164.92 &  &  & $1.38 \cdot 10^{-05}$ \\
\hline
43\% & rs3761373 (1) & 21 & 42.87--42.87 &  & & $5.96 \cdot 10^{-06}$ \\
\hline
28\% & rs2575875 (10) & 9 & 107.63--107.68 & rs3905000 & & $1.04 \cdot 10^{-06}$ \\
\hline
23\% & rs12139970 (11) & 1 & 230.35--230.42 & rs4846914 & & $1.21 \cdot 10^{-05}$ \\
\hline
12\% & rs173738 (3) & 5 & 16.71--16.73 &  & & $4.77 \cdot 10^{-06}$ \\
\hline
\caption{SNP clusters found to be associated with HDL cholesterol over 100 repetitions of our procedure. Positions follow the convention of the Human Genome Build 37, as in the original data. The marginal p-values are obtained from standard univariate linear regression.} \label{table:res_HDL}
\end{longtable}
\end{center}

\subsubsection{LDL cholesterol}
\begin{center}
\begin{longtable}{
  |>{\centering\arraybackslash}m{14mm}|
  >{\centering\arraybackslash}m{25mm}|
  >{\centering\arraybackslash}m{6mm}|
  >{\centering\arraybackslash}m{22mm}|
  >{\centering\arraybackslash}m{20mm}|
  >{\centering\arraybackslash}m{20mm}|
  >{\centering\arraybackslash}m{18mm}|}
%\begin{longtable}{ |c|c|c|c|c|c|}
\hline
\shortstack{Selection \\ frequency} & \shortstack{SNP \\ (cluster size)} & Chr. 
  & Position range (Mb) & Confirmed in Willer et al. \cite{WillerConsortium2013} 
  & Found in Sabatti et al. \cite{Sabatti2009} & Marginal p-value \\
%Sel. Frequency & SNP (cluster size) & Chrom. & Pos. Range & Sel. Freq. (C.) & Confirmed \\ 
\hline
99\% & rs4844614 (34) & 1 & 207.3--207.88 & & rs4844614 &  $2.00 \cdot 10^{-9}$ \\
\hline
97\% & rs646776 (5) & 1 & 109.8--109.82 & rs629301 & rs646776&  $2.49 \cdot 10^{-9}$ \\
\hline
97\% & rs2228671 (2) & 19 & 11.2--11.21 & rs6511720 & rs11668477 &  $2.28 \cdot 10^{-9}$ \\
\hline
94\% & rs157580 (4) & 19 & 45.4--45.41 & rs4420638$^*$ & rs157580 & $3.62 \cdot 10^{-8}$ \\
\hline
92\% & rs557435 (21) & 1 & 55.52--55.72 & rs2479409 & &  $1.17 \cdot 10^{-7}$ \\
\hline
80\% & rs10198175 (1) & 2 & 21.13--21.13 & rs1367117$^*$& rs693$^*$&  $5.05 \cdot 10^{-7}$ \\
\hline
76\% & rs10953541 (58) & 7 & 106.48--107.3 & & &  $3.75 \cdot 10^{-6}$ \\
\hline
62\% & rs6575501 (1) & 14 & 95.64--95.64 & & &  $2.32 \cdot 10^{-6}$ \\
\hline
41\% & rs1713222 (45) & 2 & 21.11--21.53 & rs1367117 & rs693 &  $4.99 \cdot 10^{-11}$ \\
\hline
40\% & rs2802955 (1) & 1 & 235.02--235.02 & rs514230$^*$ & &  $2.27 \cdot 10^{-1}$ \\
\hline
37\% & rs17129799 (23) & 11 & 96.85--97 & &&  $4.84 \cdot 10^{-6}$ \\
\hline
36\% & rs174450 (16) & 11 & 61.55--61.68 & rs174546 & rs1535 &  $9.96 \cdot 10^{-7}$ \\
\hline
26\% & rs905502 (1) & 8 & 3.13--3.13 & & &  $1.30 \cdot 10^{-4}$ \\
\hline
25\% & rs9696070 (6) & 9 & 89.21--89.24 & & &  $1.26 \cdot 10^{-5}$ \\
\hline
23\% & rs166152 (19) & 16 & 29.04--29.33 & & &  $4.29 \cdot 10^{-5}$ \\
\hline
19\% & rs12427378 (43) & 12 & 50.43--51.31 & & &  $3.69 \cdot 10^{-6}$ \\
\hline
\caption{SNP clusters found to be associated with LDL cholesterol. Other details as in caption of Table \ref{table:res_HDL}.} \label{table:res_LDL}
\end{longtable}
\end{center}

\subsubsection{Triglycerides}
\begin{center}
\begin{longtable}{
  |>{\centering\arraybackslash}m{14mm}|
  >{\centering\arraybackslash}m{25mm}|
  >{\centering\arraybackslash}m{6mm}|
  >{\centering\arraybackslash}m{22mm}|
  >{\centering\arraybackslash}m{20mm}|
  >{\centering\arraybackslash}m{20mm}|
  >{\centering\arraybackslash}m{18mm}|}
%\begin{longtable}{ |c|c|c|c|c|c|}
\hline
\shortstack{Selection \\ frequency} & \shortstack{SNP \\ (cluster size)} & Chr. 
  & Position range (Mb)& Confirmed in Willer et al. \cite{WillerConsortium2013} 
  & Found in Sabatti et al. \cite{Sabatti2009} & Marginal p-value \\
\hline
94\% & rs10096633 (19) & 8 & 19.73--19.94 & rs12678919 & rs10096633 & $7.47 \cdot 10^{-8}$ \\
\hline
91\% & rs676210 (45) & 2 & 21.11--21.53 & & rs673548  & $2.00 \cdot 10^{-7}$ \\
\hline
62\% & rs2304130 (37) & 19 & 19.28--19.87 & rs10401969 &   & $3.91 \cdot 10^{-6}$ \\
\hline
25\% & rs2907632 (13) & 17 & 52.86--52.95 & &   & $5.69 \cdot 10^{-6}$ \\
\hline
\caption{SNP clusters found to be associated with triglycerides. Other details as in caption of Table \ref{table:res_HDL}.} \label{table:res_TRIG}
\end{longtable}
\end{center}

%%% Local Variables:
%%% mode: latex
%%% TeX-master: "../../main"
%%% End:

\subsubsection{Height}
\begin{center}
\begin{longtable}{
  |>{\centering\arraybackslash}m{14mm}|
  >{\centering\arraybackslash}m{25mm}|
  >{\centering\arraybackslash}m{6mm}|
  >{\centering\arraybackslash}m{22mm}|
  >{\centering\arraybackslash}m{20mm}|
  >{\centering\arraybackslash}m{20mm}|
  >{\centering\arraybackslash}m{18mm}|}
\hline
\shortstack{Selection \\ frequency} & \shortstack{SNP \\ (cluster size)} & Chr. 
  & Position range (Mb)& Confirmed in Wood et al. \cite{Wood2014}
  & Confirmed in Marouli et al. \cite{Marouli2017} & Marginal p-value \\ 
%Sel. Frequency & SNP (cluster size) & Chrom. & Pos. Range & Sel. Freq. (C.) & Confirmed \\ 
\hline
68\% & rs2814982 (120) & 6 & 34.17--35.45 & rs12214804 & rs2814982 & $1.33 \cdot 10^{-7}$ \\
\hline
46\% & rs2882676 (5) & 15 & 89.39--89.4 & & rs2882676 & $2.73 \cdot 10^{-6}$ \\
\hline
31\% & rs6763931 (14) & 3 & 141.04--141.34 & rs724016 & rs724016$^*$ & $4.00 \cdot 10^{-6}$ \\
\hline
12\% & rs10769671 (17) & 11 & 6.19--6.28 & && $6.37 \cdot 10^{-6}$ \\
\hline
\caption{SNP clusters found to be associated with height. Other details as in caption of Table \ref{table:res_HDL}.} \label{table:res_height}
\end{longtable}
\end{center}

%%% Local Variables:
%%% mode: latex
%%% TeX-master: "../../main"
%%% End:

\subsubsection{Crohn's disease}
% P-values with the Cochrana-Armitage trend test
\begin{center}
\begin{longtable}{
  |>{\centering\arraybackslash}m{14mm}|
  >{\centering\arraybackslash}m{26mm}|
  >{\centering\arraybackslash}m{6mm}|
  >{\centering\arraybackslash}m{22mm}|
  >{\centering\arraybackslash}m{19mm}|
  >{\centering\arraybackslash}m{19mm}|
  >{\centering\arraybackslash}m{14mm}|
  >{\centering\arraybackslash}m{16mm}|}
\hline
\shortstack{Selection \\ frequency} & \shortstack{SNP \\ (cluster size)} & Chr. 
& Position range (Mb)& Confirmed in Franke et al. \cite{Franke2010} & Found in WTCCC \cite{WTCCC2007} & Found in Candes et. al \cite{candes2016} & Marginal p-value\\
\hline
100\% & rs11209026 (2) & 1 & 67.31--67.42 & rs11209026 & rs11805303 & 100\%& $2.57 \cdot 10^{-21}$ \\
\hline
99\% & rs6431654 (20) & 2 & 233.94--234.11 & rs3792109 & rs10210302 & 100\%& $1.44 \cdot 10^{-14}$ \\
\hline
98\% & rs6688532 (33) & 1 & 169.4--169.65 &  & rs12037606 & 90\% & $3.48 \cdot 10^{-8}$ \\
\hline
97\% & rs17234657 (1) & 5 & 40.44--40.44 & rs11742570 & rs17234657 & 90\% & $8.06 \cdot 10^{-13}$ \\
\hline
95\% & rs11805303 (16) & 1 & 67.31--67.46 & rs11209026 & rs11805303 & 100\%&  $5.22 \cdot 10^{-14}$ \\
\hline
91\% & rs7095491 (18) & 10 & 101.26--101.32 & rs4409764 & rs10883365 & 100\%&  $2.81 \cdot 10^{-7}$ \\
\hline
91\% & rs3135503 (16) & 16 & 49.28--49.36 & rs2076756 & rs17221417 & 90\%&  $9.55 \cdot 10^{-11}$ \\
\hline
81\% & rs7768538 (1145) & 6 & 25.19--32.91 & rs1799964 & rs9469220 & 60\%&  $5.83 \cdot 10^{-9}$ \\
\hline
80\% & rs6601764 (1) & 10 & 3.85--3.85 &  & rs6601764 & 100\%&  $1.83 \cdot 10^{-8}$ \\
\hline
75\% & rs7655059 (5) & 4 & 89.5--89.53 &  &  & 40\%&  $2.14 \cdot 10^{-7}$ \\
\hline
73\% & rs6500315 (4) & 16 & 49.03--49.07 & rs2076756 & rs17221417 & 60\%&  $5.73 \cdot 10^{-7}$ \\
\hline
72\% & rs2738758 (5) & 20 & 61.71--61.82 & rs4809330 &  & 60\%&  $2.64 \cdot 10^{-6}$ \\
\hline
70\% & rs7726744 (46) & 5 & 40.35--40.71 & rs11742570 & rs17234657 & 50\%&  $7.24 \cdot 10^{-13}$ \\
\hline
68\% & rs11627513 (7) & 14 & 96.61--96.63 &  &  & 80\%&  $6.70 \cdot 10^{-6}$ \\
\hline
66\% & rs4246045 (46) & 5 & 150.07--150.41 & rs7714584 & rs1000113 & 50\%&  $2.00 \cdot 10^{-8}$ \\
\hline
62\% & rs9783122 (234) & 10 & 106.43--107.61 &  &  & 80\%&  $1.69 \cdot 10^{-4}$ \\
\hline
61\% & rs6825958 (3) & 4 & 55.73--55.77 &  &  & 30\%&  $3.54 \cdot 10^{-5}$ \\
\hline
56\% & rs4692386 (1) & 4 & 25.81--25.81 &  &  & 40\%&  $1.31 \cdot 10^{-6}$ \\
\hline
56\% & rs4263839 (23) & 9 & 114.58--114.78 &  &  & 30\%&  $3.16 \cdot 10^{-5}$ \\
\hline
54\% & rs2390248 (13) & 7 & 19.8--19.89 &  &  & 50\%&  $4.53 \cdot 10^{-7}$ \\
\hline
51\% & rs10916631 (14) & 1 & 220.87--221.08 &  &  & 40\%&  $5.41 \cdot 10^{-5}$ \\
\hline
49\% & rs4437159 (4) & 3 & 84.8--84.81 &  &  & 60\%&  $5.42 \cdot 10^{-5}$ \\
\hline
48\% & rs9469615 (2) & 6 & 33.91--33.92 &  &  & 30\%&  $1.13 \cdot 10^{-5}$ \\
\hline
45\% & rs10761659 (53) & 10 & 64.06--64.41 & rs10761659 & rs10761659 & 10\%&  $2.55\cdot 10^{-6}$ \\
\hline
42\% & rs2836753 (5) & 21 & 39.21--39.23 &  &  & 30\%&  $1.43 \cdot 10^{-6}$ \\
\hline
39\% & rs6743984 (23) & 2 & 230.91--231.05 & rs7423615 &  & 10\%&  $3.79 \cdot 10^{-6}$ \\
\hline
38\% & rs2279980 (20) & 5 & 57.95--58.07 &  &  & 10\%&  $1.08 \cdot 10^{-6}$ \\
\hline
35\% & rs7186163 (6) & 16 & 49.2--49.25 & rs2076756 & rs17221417 & 50\%&  $7.29 \cdot 10^{-8}$ \\
\hline
32\% & rs16857006 (1) & 2 & 11.1--11.1 &  &  &&  $2.30 \cdot 10^{-3}$ \\
\hline
30\% & rs7807268 (5) & 7 & 147.65--147.7 &  &  & 10\%&  $2.57 \cdot 10^{-5}$ \\
\hline
27\% & rs4807569 (2) & 19 & 1.07--1.08 & rs740495 &  &&  $2.06 \cdot 10^{-5}$ \\
\hline
24\% & rs3779585 (2) & 7 & 90.36--90.38 &  &  &&  $7.40 \cdot 10^{-6}$ \\
\hline
23\% & rs12529198 (31) & 6 & 5.01--5.1 &  &  &&  $1.08 \cdot 10^{-6}$ \\
\hline
22\% & rs7497036 (19) & 15 & 72.49--72.73 & &  &&  $2.04 \cdot 10^{-4}$ \\
\hline
20\% & rs4959830 (11) & 6 & 3.36--3.41 & rs17309827 &  & 10\%&  $9.47 \cdot 10^{-7}$ \\
\hline
15\% & rs13282050 (8) & 8 & 69.3--69.31 &  &  &&  $3.64 \cdot 10^{-5}$ \\
\hline
15\% & rs1451890 (26) & 15 & 30.92--31.01 & &  &&  $1.23 \cdot 10^{-5}$ \\
\hline
14\% & rs2814036 (5) & 1 & 163.94--164.07 &  &  &&  $9.31 \cdot 10^{-7}$ \\
\hline
14\% & rs7759649 (2) & 6 & 21.57--21.58 & rs6908425$^*$ &  & 40\%&  $1.01 \cdot 10^{-4}$ \\
\hline
14\% & rs4870943 (10) & 8 & 126.59--126.62 & rs4871611 &  &&  $1.46 \cdot 10^{-6}$ \\
\hline
11\% & rs10923347 (1) & 1 & 117.83--117.83 &  &  &&  $9.54 \cdot 10^{-4}$ \\
\hline
10\% & rs4438299 (30) & 16 & 60.01--60.32 &  &  &&  $7.07 \cdot 10^{-5}$ \\
\hline
\caption{SNP clusters found to be associated with Crohn's disease over 100 repetitions of our procedure. Positions follow the convention of the Human Genome Build 35, as in the original data. The marginal p-values are obtained from the Cochran–Armitage test for trend \cite{Armitage1955}.} \label{table:res_chrons}
\end{longtable}
\end{center}

\end{document}